\documentclass[12pt,doublespace]{JHEP3}

\usepackage{amsmath,amssymb,amsfonts,xspace} 

\usepackage{epsfig,supertabular,longtable}
 
\newcommand{\be}{\begin{equation}}
\newcommand{\ee}{\end{equation}}
\newcommand{\beq}{\begin{eqnarray}}
\newcommand{\eeq}{\end{eqnarray}}
\newcommand{\ben}{\begin{eqnarray}\displaystyle}
\newcommand{\een}{\end{eqnarray}}
\newcommand{\bea}[2]{\be\label{#2}\begin{array}{#1}}
\newcommand{\eea}{\end{array}\ee}

\def\Tr{\,{\rm Tr}\, }
\def\det{\,{\rm det}\, }

\def\sign{{\rm sign}}

\def\({\left(}
\def\){\right)}
\def\[{\left[}
\def\]{\right]}
\def\p{\partial}

\def\11{1\!\! 1}

   \def\CF {{\cal F}}

   \def\CI {{\cal I}}

\newcommand{\bZ}{\mathbb{Z}}

\newcommand{\CP}{\mathbb{P}}

\newcommand{\kk}{k}

\newcommand{\de}{\mathrm{d}}

\newcommand{\I}{\mathrm{i}}

\newcommand{\cQ}{\mathcal{Q}}

\newcommand{\cF}{\mathcal{F}}

\newcommand{\cC}{\mathcal{C}}
\newcommand{\cS}{\mathcal{S}}

\newcommand{\cM}{\mathcal{M}}

\newcommand{\cN}{\mathcal{N}}

\newcommand{\cO}{\mathcal{O}}

\newcommand{\IR}{\mathbb{R}}
\newcommand{\IC}{\mathbb{C}}
\newcommand{\IZ}{\mathbb{Z}}

\newcommand{\IP}{\mathbb{P}}

\newcommand{\MM}{{\cal M}}
\newcommand{\NN}{{\cal N}}

\def\varpi{t}

\newcommand{\nn}{\nonumber}
\newcommand{\kahler}{{K\"ahler}\xspace}

\def\bse{\begin{subequations}}
\def\ese{\end{subequations}}

\def\qli2{{\bf E}}

\newcommand{\tot}{\Omega}

\def\ea#1\ea{\begin{align}#1\end{align}}

\usepackage{graphicx}
\usepackage[arrow,curve,matrix,arc]{xy}

\numberwithin{equation}{section}
\newcommand{\one}{S}

\newcommand{\refb}{\eqref}
\newcommand{\wt}{\widetilde}
\newcommand{\wh}{\widehat}
\newcommand{\wtQ}{\widetilde{Q}}
\newcommand{\non}{\nonumber}
\newcommand{\cME}{\cM_{\rm amb}}

\title{From Black Holes to Quivers}

\preprint{arXiv:1207.2230v4\\
Bonn-TH-2012-14\\
CERN-PH-TH/2012-191\\
HRI/ST/1204}

\author{Jan Manschot$^{1,2}$, Boris Pioline$^{3,4}$, Ashoke Sen$^{5}$
\\
$^1$ {\it Bethe Center for Theoretical Physics, Physikalisches Institut, Universit\"at Bonn, \\
53115 Bonn, Germany}\\

$^2$ {\it Max Planck Institute for Mathematics, 53111 Bonn, Germany} \\

$^3$ {\it CERN PH-TH,
Case C01600, CERN, CH-1211 Geneva 23, Switzerland}\\

$^4$ {\it Laboratoire de Physique Th\'eorique et Hautes
Energies, CNRS UMR 7589, \\
Universit\'e Pierre et Marie Curie,
4 place Jussieu, 75252 Paris cedex 05, France} \\

$^5$ Harish-Chandra Research Institute,
Chhatnag
Road, Jhusi, Allahabad 211019, India
\\

\vspace*{2mm} {\tt e-mail: \email{
manschot@uni-bonn.de, pioline@lpthe.jussieu.fr,
sen@hri.res.in}
} \vspace*{-3mm}

}

\abstract{
Middle cohomology states on the Higgs branch of supersymmetric quiver quantum mechanics - also known as pure Higgs states - have recently emerged as possible microscopic candidates for single-centered black hole
micro-states, as they carry zero angular momentum and appear to be robust under 
wall-crossing. 
Using the connection between quiver quantum mechanics on the Coulomb branch
and the quantum mechanics of multi-centered black holes, 
we propose a general algorithm for reconstructing the full moduli-dependent 
cohomology of the moduli space of an arbitrary quiver, in terms of the BPS 
invariants of the pure Higgs states. 
We analyze many examples of quivers with loops, including all cyclic Abelian quivers and several examples with two loops or non-Abelian gauge groups, and provide supporting evidence for this proposal. 
We also develop methods to count pure Higgs states directly.}

\begin{document}

\section{Introduction and Summary}  \label{sintro}

$\NN=2$ supersymmetric string theories are known to possess
supersymmetric configurations involving multiple black 
holes \cite{0005049,0206072,0304094,0702146}. An interesting
problem is to compute the spectrum of BPS states of this
multi-centered configuration. A useful quantity that encodes information
about not only the total number of states but also the angular momentum
$J_3$ carried by the states is the refined `index' $\Omega_{\rm ref}(\gamma;y,z)
\equiv \Tr'(-y)^{2J_3}$. Here the trace is taken over
all states carrying a given  total electric and magnetic charges, collectively
denoted by $\gamma$, after factoring out the trace over the degrees of freedom
associated with the center of mass of the system. 
The parameter $z$ refers to the
dependence on the values of the moduli
at spatial infinity, which is governed by the 
well-known wall-crossing formulaes \cite{0811.2435,1011.1258,1103.0261}. 
For $y=1$, $\Omega_{\rm ref}(\gamma;1,z)$ reduces to 
the usual index $\Tr'(-1)^F$, or equivalently
the second helicity supertrace \cite{9611205,9708062}.\footnote{
While the index $\Tr'(-1)^F$ is protected and depends only on the values of
the vector multiplet scalars $z$ at infinity, the refined index $\Omega_{\rm ref}$ 
is not and may depend on both the vector multiplet (VM) and hypermultiplet (HM) scalars. In particular, it need not be the same at strong and weak coupling, yet it is expected that the
dependence on the VM scalars obeys the motivic wall crossing formula of \cite{0811.2435,1011.1258,1103.0261} in both regimes \cite{1011.1258}. In this 
paper as in \cite{1011.1258,1103.1887}, we  work at fixed value of the HM scalars. In $\cN=2$ gauge theories, a variant of the refined index known as the protected spin character
does exist,  thanks to the existence of a $SU(2)$ R-symmetry \cite{Gaiotto:2010be}.
 We expect that our formulae also apply in this case with $\Omega_{\rm ref}$ replaced by the protected spin character.}

Recently, drawing inspiration from various other related
studies \cite{0807.4556,0906.0011,1011.1258,1103.0261,1102.1729,1107.0723}, 
we proposed a specific formula for the refined index
$\Omega_{\rm ref}(\gamma;y,z)$ carried by a multi-centered black hole
system in terms of the refined index $\Omega^{\rm S}_{\rm ref}(\alpha;y)$
of single centered black holes \cite{1103.1887}. One of the virtues of 
this formula is that it incorporates the full dependence on the asymptotic moduli 
consistently with wall-crossing, leaving only moduli-independent coefficients 
$\Omega^{\rm S}_{\rm ref}(\alpha;y)$ to be determined. On the other hand, at a fixed
point in moduli space,  the formula does not directly 
give any information on $\Omega_{\rm ref}(\gamma;y)$, since the
number of input variables -- the single centered refined index
$\Omega^{\rm S}_{\rm ref}(\alpha;y)$ for each charge vector $\alpha$ -- is equal
to the number of quantities to be computed -- the
total index 
 $\Omega_{\rm ref}(\gamma;y)$ for each charge vector $\gamma$.
 However the formula becomes significant when combined with the
 observation that single centered supersymmetric black holes must carry
 strictly zero angular momentum \cite{0903.1477,1008.4209,1009.3226}
 and therefore $\Omega^{\rm S}_{\rm ref}(\alpha;y)$ must be independent
 of $y$. This drastically reduces the number of input parameters to a single constant for each charge vector $\alpha$, in terms of which the formula of
 \cite{1103.1887}  expresses the refined indices $\Omega_{\rm ref}(\gamma;y,z)$. 
 This gives predictions
 for the $y$ dependence of the 
 refined index of multi-centered black hole configurations in $\NN=2$
 supersymmetric string theories, which could be tested if the same index was
 computable by other means.
 
 Unfortunately at present we do not have an independent way of computing 
 the refined index of general 
 multi-centered black hole configurations in $\NN=2$ supersymmetric string
 theories. However we can construct a subset of these black hole micro-states
 in type II string theory compactified on Calabi-Yau spaces 
 as bound states of elementary D-branes wrapped on various cycles of the internal space.
When the central charges of the constituents nearly align,  
 the dynamics of this system is described by an $\NN=4$ supersymmetric
 quiver quantum 
 mechanics \cite{0206072} containing vector and chiral multiplets characterized by
 a superpotential and a set of Fayet-Iliopoulos (FI) 
 parameters. When the FI parameters are large the vector multiplets can be integrated
 out and the dynamics is described
 by an effective theory for the chiral multiplets. The refined index $\Omega_{\rm ref}(\gamma;y)$ 
 is determined by the Poincar\'e polynomial of the moduli space $\cM$ of classical
 vacua of this effective theory, also known as the Higgs branch, which 
 can often be computed explicitly. 
 
 On the other hand when the FI parameters are small, the dynamics of the theory can be
 described by an effective theory of the vector multiplets, with the chiral multiplets
 integrated out. This effective theory -- known as the Coulomb branch theory --
 turns out to be identical to the quantum mechanics of  
 multi-centered black holes. Naively, if $r$ is the total rank of the gauge 
 group of the quiver quantum mechanics, one would expect the 
 Coulomb branch theory  to describe a system of $r$ black holes (some of
 which could be identical), with each center carrying charge $\alpha$ associated
 to one of the nodes and unit degeneracy $\Omega^S_{\rm ref}(\alpha)=1$. 
 While this is indeed so for quivers without closed loops,  this prescription however fails
 to reproduce the full spectrum computed from the Higgs branch
 in cases where  the moduli space of the Coulomb branch has singularities associated with so called
 scaling solutions, where three or more centers can come arbitrarily close
 to each other \cite{0702146,1205.5023}. In such cases,
 the general formula proposed in \cite{1103.1887} allows 
 to compensate for this failure by adding to the
 naive Coulomb branch result contributions from multi-black hole configurations
 with less than $r$ centers, with the new centers carrying composite charges.
 The contribution from these new configurations are parametrized by the
 single centered black hole indices $\Omega^{\rm S}_{\rm ref}(\alpha)$
 carried by the new centers. This general formula can then be compared 
 with the Poincar\'e polynomial of the Higgs
 branch. Again, since $\Omega_{\rm ref}(\gamma;y)$ is not 
 protected,  there is {\it a priori}  no guarantee
 that the Higgs and the Coulomb branch results for $\Omega_{\rm ref}(\gamma;y)$ should
 agree. 
 Nevertheless our analysis of several examples shows that the two results do agree,
 indicating that this quantity is more robust than what naive arguments based on
 supersymmetry would suggest.
 
{}From the description given above it is clear that in order to be able to carry
out the computation of $\Omega_{\rm ref}(\gamma;y)$ on the Coulomb branch,
there should exist a notion of `single-centered  micro-states', 
 which carry zero angular momentum and whose 
 existence is  independent of the moduli at  infinity, such that their refined index  
 $\Omega^{\rm S}_{\rm ref}(\alpha)$ be independent of both 
 $y$ and $z$. Recent work indicates that this role may be taken by 
 a subset of micro-states described 
 by the middle cohomology\footnote{For brevity, we use  the phrase `middle cohomology' 
 to refer to the part of $H^{d}(\cM,\IZ)$ which is invariant under $SU(2)$ Lefschetz rotations, where $d$ is the complex dimension of $\cM$.} of the Higgs branch \cite{1205.5023,1205.6511}. Indeed, such states are invariant under the Lefschetz $SU(2)$ action on the total cohomology $H^*(\cM,\IZ)$, which realizes spatial rotations in real space. Moreover, these states appear to be robust  under deformation of the superpotential and under wall-crossing,  unlike the rest of the cohomology  which jumps across walls of marginal stability. 
 In fact, it has been observed,  in the special case of
 the three-node quiver with a loop, that the complement of the middle cohomology is 
 in one-to-one correspondence\footnote{For quivers without loops, such that all charge vectors lie
 in a two-dimensional plane, this equivalence was proven at the level of refined indices in 
 \cite{1112.2515}. We believe that the assumption that all charge vectors lie
 in a two-dimensional plane could be relaxed.}
  with states on the Coulomb branch of the quiver quantum mechanics. In 
  contrast a subset of the
  the middle cohomology states have no counterpart on the Coulomb
 side, hence deserving the name of `pure Higgs' or `intrinsic Higgs' states \cite{1205.5023,1205.6511}. The only way to incorporate these states in
 the Coulomb branch analysis is to add their contribution by hand as
 the contribution from single-centered black holes, thereby forcing us to
 identify the micro-states of single centered black holes with pure
 Higgs states. In this paper, we shall give evidence that a
 generalized version of these properties
 continues to hold in a large class of quivers, including quivers with more than one loop
 or non-Abelian gauge groups.

For the reader's convenience we shall now summarize our prediction for the
Poincar\'e polynomial of quiver moduli spaces.
Let us consider a quiver with $K$ nodes, carrying $U(N_1)\times U(N_2)\times
\cdots U(N_K)$ gauge group, and a number $\gamma_{\ell\kk}$ 
 of $(N_\ell,  \bar N_{\kk})$ representation of $U(N_\ell)\times
U(N_{\kk})$.  A negative $\gamma_{\ell\kk}$ indicates $-\gamma_{\ell\kk}
=\gamma_{\kk\ell}$ number of $(\bar N_\ell, N_{\kk})$ representation of
$U(N_\ell)\times U(N_{\kk})$.
Also let $c_\ell$ be the Fayet-Iliopoulos (FI) parameter associated with the Abelian
factor at the $\ell$-th node, subject to the condition $\sum_\ell N_\ell c_\ell=0$.
Such a quiver is pictorially represented by $\gamma_{\ell\kk}$
arrows connecting the node $\ell$ to the node $\kk$, with the arrows being
directed from $\ell$ to $\kk$ if $\gamma_{\ell\kk}$ is positive.
The construction of the quiver moduli space begins by introducing a set of complex
variables $\phi_{\ell\kk, \alpha, ss'}$ for every pair $\ell,\kk$ for which 
$\gamma_{\ell\kk}>0$. Here $\alpha$ runs over $\gamma_{\ell\kk}$ values,
$s$ is an index labelling the
fundamental representation of $U(N_\ell)$ and $s'$ is an index representing
the anti-fundamental representation of $U(N_{\kk})$. The moduli space of
classical vacua is the space spanned by these variables $\{ \phi_{\ell\kk, \alpha, ss'}\}$
subject to the following D-term and F-term constraints:
\ben \label{emodi1}
&& \sum_{\kk , s,t,s'\atop \gamma_{\ell\kk}>0} \phi_{\ell\kk, \alpha, ss'}^* \, T^a_{st} \, 
\phi_{\ell\kk,\alpha,t s'} - \sum_{\kk ,s,t,s'\atop \gamma_{\kk\ell}>0} 
\phi_{\kk\ell, \alpha, s's}^* 
\, T^a_{st} \, 
\phi_{\kk\ell,\alpha,s't} = c_\ell \, \Tr(T^a)\quad \forall \, \ell, \, a\, , \nonumber \\
&& {\p W\over \p \phi_{\ell\kk,\alpha,ss'}}=0\, .
\een
Here $T^a$'s are the generators of the $U(N_\ell)$ gauge group, and $W$ is a
gauge invariant superpotential holomorphic in the variables
$ \phi_{\ell\kk, \alpha, ss'}$. For every closed loop in the quiver we can
construct gauge invariant polynomials by taking the products
of $ \phi_{\ell\kk, \alpha, ss'}$ along the closed loop and the superpotential
$W$ is an arbitrary linear combinations of such gauge
invariant polynomials. Besides the constraints given in \eqref{emodi1},
the variables $\{ \phi_{\ell\kk, \alpha, ss'}\}$ are also subject to 
identification under the $\prod_\ell U(N_\ell)$ gauge transformations.
The resulting manifold $\cM$, which we refer to as the quiver moduli space, 
describes the space of classical vacua on the Higgs
branch of the supersymmetric quantum mechanics. 
The associated  refined index $\Tr'(-1)^{2J_3}$
is given by the Laurent polynomial
\be \label{epol}  
Q(\cM; y) = (-y)^{-d} P(\cM;-y) = \sum_{p=1}^{2d} b_p(\cM)\, (-y)^{p-d}
\ee
where $d$ is the complex 
dimension of $\cM$, the $b_p$'s are its topological Betti numbers,
and $P(\cM;t)=\sum_{p=1}^{2d} b_p(\cM)\, t^p$ is the Poincar\'e polynomial.
The analysis of this paper gives an algorithm for computing $Q(\cM;y)$  
as follows.

We first assign to each node $\ell$ a basis vector $\gamma_\ell=(0,\dots, 1,0,\dots)$ in an
abstract vector space $\IZ^K$, and introduce a 
symplectic inner product\footnote{The quadratic form $(\gamma,\gamma')$ and 
vector $\beta$ are known in the mathematical literature on quivers as the Euler form and dimension vectors. We consider the dimension vector and FI terms to be part of the data of the quiver, in deviance from mathematical practice.}
\be
\label{defEulerForm}
\langle \gamma, \gamma'\rangle\equiv(\gamma',\gamma)-(\gamma,\gamma')\ ,\quad
(\gamma,\gamma')\equiv \sum_{\ell=1}^K n_\ell n'_\ell-\sum_{\ell, \kk=1\atop
\gamma_{\ell\kk}>0 }^K n_\ell n'_\kk \gamma_{\ell\kk},
\ee
between the elements $\beta=\sum_{\ell=1}^K n_\ell \gamma_\ell$.
It follows that for the
basis vectors, $\langle \gamma_\ell, \gamma_{\kk}\rangle=\gamma_{\ell\kk}$. 
We denote by $\Gamma\subset \IZ^K$ the collection of vectors
$\beta=\sum_{\ell=1}^K n_\ell \gamma_\ell$ where $n_\ell$
are non-negative integers, and by $\cC_\beta$ 
the hyperplane $\sum_{\ell=1}^K n_\ell c_\ell=0$
in the space of real vectors $c=\sum_{\ell=1}^K c_\ell \gamma_\ell \in \IR^K$. 
To any vectors $\beta\in\Gamma$ and $c\in\cC_\beta$, we associate a quiver $\cQ(\beta,c)$
with $K$ nodes, $\gamma_{\ell\kk}$
arrows connecting the node $\ell$ to the node $\kk$, gauge group
 $U(n_1)\times U(n_2)\times \cdots U(n_K)$, and FI parameters $\{c_1,\cdots c_K\}$.
If some of the $n_\ell$'s vanish we just drop the corresponding nodes.

This construction produces a family of quivers which contains the original quiver $\cQ(\gamma,c)$
with $\gamma=\sum_\ell N_\ell \gamma_\ell$
as a special case.  Let 
$Q(\gamma;y)$ be the corresponding Laurent polynomial
introduced in \eqref{epol}.
Our conjectured
formula for $Q(\gamma; y)$, which we shall often refer to as the
Coulomb branch formula, takes the form:
\ben \label{essp1}
Q(\gamma;y) &=& \sum_{m|\gamma} 
\mu(m) m^{-1} {y - y^{-1}\over y^m - y^{-m}}
\bar Q(\gamma/m;y^m) \nonumber \\
\bar Q(\gamma;y) &=& 
\sum_{n\ge 1}\sum_{\{\alpha_i\in \Gamma\},\, \sum_{i=1}^n \alpha_i =\gamma}
{1\over {\rm Aut}(\{\alpha_1, \alpha_2,\cdots, \alpha_n\})} 
 g_{\rm ref}\left(\alpha_1, \alpha_2,\cdots, \alpha_n;y\right) \nonumber \\ &&
\prod_{i=1}^n \left\{\sum_{m_i\in\bZ\atop m_i|\alpha_i}
{1\over m_i} {y - y^{-1}\over y^{m_i} - y^{-m_i}}\, 
\left( \Omega^{\one}_{\rm ref}(\alpha_i/m_i;y^{m_i})
+ \Omega_{\rm scaling}(\alpha_i/m_i;y^{m_i})\right)\right\}
\, ,
\een
where $\mu(m)$ is the M\"obius function, 
${\rm Aut}(\{\alpha_1,\cdots \alpha_n\})$ is given by
$\prod_k s_k!$ if among the set $\{\alpha_i\}$ there are
$s_1$ identical vectors $\tilde \alpha_1$, $s_2$ identical vectors
$\tilde\alpha_2$ etc., and $m|\alpha$ means that $m$ is a common divisor of
$(n_1,\cdots , n_K)$ if $\alpha =\sum_\ell n_\ell \gamma_\ell$. The factor
$g_{\rm ref}(\alpha_1, \alpha_2,\cdots, \alpha_n;y)$, which we shall call the
`Coulomb index', 
is (the bulk contribution to) the refined index of the quantum mechanics
of $n$ charged particles. It is
equal to 1 for $n=1$ and 
\ben
 \label{epclassint}
g_{\rm ref} (\alpha_1, \dots , \alpha_n;y)
&=& (-1)^{\sum_{i<j} \alpha_{ij} +n-1}
 \left[
(y-y^{-1})^{1-n} \, \sum_{p} \,  
s(p)\, 
y^{\sum_{i<j} \alpha_{ij}\, \sign[x_j-x_i]}\right] \, \,  \hbox{for $n\ge 2$}\non\\
\quad \alpha_{ij} &\equiv& \langle \alpha_i, \alpha_j\rangle \ .
\een
Here the sum over $p$ runs over all solutions to the system of $n-1$ 
independent algebraic equations
in $n-1$ unknowns $x_2,\dots x_n$,
\be \label{epwextrema}
 \sum_{j=1\atop j\ne i}^n {\alpha_{ij}
\over |x_i-x_j|} = \hat c_i\ , 
\quad x_i\in {\mathbb R}\, , \quad \hbox{for $1\le i\le n$}\, ,
\ee
with $x_1$ fixed to any value, and  $s(p)=\pm 1$ is a certain sign given in
\refb{espnew}.
The prescriptions for enumerating the solutions to \eqref{epwextrema} 
are detailed in \S\ref{strategy}. 
The coefficients
$\hat c_i$ are determined in terms of the FI parameters $c_i$ by $\hat c_i
=\sum_\ell A_{i\ell} c_\ell$ whenever  $\alpha_i=\sum_\ell A_{i\ell}\gamma_\ell$. 
From the restriction $\sum_i \alpha_i
=\gamma$ and that $\sum_\ell N_\ell c_\ell=0$ it follows that 
$\sum_i \hat c_i=0$, as required for the consistency of the equations \eqref{epwextrema}.

The symbols $\Omega^{\rm S}_{\rm ref}$'s appearing 
in \eqref{essp1} represent the refined indices of single centered 
micro-states
and are given as follows. 
First of all we have $\Omega^{\rm S}_{\rm ref}(\gamma_\ell;y)=1$
for $1\le\ell\le K$. For any other $\beta\in \Gamma$,
$\Omega^{\rm S}_{\rm ref}(\beta;y)$
is an unknown {\it $y$-independent} 
constant. For  reason that will become clear soon we shall 
refrain from setting $\Omega^{\rm S}_{\rm ref}(\beta;y)=\Omega^{\rm S}_{\rm ref}(\beta)$
until we determine the
functions $\Omega_{\rm scaling}$. The latter are expressed
recursively in terms of $\Omega^{\rm S}_{\rm ref}$ through
\be \label{essp2}
\Omega_{\rm scaling}(\alpha;y) =
\sum_{\{\beta_i\in \Gamma\}, \{m_i\in\bZ\}\atop
m_i\ge 1, \, \sum_i m_i\beta_i =\alpha}
H(\{\beta_i\}; \{m_i\};y) \, \prod_i 
\Omega^{\one}_{\rm ref}(\beta_i;y^{m_i})
\, ,
\ee
where the functions $H(\{\beta_i\}; \{k_i\};y)$ are
determined as follows. Firstly, when the number of
$\beta_i$'s is less that three,  $H(\{\beta_i\}; \{k_i\};y)$ vanishes.
For three or more number of $\beta_i$'s, 
we note that  the expression for $Q(\sum_i k_i\beta_i;y)$ given in
\eqref{essp1} contains a term proportional to
$H(\{\beta_i\}; \{k_i\};y)\prod_i\Omega^{\rm S}_{\rm ref}(\beta_i;y^{k_i})$
arising from the choice $m=1$ in the first equation in
\eqref{essp1}, $n=1$,
$\alpha_1=\sum_i k_i\beta_i$, $m_1=1$
in the second equation in \eqref{essp1}, and 
$m_i=k_i$ in the expression for
$\Omega_{\rm scaling} (\sum_i k_i \beta_i; y)$ in eq.\eqref{essp2}.
We fix $H(\{\beta_i\}; \{k_i\};y)$ by demanding that the net
coefficient of the $\prod_i \Omega^{\rm S}_{\rm ref}(\beta_i;y^{k_i})$ in the
expression for $Q(\sum_i k_i\beta_i; y)$ is a
Laurent polynomial in $y$. This of course leaves open the possibility  of
adding to $H$ a
Laurent polynomial. This is resolved by using the minimal
modification hypothesis, which requires that $H$ must 
be symmetric under $y\to y^{-1}$ and vanish
as $y\to\infty$  \cite{1103.1887}.
We determine $H(\{\beta_i\}; \{m_i\};y)$ iteratively by
beginning with the $H$'s with three $\beta_i$'s and then 
determining successively the $H$'s with more $\beta_i$'s. 
$\Omega^{\rm S}$ and $H$ are expected 
to be independent of the
FI parameters and hence can be calculated for any
value of these parameters.
{}From the algorithm for determining $H$ described above it is clear that
one should retain the $y$-dependence of $\Omega^{\rm S}_{\rm ref}$ at intermediate stages 
to distinguish between
$\Omega^{\rm S}_{\rm ref}(\beta;y^m)$ for different values of $m$. For Abelian quivers
this is not important since only $\Omega^{\rm S}_{\rm ref}(\beta;y)$ appear in the
final expression and hence in this case we can set 
$\Omega^{\rm S}_{\rm ref}(\beta;y)=\Omega^{\rm S}_{\rm ref}(\beta)$ 
from the outset.

The Coulomb branch formula \eqref{essp1}-\eqref{essp2}
 gives an explicit algorithm for computing the Poincar\'e polynomial
of the quivers $\cQ(\beta)$ for all dimension vectors $\beta\in\Gamma$ in terms
of the constants $\Omega^{\rm S}_{\rm ref}(\beta)$. There is one such undetermined
constant for each  $\beta\in\Gamma$, representing  the number of `pure Higgs states' 
which cannot be determined by our algorithm and must be computed by other
methods.\footnote{If the quiver $\cQ(\beta)$ has no oriented closed loop then
$\Omega^{\rm S}_{\rm ref}(\beta)$ as well as $\Omega_{\rm scaling}(\beta;y)$
are expected to vanish \cite{0702146}.} It is also worth stressing 
that the Coulomb branch formula 
automatically satisfies the Kontsevich-Soibelman wall crossing 
formula \cite{0811.2435}. This follows from the result of \cite{1103.1887}
that the formula for the index given there satisfies the wall crossing formula
of \cite{1011.1258}, and the result of \cite{1112.2515} showing the equivalence
of the wall crossing formul\ae\ of \cite{0811.2435} and \cite{1011.1258}.

As is clear from the above, the main assumption in our algorithm is that there exists
a class of `single-centered black hole micro-states' which have the property
that their index $\Omega^{\rm S}_{\rm ref}(\beta)$ is independent of $y$ and
robust under wall-crossing. 
For single centered black holes the $y$-independence of the index
follows from the fact that the black hole carries 
zero angular momentum. This is 
in turn a consequence of spherical symmetry of a supersymmetric black hole
together with the fact that an extremal black hole represents a collection of
states in the microcanonical ensemble where all charges and angular
momenta are fixed\cite{0903.1477}. For the quiver the role of single centered black hole
states is played by `pure Higgs states' -- states which are visible on the
Higgs branch but not on the Coulomb branch of the supersymmetric quantum 
mechanics \cite{0702146,1205.5023,1205.6511}.
Since the quiver description is valid in a different
region in the space of coupling constants, and since the $y$ dependence of
the index is not guaranteed to be protected under a change of coupling, one
might expect that for the quiver $\Omega^{\rm S}$ may be $y$-dependent.
Nevertheless the recent studies in \cite{1205.5023,1205.6511} indicate 
that even for the quiver the $y$
independence of $\Omega^{\rm S}$ holds, which we therefore take as our
working hypothesis. Needless to say, we find that this hypothesis holds in 
all the examples that we have analyzed.

The remainder of this work is organized as follows. In \S\ref{strategy} 
we review the general formula of \cite{1103.1887}
expressing the refined index of a multi-centered black hole system in terms of the
index carried by single centered black holes. We then show how a microscopic
version of this formula can be used to compute the Betti numbers of quiver moduli spaces. 
We also suggest an extension of this formula for  computing the  Hodge numbers. 
We then review other methods for computing the  cohomology of the Higgs branch directly,
using the Lefschetz hyperplane theorem, Riemann-Roch theorem and Harder-Narasimhan recursion method. 
In sections \ref{s3node}, \ref{scyclic}, \ref{sclosed}
and \ref{snonabelian} we apply our general methods to compute the
cohomology of different kinds of quivers, and compare the results with those
obtained by other methods. \S\ref{s3node} deals with quivers with
3-nodes, \S\ref{scyclic} with cyclic quivers and \S\ref{sclosed} with
more complicated quivers in which the arrows form more that one closed
loop. In each case however we consider only $U(1)$ gauge groups at the
nodes. In \S\ref{snonabelian} we consider example of quivers with
non-abelian gauge groups.

As this manuscript was being 
prepared for publication, the preprint \cite{Lee2} appeared
on arXiv, which overlaps substantially with our 
results for cyclic quivers in \S\ref{scyclic}.

\section{Poincar\'e polynomials from pure Higgs states }  \label{strategy}

In this section we first review the formula of
\cite{1103.1887} expressing the refined index of a general multi-centered
black hole configuration in terms of the index carried by the individual centers.
We shall then argue that a microscopic version of this formula can be used 
to constrain the form of the Poincar\'e and Dolbeault polynomials of quiver
moduli spaces. We then describe
several mathematical methods for computing the cohomology of the Higgs branch directly.

\subsection{Refined index from single-centered black holes \label{secstra1}}

Let us consider a general multi-centered black hole configuration
with individual centers carrying charges $\alpha_1,\cdots \alpha_n$.
The collinear equilibrium configurations of $n$ single centered black holes
carrying charges $\alpha_1,\cdots \alpha_n$ in some lattice $\Gamma$
are given by the extrema of the Coulomb potential 
\be
\label{defhatW}
\hat W(\{x_i\}) = -\sum_{1\le i<j\le n} \alpha_{ij}\,   {\rm sign}(x_j-x_i)\, 
\ln| x_j - x_i| - \sum_{i=1}^n \hat c_i  x_i, \quad x_i\in {\mathbb R}\, ,
\ee
with respect to $x_2,\cdots x_n$ at fixed $x_1$. 
Here $\alpha_{ij}=\langle \alpha_i, \alpha_j\rangle$
denotes the Dirac-Schwinger=Zwanziger symplectic 
product between $\alpha_i$ and $\alpha_j$.
The constants $\hat c_i$ depend on the asymptotic moduli and the
charges, and satisfy $\sum_i \hat c_i=0$.
Extremizing $W$ gives a system of $n-1$ equations\footnote{For multi-centered
black holes,  the regularity of the metric and absence of time-like curves puts 
additional conditions on the solutions to \eqref{ewextrema}. We assume that no 
such restriction arises in the case of quivers, since all the charge vectors lie
in the convex cone $N_\ell\geq 0$.}
 \be \label{ewextrema}
\sum_{j\ne i} {\alpha_{ij}
\over |x_i-x_j|} = \hat c_i\  \quad i=2\dots n 
\ee
which are algebraic in the variables $x_i$ for any fixed ordering along the real axis. 
We let 
\be \label{espnew}
s(p)={\rm sign} \det M\, ,
\ee 
where $M_{\ell\kk}=\p^2 \hat 
W/\p x_\ell \p x_\kk$ for $2\le \ell,\kk\le K$ is the Hessian of
$\hat W$ with respect to the variables $x_2,\cdots x_n$ at fixed $x_1$.
Under reflection along the $x$-axis,  one has $x_i\mapsto -x_i, s(p)\mapsto (-1)^{n-1}s(p)$.

We define the Coulomb index associated to the unordered set of charges 
$\{\alpha_i\}_{i=1\dots n}$ by $g_{\rm ref} (\alpha_1; y) = 1$ for $n=1$, and 
\be \label{eclassint}
g_{\rm ref} (\alpha_1, \dots , \alpha_n;y)
= (-1)^{\sum_{i<j} \alpha_{ij} +n-1}
 \left[
(y-y^{-1})^{1-n} \, \sum_{p} \,  
s(p)\, 
y^{\sum_{i<j} \alpha_{ij}\, \sign[x_j-x_i]}\right] 
\ee
for $n\geq 2$. Here
the sum over $p$ runs over all solutions to \eqref{ewextrema}, with the following understanding:
If two or more centers carry mutually local charges (i.e. $\alpha_{ij}=0$), then the prescription of
\cite{1103.1887} is to analytically continue the charges 
away from their original values
so that they are slightly different from each other, find the set of extrema
of $W$ and then continue the charges back to their original 
values.\footnote{If $ \sum_{p} \,  
s(p)\, 
y^{\sum_{i<j} \alpha_{ij}\, \sign[x_j-x_i]}$ contains $y$-independent constant terms
then the coefficient of the constant term can some time depend on the details of how
we take the limit to the original values. However this does not affect the final
result since the functions $H$ introduced in \refb{ess2} precisely compensates
for this.}
If all centers corresponding to mutually local $\alpha_i$'s  in the original solution
are separated then the analytic continuation has no effect. However if $M$ centers
with mutually local charges coincide in the original solution, then the analytic continuation 
will separate their locations and pick one out of the $M!$ possible orderings of these centers 
along the $x$-axis. Thus, instead of analytically continuing the charges we may adopt the equivalent prescription of counting a solution with $M$ coincident mutually local charges 
only once, rather than $M!$ times. 

In addition, we exclude solutions
in which several centers carrying mutually non-local charges (i.e. $\alpha_{ij}\ne 0$)
would  coincide. Such singular solutions arise for certain choices of charges which
allow for `scaling solutions', where a subset (or all) of the charges can approach each
other at arbitrarily small distances. In the absence of such scaling solutions, 
 $g_{\rm ref} (\alpha_1, \dots , \alpha_n;y)$ computes the refined index
$\Tr (-y)^{2 J_3}$ of the quantum mechanics of $n$ charged particles interacting by
Coulomb, Lorentz, Newton and scalar exchange forces. Here $y$   is a
parameter conjugate to the angular momentum along the $z$ axis, and 
the trace is taken after factoring out the center of mass modes.  Equivalently, $g_{\rm ref} (\alpha_1, \dots , \alpha_n;y)$  computes the index of the equivariant Dirac operator on the 
space of solutions to \eqref{ewextrema}, -- a compact symplectic space equipped with an
Hamiltonian action of $SO(3)$. The result  \eqref{eclassint} then arises
by localizing with respect to the action $J_3$, whose only fixed points are the  
isolated collinear configurations above \cite{1011.1258}.

In the presence of scaling solutions, the space of solutions to \eqref{ewextrema} is non-compact, 
and there are additional non-isolated fixed points which can contribute to the equivariant index.
We shall continue to define $g_{\rm ref}$ as the contribution of the collinear 
configurations  \eqref{eclassint}, excluding the singular scaling
solutions with mutually 
non-local coincident centers.
As we shall review, 
the additional contributions from the scaling solutions
can be determined by the `minimal modification hypothesis' of 
\cite{1103.1887}.

For $n=2$, none of these issues arises, and $g_{\rm ref}(\alpha_1,\alpha_2;y)$ 
reduces to the 
character of a spin $J=\frac12(\langle\alpha_1,\alpha_2\rangle-1)$ 
representation \cite{1011.1258},
\be \label{e2node}
g_{\rm ref}(\alpha_1,\alpha_2;y) = 
\begin{cases}
(-1)^{\langle\alpha_1, \alpha_2\rangle+1} \,
{\left(y^{\langle\alpha_1, \alpha_2\rangle} - y^{-\langle\alpha_1, \alpha_2\rangle}\right) / \left(y - y^{-1}\right)}\, \,
\hbox{for $\sign\left(\langle\alpha_1, \alpha_2\rangle\right) = \sign\left(c_1\right)$}\cr
0 \, \,
\hbox{for $\sign\left(\langle\alpha_1, \alpha_2\rangle\right) = -\sign\left(c_1\right)$}\, .
\end{cases}
\ee

For a general charge vector $\gamma\in \Gamma$, the refined index $\tot_{\rm ref}(\gamma;y)=
\Tr' (-y)^{2 J_3}$ -- where the
trace is now taken over all states carrying total charge $\gamma$ after
factoring out the center of mass degrees of freedom -- is expressed 
in terms of the Coulomb indices $g_{\rm ref}$ 
via\cite{1103.1887}\footnote{In \cite{1103.1887} the formula for
$\tot_{\rm ref}(\gamma;y)$ was given for the case when the total charge
$\gamma$ is primitive. Otherwise the same formula applies to the `rational
refined index' $\bar \tot_{\rm ref}(\gamma;y)=
\sum_{m|\gamma} {1\over m} \, {y - y^{-1}\over y^m - y^{-m}}\,
 \tot_{\rm ref}(\gamma/m;y^m)$. From this we can arrive at the expression
 for $\tot_{\rm ref}(\gamma;y)$ using the inverse transformation
 $\tot_{\rm ref}(\gamma;y)=
\sum_{m|\gamma} {\mu(m)\over m} \, {y - y^{-1}\over y^m - y^{-m}}\,
\bar  \tot_{\rm ref}(\gamma/m;y^m)$.
}
\ben \label{ess1}
\tot_{\rm ref}(\gamma;y) &=& \sum_{m|\gamma} 
\mu(m) m^{-1} {y - y^{-1}\over y^m - y^{-m}}
\bar \tot_{\rm ref}(\gamma/m;y^m) \nonumber \\
\bar \tot_{\rm ref}(\gamma;y) &=& \sum_{n\ge 1}
\sum_{\{\alpha_i\in \Gamma\},\, \sum_{i=1}^n \alpha_i =\gamma}
{1\over {\rm Aut}(\{\alpha_1, \alpha_2,\cdots, \alpha_n\})} 
 g_{\rm ref}\left(\alpha_1, \alpha_2,\cdots , \alpha_n;y\right) \nonumber \\ &&
\prod_{i=1}^n \left\{\sum_{m_i\in\bZ\atop m_i|\alpha_i}
{1\over m_i} {y - y^{-1}\over y^{m_i} - y^{-m_i}}\, 
\left( \Omega^{\one}_{\rm ref}(\alpha_i/m_i;y^{m_i})
+ \Omega_{\rm scaling}(\alpha_i/m_i;y^{m_i})\right)\right\}
\, ,
\een
where $\mu(m)$ is the M\"obius
function, $\Gamma$ is the charge lattice,
$\Omega^{\one}_{\rm ref}(\beta;y)$ denotes the refined index
carried by a single centered black hole of charge $\beta$, and
\be \label{ess2}
\Omega_{\rm scaling}(\alpha;y) =
\sum_{\{\beta_i\in \Gamma\}, \{m_i\in\bZ\}\atop
m_i\ge 1, \, \sum_i m_i\beta_i =\alpha}
H(\{\beta_i\}; \{m_i\};y) \, \prod_i 
\Omega^{\one}_{\rm ref}(\beta_i;y^{m_i})
\, ,
\ee
for some function $H(\{\beta_i\}; \{m_i\};y)$. 
We determine $H(\{\beta_i\}; \{m_i\};y)$ by requiring that the
coefficient of the $\prod_i \Omega^{\rm S}_{\rm ref}(\beta_i;y^{m_i})$ in the
expression for $\Omega_{\rm ref}(\sum_i m_i\beta_i; y)$ is a
Laurent polynomial in $y$.
The ambiguity of adding to $H$ a
Laurent polynomial is resolved by using the `minimal
modification hypothesis', which requires that $H$ must 
be symmetric under $y\to y^{-1}$ and vanish
as $y\to\infty$ (and hence also as $y\to 0$). 
Alternatively, one could absorb $ \Omega^{\one}_{\rm ref}(\alpha_i;y)$
into $\Omega_{\rm scaling}(\alpha_i;y)$ at the cost of allowing $H$ to take
a finite value as $y\to\infty$.

Concretely, suppose
that the net coefficient of the monomial 
$\prod_i \Omega^{\rm S}_{\rm ref}(\beta_i;y^{k_i})$ in the
expression for $\Omega_{\rm ref}(\sum_i k_i\beta_i; y)$ is given by
$f(y)+ H(\{\beta_i\}; \{k_i\};y)$ where $f(y)$ is a known function, 
with Laurent series expansion $\sum_{n<N} f_n y^{-n}$ around $y=0$. 
It is easy to check that 
\be \label{eHexplicit}
H(\{\beta_i\}; \{k_i\};y) = f_0 + \sum_{n\ge 1} f_n \left(y^{-n}+ y^n\right)- f(y)
\ee
is the unique solution to the conditions stated above. This may be rewritten as a contour
integral around $u=0$,
\be
\label{uint}
H(\{\beta_i\}; \{k_i\};y) = \oint \frac{\de u}{2\pi\I} \frac{(1/u-u)\, f(u)}{(1-uy)(1-u/y)} - f(y) \ .
\ee

We determine $H(\{\beta_i\}; \{m_i\};y)$ iteratively by
beginning with the $H$'s with least possible number of $\beta_i$'s
(three) and then determining successively the $H$'s with larger
number of $\beta_i$'s. Physically $\Omega_{\rm scaling}$ and hence
$H$ represent the correction to the index
due to the presence of scaling solutions \cite{0702146}.

In the formulae  \eqref{ess1}, \eqref{ess2} the $\Omega^{\rm S}_{\rm ref}(\gamma;y)$,
representing the index carried by the single centered black holes, must be
independent of $y$ since  single centered BPS black holes carry zero
angular momentum\cite{0903.1477}.
Furthermore $H$ and $\Omega^{\rm S}$ are expected to be
independent of the values of the parameters $\hat c_i$, as the 
jumps of the refined index across walls of marginal stability in  the space
of the parameters $\hat c_i$ is already captured  by the Coulomb indices
$g_{\rm ref}$ \cite{1103.1887}.

\subsection{A Coulomb branch formula for  quiver Poincar\'e polynomials}
\label{scoulombquiver}

In the weak coupling limit, the dynamics of multi-centered black holes is
described by a quantum mechanics with $\cN=4$ supersymmetry, whose
matter content is captured by a certain quiver  \cite{0206072}. For 
$N_\ell$ centers of charge $\gamma_\ell$ for
$\ell=1,\cdots K$, the corresponding quiver has $K$ nodes labelled by the
integer $\ell$, with a complex vector space of dimension $N_\ell$ attached
to the node $\ell$ and $|\gamma_{\ell\kk}|$ 
arrows connecting the node $\ell$ to the node $\kk$ if $\gamma_{\ell\kk}>0$,
or connecting the node $\kk$ to the node $\ell$ if $\gamma_{\ell\kk}<0$.
The nodes represent $U(N_\ell)$ vector multiplets, while
the arrows represent $|\gamma_{\ell\kk}|$
chiral multiplets $\phi_{\ell\kk, \alpha, ss'}$
in the bifundamental of $U(N_\ell)\times U(N_\kk)$ (or its
complex conjugate, if $\gamma_{\ell\kk}<0$). 
In addition, to each node  we associate a constant $c_\ell$
labelling the coefficient of the FI parameter for the $U(1)$ factor in $U(N_\ell)$, such that   
$\sum_\ell N_\ell c_\ell=0$.
Finally, the superpotential $W$ for the chiral multiplets is a generic sum of gauge-invariant
monomials associated to each oriented loop in the quiver.

At low energies, the supersymmetric quantum mechanics admits two different effective
descriptions, the Coulomb branch description, which is valid in the region where the vevs
of the vector multiplet scalars are large, so that the chiral multiplets can be 
integrated out, and the Higgs branch description, valid in the region where the vevs
of the chiral multiplets are large, so that the vector multiplets can be integrated out.
In this subsection we shall focus on the Coulomb branch description. 
The flat directions of the potential on the Coulomb branch turn out to reproduce the 
moduli space of supersymmetric configurations of multi-centered 
BPS black holes in $\cN=2$
supergravity, with the FI parameters determined by the values of the scalar fields at spatial
infinity  \cite{0206072}. 
This allows us to borrow the results on multi-centered black hole
quantum mechanics, reviewed in the previous section, to analyze the
refined index of the quiver quantum mechanics on the Coulomb branch.
Using \refb{epol}, we can then use this result to make predictions
for the Poincar\'e polynomial 
of the moduli space of quivers on the Higgs branch. This can then be compared
with a direct computation of the same Poincar\'e polynomial using methods to
be discussed in \S\ref{shiggsresult}-\ref{srepresent}.

We denote by $Q(\gamma;y)$ the refined index of the supersymmetric quantum mechanics
associated with the quiver $\cQ(\gamma,c)$ for 
$\gamma=\sum_\ell N_\ell\gamma_\ell$. 
 The refined index $Q(\gamma;y)$ can be computed using 
 the results of \S\ref{secstra1}
 with the following understanding:
\begin{enumerate}

\item The role of the charge lattice is now played by the set of
charges of the form $\sum_\ell m_\ell \gamma_\ell$ with 
$m_\ell\in \IZ$. Furthermore only
charge vectors with non-negative $m_\ell$ can appear in the sums over
charge vectors in \eqref{ess1}, \eqref{ess2}.

\item The index $\Omega_{\rm ref}(\sum_{\ell} N_\ell\gamma_\ell;y)$ appearing on the l.h.s of \eqref{ess1} is interpreted as the Laurent  
polynomial \eqref{epol2} of the Higgs branch of the 
quiver with gauge group $U(N_\ell)$ at the $\ell$th node.

\item  In evaluating the `Coulomb index' $g_{\rm ref}\left(\alpha_1,\cdots \alpha_n;y\right)$ appearing in \eqref{ess1},  the parameters $\hat c_i$ are determined in terms of the FI parameters through $\hat c_i = \sum_\ell A_{i\ell} c_\ell$ whenever $\alpha_i=\sum_\ell A_{i\ell} \gamma_\ell$. 
 Since only combinations such that 
$\sum_i \alpha_i = \gamma =\sum_\ell N_\ell \gamma_\ell$ appear, the condition 
$\sum_i\hat c_i = \sum_\ell N_\ell c_\ell =0$ is automatically satisfied.

\item The quantities $\Omega^{\rm S}_{\rm ref}(\beta;y)$ 
appearing on the r.h.s. of \eqref{ess1} and \eqref{ess2} are 
interpreted as the number of `single centered black hole micro-states' with charge 
$\beta$.
These are the `pure Higgs' or `intrinsic Higgs' states which originate from the middle
cohomology of the quiver or one of its subquivers. They are assumed 
to be $y$-independent, although it is useful to retain the dependence on $y$ to carry out the algorithm 
explained in \S\ref{secstra1}. 

\item For any of the basis vectors $\gamma_\ell$, we set 
$\Omega^{\rm S}_{\rm ref}(N\gamma_\ell;y)=1$ for $N=1$ and zero if $N>1$, since each 
node of the quiver is assumed 
to represent a single state of zero angular momentum 
(using the formulas in \ref{srepresent}, one finds that also
mathematically the generalized DT-invariant vanishes for
quiver representations with dimension $N$ at the node $\ell$ and
dimension 0 at the other nodes).


\end{enumerate}
It is important to note that even though the quiver quantum mechanics
maps to the quantum mechanics describing the dynamics of multiple
black holes, for an actual system of black holes
the quiver quantum mechanics counts only part of the 
black hole micro-states. In  particular if 
we associate to each node of the quiver an
elementary constituent carrying charge $\gamma_\ell$, then the
quiver quantum mechanics counts only a subset of the micro-states
 which carry a total charge 
$\gamma=\sum_\ell N_\ell\gamma_\ell$,
-- namely those which can be 
built from elementary constituents carrying charges $\gamma_1,\cdots, \gamma_K$.
Other black hole micro-states with the same total charge could arise from 
bound states of other elementary constituents described by different types of quivers.
Similarly $\Omega^{\rm S}_{\rm ref}(\beta;y)$
need not count all states of a single centered black hole
of charge $\beta$, but counts only those states which can be built from
the elementary constituents carrying charges $\gamma_1,\cdots \gamma_K$.
For the quiver, $\Omega^{\rm S}_{\rm ref}(\beta;y)$ represent contribution
from states which are elementary from the point of view of the Coulomb
branch but are composite from the point of view of the Higgs branch
except when $\beta$ coincides with one of the basis vectors $\gamma_\ell$.

This microscopic re-interpretation of the formulae \eqref{ess1},\eqref{ess2} leads to 
the algorithm for computing $Q(\cM;y)$ 
summarized in \S\ref{sintro}. Like its macroscopic
counterpart \cite{1103.1887}, this procedure leaves the `single-centered micro-state 
degeneracies' $\Omega^{\rm S}_{\rm ref}(\beta)$ for $\beta=\sum_\ell m_\ell\gamma_\ell$ undetermined. These can be fixed by independently  computing the Euler character of
the corresponding quiver moduli space, using techniques discussed in \S\ref{sec_abcohom}
and \S\ref{srepresent}, and comparing with the prediction of \eqref{ess1} at $y=1$,
where the Poincar\'e polynomial reduces to the Euler character. 
Alternatively it can be determined in terms of the middle cohomology
of the quiver with $U(m_\ell)$ gauge group at the $\ell$-th node by comparing the
$y$-independent terms on both sides of \eqref{ess1}
for $\gamma=\sum_\ell m_\ell\gamma_\ell$. Once the coefficients 
 $\Omega^{\rm S}_{\rm ref}(\gamma)$ have been determined, we can
use \eqref{ess1} to determine the Poincar\'e polynomial of a quiver with arbitrary
gauge groups $\prod_\ell U(N_\ell)$.

\subsection{A Coulomb branch formula for Hodge numbers} \label{shodgesec}

So far we have focused on the Poincar\'e polynomial of
the quiver moduli space $\cM$, i.e. on the topological Betti numbers $b_i(\cM)$. However,
since $\cM$ is  a complex \kahler manifold its cohomology admits a Dolbeault decomposition 
\be
H^*(\cM,\IZ) = \sum_{p,q=0}^d H^{p,q}(\cM,\IZ)\ ,
\ee
We shall now give an algorithm for computing the Hodge numbers $h_{p,q}(\cM)=\dim
H^{p,q}(\cM,\IZ)$,
generalizing the prescription in the previous subsection.

Let us define the Dolbeault polynomial as the Laurent polynomial  in two
variables 
\be \label{ehodge1}
\wtQ(\cM;y,t) =  \sum_{p,q} h_{p,q}(\cM)\, (-y)^{p+q-d} t^{p-q} \ .
\ee
Using the standard symmetries of the Hodge numbers $h_{p,q}=h_{d-p,d-q}=h_{q,p}$, one has
\be
\wtQ(\cM;y,t) = \wtQ(\cM;1/y,1/t) = \wtQ(\cM;y,1/t)\ .
\ee
For $t=1$, $\wtQ(\cM;y,t)$ reduces to the  Laurent  
polynomial $Q(\cM;y)$ introduced in \refb{epol}, while 
for $t=1/y$,  it reduces to 
\be 
\label{PoincaHirz}
\wtQ(\cM;y,1/y) = (-y)^{-d}  \chi(\cM;y^2)\ ,  
\ee
where $\chi(\cM;v)$ is the Hirzebruch polynomial \cite{Hirzebruch:1995}
\be
\label{defhirz}
\chi(\cM;v) = \sum_{p,q} (-1)^{p+q}\, v^q\, h_{p,q} (\cM) = v^d\, \chi(\cM;1/v)\, .
\ee
For $y=t=1$, $\wtQ(\cM;y,t)$ 
reduces to $(-1)^d$ times the Euler number $\chi(\cM)$. Finally
$\wtQ(\cM;y,t)$ is related to the Hodge polynomial $H(u,v)\equiv 
h_{p,q}u^p v^q$ via $\wtQ(\cM;y,t)=(-y)^{-d}H(-yt, -y/t)$.  

The parameter $t$ is a chemical potential conjugate to the quantum
number $I_3\equiv p-q$, which can be viewed as the remnant of a $SU(2)_R$ symmetry 
in the full quiver quantum mechanics, before integrating out the vector multiplets
to reach the Higgs branch description. For $t=y$ (or $t=1/y$), the Dolbeault polynomial 
$\wtQ(\cM;y,t)$ at $t=y$ is therefore identified with  the protected spin character $\Tr' (-1)^{2J_3} y^{2(I_3+J_3)}$ \cite{Gaiotto:2010be}. On the other hand, there is much evidence\footnote{
This was observed in the context of 2-centered black holes in \cite{0206072}, and elevated in the context of $N=2$ gauge theories to 
the `no exotics' conjecture of \cite{Gaiotto:2010be}. As we shall see in the next subsection, this is consistent with the fact that the Coulomb branch accounts for all non-middle cohomology states on the Higgs branch, which necessarily have $p=q$. } that all
states associated with the quantization of multi-centered black holes 
are singlets under $I_3$.
As a result the functions $g_{\rm ref}$
and consequently $H$ are independent of the parameter $t$
conjugate to $I_3$. Thus the only possible
source of $t$ dependence is in the index for `pure Higgs states'. This motivates the 
conjecture that the Dolbeault polynomial 
$\wtQ(\gamma;y, t)$ for a general quiver  $\cQ(\gamma)$  
is given by the same formula as in
\S\ref{sintro} with the replacement
$\Omega^{\rm S}_{\rm ref}(\alpha/m) \to \wt \Omega^{\rm S}_{\rm ref}(\alpha/m;t^m)$:
\ben \label{ehodge2}
\wt Q(\gamma;y,t) &=& \sum_{m|\gamma} 
\mu(m) m^{-1} {y - y^{-1}\over y^m - y^{-m}}
\bar {\wt Q}(\gamma/m;y^m,t^m) \nonumber \\
\bar {\wt Q}(\gamma;y,t) &=& 
\sum_{n\ge 1}\sum_{\{\alpha_i\in \Gamma\},\, \sum_{i=1}^n \alpha_i =\gamma}
{1\over {\rm Aut}(\{\alpha_1, \alpha_2,\cdots, \alpha_n\})} 
 g_{\rm ref}\left(\alpha_1, \alpha_2,\cdots, \alpha_n;y\right) \nonumber \\ &&
\prod_{i=1}^n \left\{\sum_{m_i\in\bZ\atop m_i|\alpha_i}
{1\over m_i} {y - y^{-1}\over y^{m_i} - y^{-m_i}}\, 
\left( \wt \Omega^{\one}_{\rm ref}(\alpha_i/m_i;t^{m_i})
+ \wt \Omega_{\rm scaling}(\alpha_i/m_i;y^{m_i},t^{m_i})\right)\right\}
\, ,
\non \\
\wt\Omega_{\rm scaling}(\alpha; y,t) &=&
\sum_{\{\beta_i\in \Gamma\}, \{m_i\in\bZ\}\atop
m_i\ge 1, \, \sum_i m_i\beta_i =\alpha}
H(\{\beta_i\}; \{m_i\};y) \, \prod_i 
\wt\Omega^{\one}_{\rm ref}(\beta_i;t^{m_i})
\, .
\een
Thus we can parametrize $\wtQ(\gamma;y,t)$ in terms of the unknown functions
$\wt\Omega^{\rm S}_{\rm ref}(\alpha;t)$. These in turn can be determined if we know
the Hirzebruch polynomial $\wt Q(\gamma;y,1/y)$ given in
 \eqref{defhirz} for each $\gamma$. The functions
$H(\{\beta_i\}; \{m_i\};y)$ appearing in \refb{ehodge2}
are the same  as before and so need
not be determined again. For this reason we have dropped the $y$ dependence in the
arguments of $\wt\Omega^{\rm S}_{\rm ref}$ from the outset.

Note that since $\wt Q(\gamma;y,t)$ is not protected, we cannot prove that
this quantity computed from the Coulomb branch description must match the
Higgs branch result. In that sense \refb{ehodge2} should again be taken as a
conjecture for the Dolbeault polynomial of the Higgs branch moduli space.
Since all quantities except the $\wt\Omega^{\rm S}_{\rm ref}(\alpha,t)$'s on the
right hand side of this expression are computable, this formula 
can be tested by independently computing the Dolbeault polynomial of the 
Higgs branch moduli space for simple quivers.

\subsection{Quiver Poincar\'e polynomial from the Higgs branch analysis}
\label{shiggsresult}
\label{sec_abcohom} 

The analysis of \S\ref{scoulombquiver} gives a specific algorithm for computing
the Poincar\'e polynomial of quiver moduli space. In order to test this formula we
need an independent determination of the Poincar\'e polynomial. In this subsection
we shall outline the procedure for doing this, generalizing methods used in \cite{0702146,1205.5023,1205.6511}. 

The classical moduli space $\cM$ on the Higgs branch, or quiver moduli space for brevity, is 
described by the D-term and F-term 
conditions \eqref{emodi1}, subject to identifications under $\prod_{\ell} U(N_\ell)$ gauge
transformations. 
In general, $\cM$ is a complex \kahler manifold of complex dimension 
\be
d = 1 - \sum_{\ell} N_l^2 + \sum_{\gamma_{\ell\kk}>0} \gamma_{\ell \kk}\, N_\ell N_\kk - f\ ,
\ee
where $f$ is the number of independent F-term conditions ($f=0$ for quivers without loop).
The BPS states in the supersymmetric quantum mechanics are identified as classes in the total cohomology $H^*(\cM,\IZ)$.  The Lefschetz operators~\cite{GriffithsHarris}
\be
J_+\cdot h=\omega \wedge h\ , \quad 
J_- = \omega \,\llcorner\, h\ ,\quad J_3\cdot h = \tfrac12 (n-d) h\ ,
\ee
where $\omega$ is the \kahler form, $\llcorner=*\wedge *$ where $*$
denotes Hodge star operation, and $n$ is the degree of the differential form $h$,
generate an action of $SU(2)$ on $H^*(\cM,\IZ)$ which is identified as $SO(3)$ 
rotations in space-time  \cite{0206072}. The refined index 
$\Tr'(-y)^{2J_3}$ of the supersymmetric  quantum mechanics on the Higgs branch is given by 
\be
\label{epol2}
Q(\cM;y)\equiv
\sum_{p=1}^{2d} b_p(\cM) (-y)^{p-d}
\ee
where $b_p(\cM)=\dim H^p(\cM,\IZ)$ is the $p$-th  Betti number. 
In the rest of this subsection
we shall restrict our analysis to Abelian
quivers, \i.e.\ with $U(1)$ gauge groups at each node, deferring
a discussion of non-Abelian quivers to \S\ref{srepresent}. 


For Abelian quivers, the moduli space of classical vacua $\cM$ is 
obtained by first  using part of  the  F-term constraints to set some of the
variables $\phi_{\ell k,\alpha,ss'}$ to zero, and then using the standard 
relation between \kahler quotients and algebro-geometric quotients\footnote{i.e. the equivalence between, on the one hand, the space of solutions
of the D-term constraints modulo the compact gauge group $G=U(1)^K$ and on the other hand,
the quotient of the semi-stable locus by the complexified gauge group $G_\IC=(\IC^\times)^K$.}
to solve the  D-term constraints for the remaining variables. As a result,
the quiver moduli space  $\cM$ is generally obtained as 
a complete intersection of $k$ hypersurfaces inside a product $\cME$ of complex 
projective spaces,
corresponding to  the F-term constraints which are not trivially 
solved in the first step.  Provided each of the F-term constraints
arises as the zero locus of the section of a positive line
bundle\footnote{We shall also encounter examples where some
of the F-term constraints are not given by sections of line bundles with strictly positive
curvature  -- this happens {\it e.g.} when some of the
constraints are independent of the coordinates of some projective space 
in $\cME$.
Even though \eqref{hypthm} no longer holds for $\cM$ directly, one can still
regard $\cM$ as a product  of
manifolds for which \refb{hypthm} holds. In this case one may
parametrize  the lack of knowledge of the middle cohomology of each
of the manifolds in the product by unknown constants, and
determine the $y$ dependence of $Q(\cM;y)$ in terms of these constants. 
As we shall see in many
examples later, knowing the $y$-dependence of $Q(\cM;y)$ will allow
us to test the general algorithm for computing $Q(\gamma;y)$ given in
\S\ref{sintro}, and the unknown constants mentioned above
will be in one to one correspondence to the constants $\Omega^{\rm S}_{\rm ref}
(\gamma)$
which appear in the formul\ae\ in \S\ref{sintro}.} 
over $\cME$, it follows from the Lefschetz hyperplane theorem \cite{GriffithsHarris}
that the Betti numbers $b_p(\cM)$ for $p$ not equal to the complex dimension $d$
of $\cM$ are given in terms of the Betti numbers of the ambient space by
\be
\label{hypthm}
b_p(\cM) = \begin{cases} b_p(\cME) & p<d \\
b_{2d-p} (\cME) & p>d\ ,
\end{cases}
\ee
leaving the middle cohomology undetermined. 
The Betti numbers $b_p(\cME)$ on the other hand are given by
\be \label{ebpme}
\sum_p b_p (\cME) \, y^p = \prod_{\ell=1}^n {1 - y^{2a_\ell}\over 1 - y^2}\, .
\ee
Eq.\refb{hypthm}, \refb{ebpme} allows us to determine the $y$-dependence of
$Q(\cM;y)\equiv (-y)^{-d} \sum_p b_p (-y)^p$, but leaves undetermined
the constant term in $Q(\cM;y)$. More generally, the Lefschetz hyperplane theorem
ensures that the Hodge numbers $h_{p,q}(\cM)$ for $p+q\neq d$ are inherited from
the ambient space, and therefore that they vanish unless $p=q$. As a result, 
the Dolbeault polynomial is then a sum
\be \label{edecompA}
\wtQ(y,t) = \wtQ_{\rm amb}(y) + \wtQ_{\rm mid}(t)
\ee
of a $t$-independent piece, coming from
the cohomology of the ambient projective space, and a $y$-independent piece,
coming from the middle cohomology. 
In general however the quiver moduli space is more complicated, {\it e.g.}
given by the product of manifolds for each of which \refb{edecompA} holds.
Our conjecture of \S\ref{shodgesec} is expected to reproduce correctly
the Dolbeault polynomial in all such cases.


In order to complete the computation of the Poincar\'e and Dolbeault polynomials, we need
to  evaluate the contribution of the middle cohomology. This can be easily obtained from
the Euler number $\chi(\cM)$, equal up to a sign to the value of  $Q(\gamma;y)$ at $y=1$, or from the Hirzebruch polynomial $\chi(\cM,v)$, related to the Dolbeault polynomial at $t=1/y$ 
via \eqref{PoincaHirz}. Both of them can be computed using the Riemann-Roch theorem,
as follows. Suppose as before that the quiver moduli space $\cM$
is given by the complete intersection
of $k$ hypersurfaces in $\cME=\CP^{a_1-1}\times \cdots \times \CP^{a_n-1}$.
The Riemann-Roch 
theorem (see e.g. \cite{Hirzebruch:1995}) expresses the Euler characteristics $\chi(\cM)$ as 
the coefficient of the top form $J_1^{a_1-1}\dots J_n^{a_n-1}$ in the Laurent
expansion around $J_\ell=0$ of the rational function
\be
\label{RRthm}
\prod_{\ell=1}^n (1+J_\ell)^{a_\ell} \, \prod_{j=1}^k 
\frac{ d_1^{(j)} J_1+\dots +d_n^{(j)}  J_n}
{1+d_1^{(j)}  J_1+\dots +d^{(j)} _n J_n}\ ,
\ee
where $d^{(j)}_\ell$ is the degree of the algebraic equation defining the $j$-th  hypersurface
with respect to the homogeneous coordinates on $\CP^{a_\ell-1}$. Equivalently, $\chi(\cM)$ can be obtained as a contour integral around $J_\ell=0$,
\be
\label{Leeint}
\chi(\cM)= \oint\prod_{\ell=1}^{n} \frac{\de J_\ell}{2\pi\I} \, 
\prod_{\ell=1}^{n} \left(\frac{1+J_\ell}{J_\ell}\right)^{a_\ell}\,
\prod_{j=1}^k 
\frac{ d_1^{(j)} J_1+\dots +d_n^{(j)}  J_n}
{1+d_1^{(j)}  J_1+\dots +d^{(j)} _n J_n}\ .
\ee
For example, for $\cM=\cME=\CP^{a-1}$, the Euler number is given by 
\be
\label{chicp}
\chi(\CP^{a-1}) = \oint \frac{\de J}{2\pi\I}\, \left(
\frac{1+J}{J}\right)^a = 
{1\over 2\pi\I}\, \oint \frac{\de x}{(1-x)^2\, x^a}= a\ ,
\ee
where, in the second equality, we changed variable to $J=x/(1-x)$. For a less trivial
example, consider a complete intersection $\cM_{n,d_1,\dots d_k}$ of $k$ hypersurfaces of degree $d_1,\dots d_k$ inside $\CP^{n+k}$. Using \eqref{Leeint}, we find that  
 the Euler number is given by 
\be
\label{chicpk}
\begin{split}
\chi(\cM_{n,d_1,\dots d_k}) =& \oint \frac{\de J}{2\pi\I} \left( \frac{1+J}{J}\right)^{n+k+1} 
\prod_{j=1}^k  \frac{d_j\, J}{1+d_j J}\\ 
=& \oint \frac{\de x}{2\pi\I \, (1-x)^2\, x^{n+k+1}}
\prod_{j=1}^k  \frac{d_j\, x}{1+(d_j-1) x}
\end{split}
\ee
where we used the same change of variable. The integral can be computed by introducing
the generating function
\be
Z(d_1,\dots d_k;z)\equiv \sum_{n=0}^{\infty} \chi(\cM_{n,d_1,\dots d_k};v)\, z^{n+k} = 
\oint \frac{\de x}{2\pi\I \, (1-x)^2}\, \frac{z^k}{x-z}
\prod_{j=1}^k  \frac{d_j}{1+(d_j-1) x}\, .
\ee
The integral picks up the residue at $x=z$, leading to the simple result
\be
Z(d_1,\dots d_k;z)= \frac{1}{(1-z)^2}\, 
\prod_{j=1}^k  \frac{d_j\, z}{1+(d_j-1) z}\ .
\ee

As explained in \cite{Hirzebruch:1995}, the Hirzebruch polynomial \eqref{defhirz} can be 
similarly obtained from the Riemann-Roch theorem, by performing 
in \refb{Leeint}
the replacements
\be \label{edefrvj}
\frac{J_\ell}{J_\ell+1} \to R_v(J_\ell), \quad 
\frac{ d_1^{(j)} J_1+\dots +d_n^{(j)}  J_n}
{1+d_1^{(j)}  J_1+\dots +d^{(j)} _n J_n} 
\to R_v(d_1^{(j)} J_1+\dots +d_n^{(j)}  J_n), \qquad 
\ee
in \eqref{Leeint}. Here $R_v(J)$ is a function of $J$ which reduces to $x=J/(J+1)$ at $v=1$,
\be
R_v(J) = \left\{\frac{1-v}{1-e^{-(1-v)J}}+v\right\}^{-1}\ .
\ee
As in \eqref{chicp}, it is useful to change variable from $J$ to $R=R_v(J)$ using
\be
\label{JtoR}
\de J =\frac{ \de R}{(1-R)(1-v R)} \ ,\quad J=\frac{1}{1-v}\log\frac{1-v R}{1-R}\ .
\ee
For example,   the Hirzebruch polynomial of  $\CP^{a-1}$ is given by 
\be
\chi(\CP^{a-1};v) = \oint \frac{\de J}{2\pi\I [R_v(J)]^a} = 
\oint \frac{\de R}{2\pi\I (1-R)(1-v R) R^a} =
 \frac{1-v^a}{1-v} \ ,
\ee
in agreement with the fact that $h_{p,p}=1$ for $0\leq p\leq a-1$, $h_{p,q}=0$ for $p\neq q$. 
 
 Returning to the example discussed in \eqref{chicpk}, 
 the Hirzebruch polynomial of the complete intersection $\cM_{n,d_1,\dots d_k}$
is given by
\be
\begin{split}
\chi(\cM_{n,d_1,\dots d_k};v) =& \oint \frac{\de J}{2\pi\I\, [R_v(J)]^{n+k+1}} 
\prod_{j=1}^k  R_v[ d_j J ] \\
=&  \oint \frac{\de R}{2\pi\I \, (1-R)(1-v R)R^{n+k+1}}  \prod_{j=1}^k
\frac{(1-v R)^{d_j}-(1-R)^{d_j}}{(1-v R)^{d_j}-v(1-R)^{d_j}}\ .
\end{split}
\ee
The generating function of these polynomials is obtained by summing
up the geometric series and picking up the residue at $R=z$,  leading to  
\be
\label{Zhir}
Z(d_1,\dots d_k;z,v)\equiv \sum_{n=0}^{\infty} \chi(\cM_{n,d_1,\dots d_k};v)\, z^{n+k} = 
\frac{1}{(1-z)(1-v z)}\prod_{j=1}^k \frac{(1-v z)^{d_j}-(1-z)^{d_j}}{(1-v z)^{d_j}-v(1-z)^{d_j}}\ ,
\ee
correcting a misprint in  \cite{Hirzebruch:1995}, Thm 22.1.1. 
The manipulations shown in this example will be used in the examples
discussed  in \S\ref{s3node} onward.

 \subsection{Cohomology of quivers without superpotential and HN recursion method \label{srepresent}}

As discussed in \S\ref{shiggsresult}, the moduli
space for Abelian quivers arises as a complete intersection in an ambient space 
$\cME$ given by a product of complex projective spaces, whose cohomology is 
easily computed  via the Lefschetz hyperplane and Riemann-Roch theorems. 
For a generic non-Abelian quiver, the situation is more complicated, since 
the classical moduli space $\cM$ is defined by the solution
to D-term and F-term equations \eqref{emodi1} inside the space $\IC^N$ parametrized by
all the complex variables $\phi_{\ell\kk, \alpha, ss'}$, quotiented by the action of the 
compact group $G=\prod_\ell U(N_\ell)$. We denote by $\cM_0$ the space of
solutions to the D-term equations only, quotiented by the action of $G$. Since the
F-term equations are gauge invariant, they descend to the quotient, hence $\cM$ is a 
submanifold of $\cM_0$. The cohomology of $\cM$ can  in principle be computed 
from the cohomology of the ambient space $\cM_0$ using the methods
described  
in \S\ref{shiggsresult}. 
In this section we shall describe a general procedure for computing the
cohomology of $\cM_0$.

By the usual equivalence between \kahler quotients and algebro-geometric quotients, $\cM_0$ is isomorphic to the quotient of the semi-stable locus $\cS\subset \IC^N$ by the action of the 
complexified  gauge group $G_{\IC}=\prod_\ell GL(N_\ell,\IC)$ \cite{Kirwan:1984}. Recall 
that the semi-stable locus is defined as the set of points $x$ for which there exists an
homogeneous non-constant $G_{\IC}$-invariant polynomial $F_x(X)$ such that $F_x(x)\neq 0$.
Roughly speaking, semi-stable points are those where the complexified gauge group is 
broken to a finite group. This implies in particular that $\cM_0$ is a projective variety, which 
contains $\cM$ as a complex submanifold. 

The space $\cM_0$ is, in effect, the classical quiver moduli space when the superpotential is tuned to zero. When the quiver contains closed loops which admit scaling configurations,
this space is non-compact, and hence one might wonder whether its
Poincar\'e polynomial  is well-defined. 
Typically, this non-compactness arises due to a certain number of
independent variables $\phi^{\rm NC}_i, i=1\dots N'$ which are allowed to vary over 
$\IC^{N'}$ without restrictions, and hence can become arbitrary large consistently with the D-term constraints. For fixed, finite values of $\phi^{\rm NC}_i$, the remaining variables $\phi^{\rm C}_j, j=1\dots N-N'$ take values in a compact space $\cME(\phi^{\rm NC})$, 
trivially fibered over $\IC^{N'}$,
such that the fiber never degenerates. 
The Poincar\'e polynomial $P_{\cM_0}(y)$ 
of $\cM_0$ is then given by the product of the Poincar\'e
polynomial $P_{\cME}(y)$ of the compact space 
$\cME(\phi^{\rm NC})$ (evaluated for example at 
$\phi^{\rm NC}_i=0$) and the Poincar\'e polynomial of the complex plane labelled by the free variables. For the latter we have $b_0=1$ and $b_p=0$ for $p>0$, and hence the Poincar\'e polynomial is 1. Thus, 
\be \label{eqmey}
Q(\cM_0; y) \equiv (-y)^{-d_0} P_{\cM_0}(-y) =
(-y)^{-d_0} P_{\cME}(-y)  =(-y)^{-d_0+d_E}Q(\cME;y)\, ,
\ee
where $d_0$ and $d_E$ are complex dimensions of $\cM_0$ and $\cME$
respectively.
$Q(\cME;y)$ is invariant under $y\to 1/y$ but in general 
$Q(\cM_0; y)$ is not
invariant under $y\to 1/y$, since Poincar\'e duality does not  
hold for non-compact spaces.
In the absence of scaling solutions (in particular, for quivers without loops), 
the moduli space
$\cM_0$ is compact, and this issue does not arise. 

To compute $Q(\cM_0; y)$, we shall 
apply the Harder-Narasimhan (HN) recursion 
method.
This method was originally established for stable vector bundles over Riemann surfaces 
\cite{Harder:1975, Atiyah:1982fa} and recently
 applied to semi-stable sheaves over rational complex surfaces \cite{Manschot:2011ym}.
For arbitrary quivers without oriented closed loops, the  HN recursion
method was developed by Reineke, culminating in a general formula for
the Poincar\'e polynomial of the quiver moduli space  
with any primitive dimension vector \cite{Reineke:2002}. The method
was later generalized to quivers with oriented closed loops but vanishing 
superpotential by \cite{Reineke:2007}. 
We find strong evidence that the HN recursion method does indeed produce the
cohomology of the ambient space $\MM_0$.  

Instead of explaining the logic behind the method, we shall just give the
algorithm  that derives from it.
We introduce the following notations: for the quiver $\cQ(\gamma)$ associated 
to the charge vector $\gamma$  we define $Q_0(\gamma;y)
=Q(\cM_0; y)$, and, as in \eqref{essp1}, 
$\bar Q_0(\gamma;y)$  by  
\be \label{edefbarq0}
\bar Q_0(\gamma;y)=\sum_{m|\gamma} m^{-1}
(y-y^{-1}) (y^m - y^{-m})^{-1} Q_0(\gamma/m;y^m)\, .
\ee
From this data we construct a new set of invariants 
$\CI(\gamma;w)$ 
via the relations \cite{MR2357325,Joyce:2008pc,Joyce:2009xv}, 
\be
\label{eq:stackinv}
\CI(\gamma;w)=\sum_{\sum_{i=1}^k \alpha_i=\gamma,\atop
  \mu(\alpha_i)=\mu(\gamma)} \frac{1}{k!}\,\prod_{i=1}^k\left( 
  \frac{\bar Q_0(\alpha_i;-w^{-1})}  
  {w-w^{-1}} \right)\ ,
\ee
where $\mu(\beta)$ is the `slope' of the dimension vector $\beta=\sum_\ell n_\ell \gamma_\ell$, defined by 
 \be \label{efisub}
\mu(\beta)\equiv \frac{\sum_\ell c_\ell n_\ell}{\sum_\ell n_\ell}\, .
\ee
Recall that the parameters $c_\ell$ are chosen to satisfy $\sum_\ell c_\ell N_\ell=0$,
hence $\mu(\gamma)=0$.
Finally, for any dimension vector $\beta$ and  ordered set of dimension vectors 
$\{\alpha_i\}$ we define\footnote{The function $h(\beta;w)$ counts the number of
quiver representations over finite fields, and is sometimes known as the counting
function. The parameter $w$ is related to $y$ by $w=-1/y$.}
\be
\label{eq:totalset} 
h(\beta;w)=\frac{w^{-(\beta,\beta)}}{\prod_{\ell=1}^K \prod_{j=1}^{n_\ell}(1-w^{-2j})}\ ,\quad
\CF(\{\alpha_i\};w) =w^{-\sum_{i<j}\left<\alpha_i,\alpha_j\right>}\prod_{i=1}^k\CI(\alpha_i;w)
\, ,
\ee 
where $(\gamma,\gamma)$ and $\langle \gamma,\gamma'\rangle$
are the Euler form and antisymmetric form defined in \eqref{defEulerForm}.
The HN method states that the invariants $\CI(\gamma;w)$ satisfy the 
 relation \cite{Harder:1975, Atiyah:1982fa, Kirwan:1984}
\be 
\label{eq:CIrecurs}
\CI(\gamma;w)=h(\gamma; w)-
\sum_{k\ge 2} \sum_{\sum_{i=1}^k\alpha_i=
\gamma,\atop \mu(\alpha_i)>\mu(\alpha_{i+1})}\CF(\{\alpha_i\};w).
\ee
We can solve these relations recursively to find $\CI(\gamma;w)$. The solution
to the recursion \eqref{eq:CIrecurs} is in fact given in terms of the 
`counting functions' $h(\beta;w)$ by  \cite{Reineke:2002}
\be
\label{eq:solrecursion} 
\CI(\gamma;w)=  \sum_{{\alpha_1+\dots +\alpha_k=\gamma, k\geq
      1\atop   \mu(\sum_{j=1}^m \alpha_j)>\mu(\gamma),\, m=1,\dots, k-1
}} (-1)^{k-1}\, w^{-\sum_{i<j}\left< \alpha_i,\alpha_j\right>} \prod_{j=1}^k h(\alpha_j,w).
\ee 
Using \eqref{eq:stackinv}
we can then find $\bar Q_0(\gamma;y)$ and hence
$Q(\cM_0; y)=Q_0(\gamma;y)$. 
In particular, for a one-node quiver with dimension $N$ and no arrow, we find $Q_{\cM_0}=1$ 
if $N=0$ and 0 if $N>1$. This reproduces  the fact mentioned in \S\ref{scoulombquiver}, item 5, that $\Omega_{\rm ref}(N\gamma_\ell;y)=\delta_{1,N}$ for any basis vector $\gamma_\ell$ of any quiver.

To further illustrate this method, let us consider 
the simplest example, the Kronecker quiver with 2 nodes 
and $\gamma_{12}=a>0$ arrows from node 1 to 2,
\be
\begin{xy} 0;<1pt,0pt>:<0pt,-1pt>:: 
(0,0) *+{1} ="0",
(98,0) *+{2} ="1",
"0", {\ar|*+{\scriptstyle a}"1"},
\end{xy}
\ee
For $U(1)$ gauge groups at each node, $\gamma=\gamma_1+\gamma_2$.
Eq. \eqref{eq:totalset} then gives
\be
h(\gamma;w)=\frac{w^{a}}{(w-w^{-1})^2}\ ,\quad
\CI(\gamma_\ell;w) = h(\gamma_\ell;w)= \frac{1}{w-w^{-1}}\ .
\ee 
For $c_1 > 0, c_2<0$ we
have $\mu(\gamma_2)<0<\mu(\gamma_1)$ and hence
the sum over $\{\alpha_i\}$ in \eqref{eq:CIrecurs} runs over
the ordered pair $\{\alpha_1,\alpha_2\}=\{\gamma_1, \gamma_2\}$.
Eq. \eqref{eq:CIrecurs} now gives
\be
\CI(\gamma_1+\gamma_2;w) = (w^a-w^{-a})/(w-w^{-1})^2
\ee
and hence
from \eqref{edefbarq0}
and \eqref{eq:stackinv} we get 
\be
Q_0(\gamma_1+\gamma_2;y)
= (-1)^{a+1}(y^a - y^{-a})/(y - y^{-1})\ , \quad \hbox{for $c_1 > 0$} \, . 
\ee
 On the other hand if
$c_1<0$, $c_2>0$, then the sum over $\{\alpha_i\}$ in \eqref{eq:CIrecurs} runs over
the ordered pair $\{\alpha_1,\alpha_2\}=\{\gamma_2, \gamma_1\}$.
Eq. \eqref{eq:CIrecurs} now gives
$\CI(\gamma_1+\gamma_2;w) =0$ and hence from \eqref{edefbarq0}
and \eqref{eq:stackinv} we get $Q_0(\gamma_1+\gamma_2;y)=0$. 
These results agree with the fact that the quiver moduli space is $\CP^{a-1}$
for $c_1>0$, and empty otherwise. The Poincar\'e polynomials for arbitrary 
dimension vectors $n_1 \gamma_1+n_2\gamma_2$ (including non-primitive
vectors with ${\rm gcd}(n_1,n_2)>1$) can be obtained by 
iterating this procedure.
Note that in this case there is no distinction between the embedding space
$\MM_0$ and the actual moduli space $\MM$.

\section{Abelian three-node quiver\label{s3node}}

In this section we shall illustrate the general algorithm outlined in
\S\ref{sintro} for the case of a quiver with three nodes, each carrying
a $U(1)$ gauge group, with $(a,b,c)$ arrows as depicted below:
\be
\label{xy3node}
\begin{xy} 0;<1pt,0pt>:<0pt,-1pt>:: 
(38,0) *+{1} ="0",
(79,67) *+{2} ="1",
(0,69) *+{3} ="2",
"0", {\ar|*+{a}"1"},
"2", {\ar|*+{c}"0"},
"1", {\ar|*+{b}"2"},
\end{xy}
\ee
We consider both the  the cyclic case,  where the
arrows form a loop (i.e. $a,b,c$ all of the same sign), 
and the acyclic case, where the orientation
of the arrows does not allow for any loop. The acyclic case has no pure Higgs states, 
\i.e.\ $\Omega^{\rm S}_{\rm ref}(\gamma_1+\gamma_2+\gamma_3;y)=0$.
Subject to certain inequalities on the number of arrows, 
the cyclic case has a non-trivial middle cohomology, which can be exponentially
large \cite{0702146}. It is a special case  of the more general cyclic quivers analyzed
in \S\ref{scyclic}. All of these cases allow us to test the algorithm of \S\ref{sintro}.

\subsection{Identifying the contributing collinear configurations}  \label{s3.1cou} 

We start by analyzing the Coulomb branch of the quiver, using the localization techniques
of \cite{1011.1258,1103.1887}. According to the prescription of \S\ref{strategy}, we choose
a three vectors  $\gamma_1, \gamma_2, \gamma_3$ such that 
\be \label{esix}
a=\langle \gamma_1, \gamma_2\rangle, \quad b = 
\langle \gamma_2, \gamma_3\rangle, \quad
c=\langle \gamma_3, \gamma_1\rangle\, , 
\ee
By permuting the nodes and/or flipping the sign of all $\gamma_{\ell k}$'s
and $c_\ell$'s, operations which leave \refb{epwextrema} unchanged, we can 
take
 the FI parameters to
 satisfy\footnote{The $c_i$'s are related to the parameters $\theta_i$ in
\cite{1205.5023} (or $\zeta_i$'s in \cite{1205.6511}) 
as $c_i=-\theta_i=-\zeta_i$.}
\be \label{eeight}
c_1>0, \quad c_2> 0, \quad c_3=-c_1-c_2<0\, ,
\ee
as in \cite{1205.5023}. 
We shall avoid situations where one of the $c_\ell$'s vanish since this may lie on a
wall of marginal stability.
Our goal in this subsection will be to identify solutions to \eqref{epwextrema}
which contribute to
$g_{\rm ref}(a,b,c;y )\equiv g_{\rm ref} (\gamma_1, \gamma_2, \gamma_3; y)$
for the above values of the FI parameters. We shall also determine the sign
$s(p)$ associated with these solutions via \refb{espnew}.

According to  \S\ref{strategy}, we need to enumerate the permutations $p$ of (123) for which
solutions to \eqref{epwextrema} exist. Using symmetry under reversal of the $x$-axis, 
we only need to examine three permutations: (123), (213) and (132). We first 
consider the case where none of the multiplicities $a,b,c$ vanish. Let us define
\be \label{enew1}
z_1 = a/|x_1-x_2|, \quad z_2 = b/|x_2-x_3|, \quad z_3 = c/|x_1- x_3|,
\ee
and 
\be \label{enew2}
\sigma_\ell = \sign (x_{\ell+1}-x_\ell) \quad \hbox{for $1\le \ell\le 3$}, \qquad
x_4\equiv x_1\, .
\ee
The equations \refb{epwextrema} determine $z_1,z_2$ in terms of $z_3$ through
\be \label{enew3}
z_1 = z_3 + c_1, \quad z_2=z_3 + c_1 + c_2\, ,
\ee
while $z_3$ is determined by the requirement that  $(x_2-x_1)+(x_3-x_2)+(x_1-x_3)=0$, 
\be \label{enew4}
f(z_3) =0, \quad f(z_3) \equiv {a\,\sigma_1\over z_3+c_1}+ {b\,\sigma_2\over
z_3+c_1+c_2} + {c\,\sigma_3\over z_3}\, .
\ee
Thus for any given ordering specified by the choice of $\sigma_i=\pm 1$ the solutions of \eqref{epwextrema} are in one-to-one correspondence with the
zeros of $f(z_3)$, subject to the  inequalities
\be \label{enew5}
a/(z_3+c_1)>0, \quad b/(z_3+c_1+c_2) >0, \quad c/ z_3 > 0\, ,
\ee
which follow from \refb{enew1} and \refb{enew3}. 
Finally, the determinant of the
Hessian  of $\hat W$ is given by
\be \label{enew6}
\det M =  - \frac{a\, b\, c\, \sigma_1\sigma_2\sigma_3}{ (x_1-x_2)^2 (x_2-x_3)^2
(x_1-x_3)^2} f'(z_3) \, .
\ee
{}From this
we see that any solution to $f(z_3)=0$ contributes a term 
\be 
(-1)^{a+b+c} \, (y-y^{-1})^{-2} \, s(p) \, y^{2  J_3(p)}\, 
\ee
to the Coulomb index $g_{\rm ref}(a,b,c;y)$,
where
\be
2J_3(p)=a\sigma_1+b\sigma_2+c\sigma_3\ ,\quad 
s(p)=\sign[- a\, b\, c\, \sigma_1\sigma_2\sigma_3\, f'(z_3)]\ .
\ee
A short analysis shows that the conditions \eqref{enew5} allow the variable $z_3$ to take
values in an interval $z_{\rm min}<z_3<z_{\rm max}$, depending on the signs of $a,b,c$.
Within this interval, the function $f(z_3)$ may have several zeros, but since solutions contribute
with a sign $s(p)$ proportional to $f'(z_3)$, they will cancel in pairs. Thus, a necessary and sufficient condition for the ordering specified by $\sigma_i$ to contribute is that 
$f(z_3)$ should have opposite signs near the two ends of the allowed range, so that an odd number of solutions exist, in which case the sign $s(p)$ will be that of  $abc\sigma_1\sigma_2\sigma_3$, times the sign of $f(z_3)$ near the lower limit of $z_3$. Below
we tabulate the allowed range of $z_3$ as well as the signs of $f(z_3)$ at the two ends 
of the interval (other combinations of signs of $a,b,c$ are ruled out by 
the conditions \eqref{enew5}) :
\be \label{etable}
\begin{array}{|ccc|c|c|c|c|}
a & b & c & z_{\rm min} & \sign f(z)\vert_{z\to z_{\rm min}^+} 
& z_{\rm max} & \sign f(z)\vert_{z\to z_{\rm max}^-}  \\
+ & + & + & 0 & c \sigma_3 & +\infty & a\sigma_1+b\sigma_2+c \sigma_3 \\
+ & + & - & -c_1 & a\sigma_1 & 0 & -c \sigma_3 \\
- & + &- & -c_1-c_2 & b\sigma_2 & -c_1 & -a\sigma_1\\
- & - & - & -\infty & -(a\sigma_1+b\sigma_2+c \sigma_3) & -c_1-c_2 & -b \sigma_2
\end{array}
\ee
Using this table, it is straightforward to show that the ordering  $p(123) = (123)$,
corresponding to $\sigma_1=\sigma_2=1$, $\sigma_3=-1$, contributes whenever
\be
\label{enew7}
a\, b>0, \quad c<a+b\ ,\quad s(p)=\sign(a)\ ,\quad 2J_3(p)=a+b-c\ ,
\ee
while the ordering $p(123)=(213)$, corresponding to $\sigma_1=\sigma_3=-1$, $\sigma_2=1$
contributes whenever
\be
\label{enew8}
a\, c>0, \quad b>a+c\ ,\quad s(p)=-\sign(a)\ ,\quad 2J_3(p)=b-a-c\ ,
\ee
Finally the ordering $p(123)=(132)$, corresponding to $\sigma_1=1$, $\sigma_2=\sigma_3=-1$,
contributes in four possible cases
\be \label{enew9}
\begin{array}{cl@{\hspace*{1cm}}cl}
 (i)& b,c>0, \quad a>b+c &
 (ii) &a,b>0, \quad c<0 \\
 (iii) & a,c<0, \quad b>0 &
(iv) &b,c<0, \quad a<b+c \\
\end{array}
\ee
with $s(p)=-1$, $2J_3(p)=a-b-c$
in all these cases.

We now consider the case where the multiplicity $a$ vanishes, and $b,c\neq 0$. In that case the
equations \refb{epwextrema} can be solved algebraically. We find that 
solutions exist only when $b>0, c<0$, and their topology depends on the sign of 
$\delta=b c_1+c c_2$. If $\delta>0$, the orderings (213) and (132) contribute
with signs $s(213)=1$, $s(132)=-1$, respectively. If $\delta<0$, the orderings (123) and (132) contribute
with signs $s(123)=1$, $s(132)=-1$, respectively. The Coulomb index $g_{\rm ref}(a,b,c;y) $
is continuous across the locus $\delta=0$, which  corresponds to a wall of threshold stability.
 
Similarly, if $b=0$ and $a,c\neq 0$, we find that solutions exist only when $a<0, c<0$,
and their topology depends on the sign of $\delta=a c_3+c c_2$. For $\delta>0$, 
the orderings $(213)$ and $(132)$ contribute, with $s(213)=1$, $s(132)=-1$, respectively.
For $\delta<0$, the orderings $(213)$ and $(123)$ contribute, with $s(213)=1$, $s(123)=-1$.

Finally, if $c=0$ and $a,b\neq 0$, we find that solutions exist only when $a>0, b>0$,
and their topology depends on the sign of $\delta=a c_3+b c_1$. For $\delta>0$, 
the orderings $(123)$ and $(213)$ contribute, with $s(123)=1$, $s(213)=-1$, respectively.
For $\delta<0$, the orderings $(123)$ and $(132)$ contribute, with $s(123)=1$, $s(132)=-1$.

At last, if two of the multiplicities vanish, one of the centers decouples from the other two and
the solutions to  \eqref{epwextrema} are no longer isolated. This case never arises when discussing non-marginal bound states, as we do in this paper.

We should also discuss the cases where one of the triangle inequalities is
saturated. In this case we can still make use of \refb{etable}, but each of 
the
entries where $(a\sigma_1+b\sigma_2+c\sigma_3)$ appears will
need modification when it vanishes since we cannot use this to determine 
the sign of $f(z_3)$ in appropriate
limits. In such cases we need to go back to the expression for $f(z_3)$
given in \refb{enew4} and keep the
subleading terms to determine the behaviour of $f(z_3)$ in the
$z_3\to\pm\infty$ limit. Take for example the case $c=a+b$ with $a,b,c>0$.
In this case $(a\sigma_1+b\sigma_2+c\sigma_3)$  vanishes for
$\sigma_1=\sigma_2=-\sigma_3$. In this case we see from 
\refb{enew4} that in the $z_3\to\infty$ limit, the sign of $f(z_3)$ is 
given by that of $c\sigma_3$. Since this is the same as the sign of
$f(z_3)$ for $z_3\to 0$ we see that this configuration does not
contribute to $g_{\rm ref}$.
Similar analysis can be done for all other configurations as well.

\subsection{Three-node quiver without loop} 

Let us now consider the case $a<0, b>0, c<0$, corresponding to 
a three-node quiver without loop. The results of the previous subsection 
show that only the orderings $213$ and $132$ contribute, leading to 
the Laurent polynomial
\ben
\label{Coul3nodenoloop}
g_{\rm ref}(\gamma_1, \gamma_2, \gamma_3;y)&=& (-1)^{a+b+c} (y-1/y)^{-2}\,\(
y^{b-a-c}+y^{a+c-b}-y^{a-c-b}-y^{b+c-a}\) \non\\
&=& (-1)^{a+b+c} \, y^{a+c-b-2} \, (1-y^2)^{-2} \left(1 - y^{-2c}\right) 
\left(1-y^{2(b-a)}\right)\, .
\een
Since there are no scaling solutions with two centers, 
\be
\label{eone}
\Omega^{\rm S}_{\rm ref}(\gamma_\ell)=1\ ,\qquad \Omega^{\rm S}_{\rm ref}(\gamma_\ell+\gamma_{\kk}) = 
\Omega_{\rm scaling}(\gamma_\ell+\gamma_{\kk};y) = 0,
\quad \hbox{for $1\le \ell<\kk\le 3$}\, .
\ee
Furthermore for a quiver without closed loop there are also no three centered
scaling solutions and hence
\be \label{ethreezero}
\Omega^{\rm S}_{\rm ref}(\gamma_1+\gamma_2+\gamma_3)
= \Omega_{\rm scaling}(\gamma_1+\gamma_2+\gamma_3;y)=0\, .
\ee
Eq.\refb{essp1} now gives
\be
Q(\gamma_1+\gamma_2+\gamma_3; y) = 
g_{\rm ref} (\gamma_1, \gamma_2, \gamma_3; y) \ .
\ee

On the other hand, the quiver moduli space is described by the D-term equations
\ben
\sum_{\gamma=1}^{|c|} |\phi_{13,\gamma}|^2 -\sum_{\alpha=1}^{|a|} 
|\phi_{21,\alpha}|^2 &=& c_1\non  \\
\sum_{\beta=1}^{b} |\phi_{23,\beta}|^2 + \sum_{\alpha=1}^{|a|} 
|\phi_{21,\alpha}|^2 &=& c_2 \, .
\een
Since the diagonal $U(1)$ acts trivially on all the fields, they define a manifold $\cM$ of
complex dimension $|a|+b+|c|-2$, which is a smooth $\CP^{|c|-1}$ bundle over 
$\CP^{|a|+b-1}$. The Poincar\'e polynomial of $\cM$ is the product of the Poincar\'e
polynomial of these two projective spaces, in perfect agreement with \eqref{Coul3nodenoloop}. 

\subsection{Three node quiver with loop \label{sec_3loop}} 

Let us now consider a three-node quiver with loop,  choosing $a>0,b>0,c>0$. 
In the case where the triangular inequalities 
\be
\label{3trian}
a<b+c,\quad b<a+c\ ,\quad c<a+b\ ,
\ee 
hold,
the analysis of \S\ref{s3.1cou} shows that only the ordering $(123)$ 
contributes, leading to  
\be \label{eks2}
g_{\rm ref} (\gamma_1, \gamma_2, \gamma_3; y) =
(-1)^{a+b+c} (y-y^{-1})^{-2} \left( y^{a+b-c} + y^{c-a-b}\right)\, .
\ee
Unlike \eqref{Coul3nodenoloop}, this is not a Laurent polynomial, as expected
since the Coulomb moduli space has scaling regions. 
Applying the prescription of \S\ref{strategy} we find
\be \label{efour}
Q(\gamma_1+\gamma_2+\gamma_3; y) 
= g_{\rm ref} (\gamma_1, \gamma_2, \gamma_3; y) + 
\Omega^{\rm S}_{\rm ref}(\gamma_1+\gamma_2+\gamma_3) + H(\{\gamma_1,\gamma_2,\gamma_3\};\{1,1,1\}; y)\, .
\ee
The unique choice of $H$, which is even under $y\to y^{-1}$, vanishes
as $y\to\infty$ and makes the right hand side of \eqref{efour} a
polynomial in $y$, $y^{-1}$ is given by\footnote{Since $H$ is independent of 
the FI parameters, the result \eqref{eks4}
can be used for any scaling subquiver of a larger quiver. }
\be \label{eks4}
H(\{\gamma_1,\gamma_2,\gamma_3\};\{1,1,1\}; y) =
\begin{cases} - 2   \,  (y-y^{-1})^{-2}\, \, \hbox{for $a+b+c$ even}\cr
(y + y^{-1}) \, (y-y^{-1})^{-2} \, \, \hbox{for $a+b+c$ odd}
\end{cases}
\ee
Substituting these in \eqref{efour} we finally get
\be
\begin{split}
 \label{efin3node}
Q(\gamma_1+\gamma_2+\gamma_3; y) &=
\Omega^{\rm S}_{\rm ref}(\gamma_1+\gamma_2+\gamma_3) + \\
+ & 
(y-1/y)^{-2}  \times 
\begin{cases}
 \left( y^{a+b-c} + y^{c-a-b} -2\right)\ , & \hbox{for $a+b+c$ even} \\
- \left( y^{a+b-c} + y^{c-a-b} -y - y^{-1}\right), & \hbox{for $a+b+c$ odd}
\end{cases} 
\end{split}
\ee
On the other hand, if $a,b,c$ are all positive but the triangle inequalities are
violated, then it follows from the analysis of \S\ref{s3.1cou} that
\ben \label{e3.21a}
 Q(\gamma_1+\gamma_2+\gamma_3; y) &=&
 \frac{(-1)^{a+b+c}}{(y-y^{-1})^{2}} 
 \begin{cases}
 y^{a+b-c} + y^{c-a-b} -y^{b+c-a} - y^{a-b-c}
 & \hbox{for $a>b+c$} \\
 y^{a+b-c} + y^{c-a-b} -y^{a+c-b} - y^{b-a-c}
& \hbox{for $b>a+c$}  \\
0 & \hbox{for $c>a+b$}
\end{cases}
\non\\
\een
Note that in these cases we have set $\Omega^{\rm S}_{\rm ref}(\gamma_1
+\gamma_2+\gamma_3)$ to zero since the Coulomb branch moduli space does
not have scaling region.
Finally when any one of the inequalities is saturated then we can use
either \eqref{efin3node} or \eqref{e3.21a} since they give the same result.

Let us now compare this result with the cohomology of the Higgs branch. Since the
loop allows for a superpotential $W=\sum_{\alpha\beta\gamma} C_{\alpha\beta\gamma} \,\phi_{12,\alpha}\phi_{23,\beta}\phi_{31,\gamma}$, the moduli space of classical vacua is
described by the F-term
\be
\partial_{\phi_{12,\alpha}} W = \partial_{\phi_{23,\beta}} W = \partial_{\phi_{31,\gamma}} W=0
\ee
and  D-term constraints
\beq
\sum_{\alpha=1}^a |\phi_{12,\alpha}|^2  - \sum_{\gamma=1}^c |\phi_{31,\gamma}|^2 &=& c_1 \nn \\
\sum_{\beta=1}^b |\phi_{23,\beta}|^2 - \sum_{\alpha=1}^a |\phi_{12,\alpha}|^2 &=& c_2 \\
\sum_{\beta=1}^c |\phi_{31,\gamma}|^2 - \sum_{\beta=1}^b  |\phi_{23,\beta}|^2 &=& c_3 
=-c_1-c_2\, .\nn
\eeq
As shown in \cite{0702146}, for generic choice of the superpotential the moduli space 
splits into three different branches, where one of set of chiral multiplets 
$\phi_{12}, \phi_{23}$ or $\phi_{31}$
vanishes. For the choice of FI terms in \eqref{eeight}, $\phi_{31}$ vanishes  identically, so that
the solution to the D-term constraints modulo gauge transformation is given by 
$\CP^{a-1}\times \CP^{b-1}$ parametrized by $\phi_{12,\alpha}$ and
$\phi_{23,\beta}$ respectively, 
upon which the  F-term conditions $\partial_{\phi_{31,\gamma}} W=0$ impose $c$ bilinear constraints. Thus, $\cM$ is a complete intersection in 
$\CP^{a-1}\times \CP^{b-1}$.
Its cohomology can be computed by the Lefschetz 
hyperplane theorem, which predicts 
\ben
Q(\cM; y) &\simeq&
(-1)^{a+b+c} y^{c-a-b+2} (1-y^2)^{-2} (1-y^{2a}) (1-y^{2b}) \non\\
&\simeq& (-1)^{a+b+c} (y-1/y)^{-2}\, y^{c-a-b} + \cO(1)\ ,
\een
where $\simeq$ denotes equality up to additive constant and positive
powers of $y$. This is
in agreement with \eqref{efin3node}. The constant $\Omega^{\rm S}_{\rm ref}(\gamma_1+\gamma_2+\gamma_3)$ in \eqref{efin3node} correspond to the 
`pure Higgs states' carrying zero angular momentum. 

We shall now
obtain the undetermined constant $\Omega^{\rm S}_{\rm ref}(\gamma_1+\gamma_2+\gamma_3)$, by computing the Euler number of $\cM$ using the Riemann-Roch theorem. This computation was first carried out in \cite{1205.5023}, generalizing the analysis
of \cite{0702146}. We shall extend these results by computing the Hirzebruch polynomial \eqref{defhirz} of the quiver moduli space, which provides finer information on the middle
cohomology.

For the three-node with loop of interest in this section, $\cM$ is a complete intersection
of codimension $c$ in the product $\CP^{a-1}\times \CP^{b-1}$. After performing the 
change of variable \eqref{JtoR}, we find that the Hirzebruch polynomial is given by 
\be
\begin{split}
\chi(a,b,c;v) = \oint &
\frac{ \de R_1}{2\pi\I\,(1-R_1)(1-v R_1)R_1^{a}}
\frac{\de R_2}{2\pi\I\, (1-R_2)(1-v R_2)R_2^{b}}\\
&\times
\left( \frac{R_1+R_2-R_1 R_2(1+v)}{1-R_1R_2 v}\right)^c 
\end{split}
\ee
To evaluate this integral, it is useful to construct the partition function
\be
\chi(x_1,x_2,x_3;y)= \sum_{a\geq 0,b\geq 0,c\geq 0} (-y)^{-a-b+c+2}
\chi(a,b,c;y^2) \, x_1^a\, x_2^b\, x_3^c\ .
\ee
Summing up the geometric series and computing the contour integral using Cauchy's theorem,
we arrive at 
\be
\begin{split}
\label{echivv}
\chi&(x_1,x_2,x_3;y)=\\
&\frac{x_1x_2(1-x_1 x_2 )}
{(1+x_1 y)(1+x_1/y)(1+x_2 y)(1+x_2/y)
\left[ 1-x_1x_2-x_2 x_3-x_1 x_3- x_1x_2x_3 (y+1/y)\right]}\, .
\end{split}
\ee

On the other hand,
denoting by $\wh Q(x_1,x_2,x_3;y,t)$ the generating function
of the Dolbeault polynomial $\wtQ(a,b,c;$ $y,t)\equiv 
\wt Q(\gamma_1+\gamma_2+\gamma_3;y,t)$,
\be \label{egenqhat}
\wh Q(x_1,x_2,x_3;y,t) = \sum_{a\ge 0, b\ge 0, c\ge 0} (x_1)^a (x_2)^b (x_3)^c
\wtQ(a,b,c;y,t)\, , 
\ee
 we find by
 using  \eqref{e3.21a}, \eqref{ehodge2} and \eqref{efin3node}
\be
\begin{split}
\wh Q&(x_1,x_2,x_3;y,t)=\wh Q^{\rm S}(x_1,x_2,x_3;t)\\
&+\frac{x_1 x_2\left\{1-x_1x_2+x_1 x_2 x_3 \left(x_1+x_2 +
    y + y^{-1}
    \right)\right\}}{(1-x_1 x_2) (1-x_1 x_3)(1-x_2 x_3) (1+x_1/y)
    (1+x_1 y) (1+x_2/y) (1+x_2 y)}\ .
\end{split}
\ee
Here $\wh Q^{\rm S}(x_1,x_2,x_3;t)$ is the generating function of
$\wt\Omega^{\rm S}_{\rm ref}(\gamma_1+\gamma_2+\gamma_3;t)$.
Note that the $t$ dependence comes only from $\wh Q^{\rm S}$.
Now according to  \eqref{PoincaHirz}, at $t=1/y$ this should reduce to
$(-y)^{-d} \chi(y^2)$. Comparing this with \eqref{echivv} we find
\be
\label{QS3t}
\wh Q^{\rm S}(x_1,x_2,x_3;t)=\frac{x_1^2 x_2^2 x_3^2}{
(1-x_1 x_2)(1-x_2 x_3)(1-x_1 x_3)[1-x_1x_2-x_2 x_3-x_1 x_3- x_1x_2x_3 (t+1/t)]}
\ee
It is striking that $\wh Q^{\rm S}(x_1,x_2,x_3;t)$ is symmetric under permutations of 
$x_1,x_2,x_3$, which implies that the middle cohomology states are robust under
wall-crossing.  This property at $t=1$ was noticed in \cite{1205.5023}.

Finally, let us test the HN recursion method described in \S\ref{srepresent}
by computing the cohomology of the quiver moduli space in the absence of a
superpotential. We still assume $a,b,c>0$ and $c_1>c_2>0$, $c_3=-c_1-c_2<0$. 
Using the fact that the  slopes are  ordered according to  
\be
\gamma_3 <\gamma_2+\gamma_3
<\gamma_1+\gamma_3<
\gamma_1+\gamma_2+\gamma_3<
\gamma_2 < \gamma_1+\gamma_2< \gamma_1\ ,
\ee
we find from \eqref{eq:solrecursion} %
\begin{eqnarray} 
&&\CI(\gamma_1;w)=\CI(\gamma_2;y)=\CI(\gamma_3;y)={1/(w - w^{-1})}, \non \\
&&\CI(\gamma_1+\gamma_2;w)=  (w^a - w^{-a})/(w - w^{-1})^2,\non\\
&&\CI(\gamma_2+\gamma_3;w)= (w^b - w^{-b})/(w - w^{-1})^2,\\
&&\CI(\gamma_1+\gamma_3;w)=0 ,\non
\een
and, from \eqref{eq:stackinv} and \refb{eq:solrecursion}
\ben \label{erep1}
&&Q_0(\gamma_1+\gamma_2+\gamma_3;y)=(-1)^{a+b+c}y^{-c} (y^a - y^{-a})(y^b -
y^{-b})/(y - y^{-1})^2  \non\\
&&\qquad \qquad \qquad \qquad \, \, \,
= (-1)^{a+b+c} y^{-a-b-c+2} (1-y^{2a}) (1-y^{2b}) (1-y^2)^{-2}\, .
\end{eqnarray}
This agrees with the fact that the embedding space $\MM_0$ is given by $\CP^{a-1} \times \CP^{b-1}\times{\mathbb C}^c$ \cite{1205.5023,1205.6511}.

\section{Cyclic quivers} \label{scyclic}

We shall now consider a generic cyclic quiver with $K$ nodes, of the 
form\be
\begin{xy} 0;<1pt,0pt>:<0pt,-1pt>:: 
(65,0) *+{1} ="0",
(129,47) *+{2} ="1",
(105,123) *+{\dots} ="2",
(25,123) *+{K-1} ="3",
(0,47) *+{K} ="4",
"0", {\ar|*+{a_1}"1"},
"1", {\ar|*+{a_2}"2"},
"2", {\ar"3"},
"3", {\ar|*+{a_{K-1}}"4"},
"4", {\ar|*+{a_K}"0"},
\end{xy}\\
\ee
We assume that each node
carries a $U(1)$ factor.  We take $\gamma_{\ell (\ell+1)} = a_\ell > 0$ for $\ell=1,\cdots K-1$,
$\gamma_{K1}= a_K > 0$, and choose the FI parameters to satisfy
\be \label{egen1}
 c_1, c_2, \cdots c_{K-1}>0, \quad c_K<0\, .
\ee
The Higgs branch of this class of quivers was analyzed in \cite{1205.6511}.
Since in this case there are no subquivers with closed loops the
analysis of both the Coulomb branch and Higgs branch simplifies. 

\subsection{Coulomb branch analysis}

The prescription of \S\ref{strategy} yields
\be \label{ecycle2aa}
Q(\gamma_1+\cdots +\gamma_K;y)
= g_{\rm ref}(\gamma_1,\cdots \gamma_K;y) +
H(\{\gamma_1,\cdots \gamma_K\}, \{1,\cdots 1\};y)
+ \Omega^{\rm S}_{\rm ref}(\gamma_1+\cdots +\gamma_K)
\, .
\ee
To evaluate the Coulomb index $g_{\rm ref}(\gamma_1,\cdots \gamma_K;y)$,
we need to find  the solutions to \eqref{epwextrema} for this system. Extending the
procedure of \S\ref{s3.1cou}, let us  define
\ben \label{egg3}
&& z_\ell\equiv \frac{a_\ell \sigma_\ell}{x_{\ell+1}-x_\ell}\ ,\quad 
\sigma_\ell\equiv \sign(x_{\ell+1}-x_\ell) \quad \hbox{for $1\le \ell\le K-1$}\, , \nonumber \\
&& 
z_K \equiv \frac{a_K \sigma_K}{x_{1}-x_K}, \quad 
\sigma_K\equiv \sign(x_{1}-x_K)\, ,
 \een
and rewrite \eqref{epwextrema} as
\be \label{egg2}
z_{\ell+1} - z_{\ell} = c_{\ell+1} \quad \hbox{for $1\le\ell\le K-2$}\ ,\qquad z_1 - z_K = c_1\, .
\ee
Since all the $a_\ell$'s are positive, this is also the case of the 
$z_\ell$'s. 
Without any loss of generality we can fix $x_1=0$. This gives
\be \label{egg4}
z_\ell=z_K + \sum_{\kk=1}^\ell c_{\kk}\ ,\qquad  x_\ell=\sum_{\kk=1}^{\ell-1} 
\frac{a_{\kk}\sigma_{\kk}}{z_\kk}\, ,
\ee
where the only unknown $z_K$ is determined by the algebraic equation
\be \label{egg5}
f(z_K)=0 \qquad \hbox{where} \qquad f(z_K)\equiv \sum_{\ell=1}^{K} \frac{a_\ell\sigma_\ell}{z_\ell}\, .
\ee
Since we assume that all $a_\ell$ ($\ell=1\dots K$) and $c_\ell$ ($\ell=1\dots K-1)$ are positive, 
the only requirement on
the solution of \eqref{egg5} is that $z_K>0$. For such a solution, 
the determinant of the Hessian $M_{\ell\kk}=\p^2 \hat W/\p x_\ell \p x_\kk$ for $2\le \ell,\kk\le K$
evaluates to 
\be \label{egg7}
\det M = - 
f'(z_K)\,
\prod_{\ell=1}^{K} \frac{a_\ell \sigma_\ell }{(x_{\ell+1}-x_\ell)^2}
\, , \quad  x_{K+1}\equiv x_1\, .
\ee
Thus, a solution to $f(z_K)=0$ contributes to $g_{\rm ref}(\gamma_1,\cdots \gamma_K;y)$ with a sign 
\be \label{egg8}
s(p)=  - \sign\left[f'(z_K)\right] \, \prod_{\ell=1}^{K} \sigma_\ell \  .
\ee
In general, the equation $f(z_K)=0$ may have several solutions in the range $0<z_K<+\infty$.
However, due to \eqref{egg8}, the contribution of these solutions to $g_{\rm ref}(\gamma_1,\cdots \gamma_K;y)$ will cancel in pairs. The only possibility for the ordering specified by $\sigma_i$
to contribute is that there should be an odd number of solutions. For this 
we need to ensure that $f(z_K)$ has opposite signs in the two
extreme limits: as $z_K\to 0$ and as $z_K\to\infty$. As long as the $c_\ell$'s
are not zero we see from \eqref{egg4} that all the $z_\ell$'s other than $z_K$
remain finite in the $z_K\to 0$ limit and hence $f(z_K)$ approaches
$a_K\sigma_K/z_K$. On the other hand as $z_K\to\infty$, we see from \eqref{egg4} that
all the other $z_\ell$'s also approach infinity keeping the difference
$z_\ell-z_K$ finite and $f(z_K)$ 
goes as ${\sum_{\ell=1}^{K}} a_\ell\sigma_\ell/ z_K$. Thus
\eqref{egg5} has an odd number of solutions if
\be \label{egg6}
\sign\left[{\sum_{\ell=1}^{K}} a_\ell\sigma_\ell\right] =
- \sigma_K \, .
\ee
As indicated above, if there is more than one solution the solutions will cancel in pairs, 
but   the sign of $f'(z_K)$ at the uncancelled solution
will be the opposite of the sign of $f(z_K)$ as  $z_K\to 0$.
Since the sign of $f(z_K)$ as $z_K\to 0$ is $\sigma_K$, we get, from \refb{egg8}, 
\be \label{sfinsig}
s(p) = \prod_{\ell=1}^{K-1} \sigma_\ell\, .
\ee
Using \eqref{eclassint} we arrive at
\be \label{egg9}
g_{\rm ref}(\gamma_1,\cdots \gamma_K; y)
= (-1)^{K-1 + \sum_{\ell} a_\ell} (y-y^{-1})^{-K+1} \, \sum_{\sigma_1=\pm 1, \sigma_2=\pm 1,
\cdots \sigma_K=\pm 1\atop 
\sign\left[{\sum_{\ell=1}^{K}} a_\ell\sigma_\ell\right] =
- \sigma_K}\,  \left(\prod_{\ell=1}^{K-1} \sigma_\ell\right)\, 
y^{\sum_{\ell=1}^K \sigma_\ell a_\ell}\, .
\ee
Inserting this result in  \eqref{ecycle2aa}, we find
\be\label{ecycle2bb}
\begin{split}
Q&(\gamma_1+\cdots +\gamma_K;y) =
\Omega^{\rm S}_{\rm ref}(\gamma_1+\cdots +\gamma_K)
+H(\{\gamma_1,\cdots \gamma_K\}, \{1,\cdots 1\};y)
 \\
& +(-1)^{K-1 + \sum_{\ell} a_\ell} (y-y^{-1})^{-K+1} \,
\sum_{\sigma_1=\pm 1, \sigma_2=\pm 1,
\cdots \sigma_K=\pm 1\atop 
\sign\left[{\sum_{\ell=1}^{K}} a_\ell\sigma_\ell\right] =
- \sigma_K}\,  \left(\prod_{\ell=1}^{K-1} \sigma_\ell\right)\, 
y^{\sum_{\ell=1}^K \sigma_\ell a_\ell} \, .
\end{split}
\ee
$H(\{\gamma_1,\cdots \gamma_K\}, \{1,\cdots 1\};y)$ is fixed uniquely
by demanding that it is symmetric under $y\to y^{-1}$, vanishes as $y\to \infty, 0$,
and that $Q(\gamma_1+\cdots +\gamma_K;y)$ is a Laurent polynomial in $y$.
It can be obtained for example using the contour integral prescription \eqref{uint},
inserting the second line of \eqref{ecycle2bb} in place of $f(y)$.
The constant 
$\Omega^{\rm S}_{\rm ref}(\gamma_1+\cdots +\gamma_K)$ 
appearing in \eqref{ecycle2bb} can be determined from the Euler characteristics
of the Higgs branch, as explained in the next subsections. 

In preparation for the analysis of the Higgs branch, let us now try to identify the negative
 powers of $y$ in \eqref{ecycle2bb}. Firstly, 
 neither $H$ nor $\Omega^{\rm S}$ contributes negative powers of $y$ in
an expansion around $y=0$ since $H$ vanishes as $y\to 0$ and $\Omega^{\rm S}$
is $y$-independent constant. To get negative powers of $y$ from the first term on the
right hand side of \eqref{ecycle2bb}, we need $\sum_{\ell=1}^K\sigma_\ell a_\ell<0$.
Due to the restriction on the $\sigma_\ell$'s in the sum, this 
implies that $\sigma_K=1$. Thus we can express \eqref{ecycle2bb} as
\be \label{eqnew}
Q(\gamma_1+\cdots +\gamma_K;y) 
\simeq (-1)^{K-1 + \sum_{\ell} a_\ell} (y-y^{-1})^{-K+1} \!\!\!\!
\sum_{\sigma_1=\pm 1, \sigma_2=\pm 1,
\cdots \sigma_{K-1}=\pm 1\atop 
{\sum_{\ell=1}^{K-1}} a_\ell\sigma_\ell + a_K<0}\,  \left(\prod_{\ell=1}^{K-1} \sigma_\ell
\right)\, 
y^{\sum_{\ell=1}^{K-1} \sigma_\ell a_\ell+a_K} \, ,
\ee
where as usual $\simeq$ denotes equality up to additive constant and positive powers
of $y$.

\subsection{Higgs branch analysis}

Now according to the analysis of \cite{1205.6511} the moduli space $\MM$
of this quiver
is a codimension $a_K$ complete intersection hypersurface in
$\CP^{a_1-1}\times \cdots \times \CP^{a_{K-1}-1}$. Thus the
complex dimension of this manifold is given by
\be \label{egen2}
d = \sum_{\ell=1}^{K-1} a_{\ell} - a_K - (K-1)\, .
\ee
By Lefschetz hyperplane theorem, the cohomology of $\MM$
coincides with that of $\CP^{a_1-1}\times \cdots \times 
\CP^{a_{K-1}-1}$ for degree less than $d$. Since the Poincar\'e
polynomial of $\CP^{n-1}$ is given by
$(1 - t^{2n}) / (1 - t^2)$, we see that the first $d-1$ powers of $t$ of the
Poincar\'e polynomial of $\MM$ is given by that of
$\prod_{\ell=1}^{K-1} \{(1 - t^{2a_\ell}) / (1 - t^2) \}$. Thus the Laurent polynomial associated
to $\MM$ is given by
\ben \label{egen3}
Q(\MM;y) &\simeq& (-y)^{-\sum_{\ell=1}^{K-1} a_{\ell} + a_K + (K-1)}
\, \prod_{\ell=1}^{K-1} \{(1 - y^{2a_\ell}) / (1 - y^2) \}
\nonumber \\
&\simeq& (-1)^{-\sum_{\ell=1}^{K} a_{\ell} }\, 
(y - y^{-1})^{-K+1} \, y^{-\sum_{\ell=1}^{K-1} a_{\ell} + a_K}\,
\prod_{\ell=1}^{K-1} (1 - y^{2a_\ell})\, ,
\een
where $\simeq$ denotes equality of terms involving negative powers
of $y$. The terms in $Q(\MM;y)$ involving positive powers of $y$ are
given by the $y\to y^{-1}$ symmetry. Now to identify terms in \eqref{egen3}
involving negative powers of $y$, we can explicitly expand the product
$\prod_{\ell=1}^{K-1} (1 - y^{2a_\ell})$, and pick up those powers of
$y$, which when multiplied by $y^{-\sum_{\ell=1}^{K-1} a_{\ell} + a_K}$, still
gives negative powers of $y$. Thus we get
\ben \label{egen4}
Q(\MM;y) &\simeq& (-1)^{-\sum_{\ell=1}^{K-1} a_{\ell} + a_K +K-1}\, 
(y - y^{-1})^{-K+1} 
\nonumber \\
&&  \times 
\sum_{\vec\sigma\atop \sigma_1 a_1 +\cdots \sigma_{K-1}a_{K-1}
+ a_K < 0} \left(\prod_{\ell=1}^{K-1} \sigma_\ell \right)\, 
 y^{\sigma_1 a_1+\cdots \sigma_{K-1} a_{K-1} + a_K}\, ,
\een
where the sum over $\vec\sigma$ runs over all $K-1$ dimensional vectors
of the form $(\pm1, \pm 1, \cdots \pm 1)$ subject to the restriction
given above. This is in perfect agreement with \eqref{eqnew}. The agreement
between the positive powers of $y$ between $Q(\MM;y)$ and
$Q(\gamma_1+\cdots +\gamma_K;y) $ then follows from the $y\to y^{-1}$
symmetry of both terms. The $H$ in \eqref{eqnew} ensures that 
$Q(\gamma_1+\cdots +\gamma_K;y)$, like $Q(\MM;y)$, is a Laurent
polynomial in $y$. Finally the $\Omega^{\rm S}_{\rm ref}(\gamma_1+\cdots +\gamma_K)$
in \eqref{eqnew} will have to be adjusted so that the constant terms
in the expressions for $Q(\MM;y)$ and $Q(\gamma_1+\cdots +\gamma_K;y) $
match.

\subsection{Middle cohomology}

Using the Riemann-Roch theorem summarized in  \S\ref{sec_abcohom}, we find that the Hirzebruch polynomial is given  by the contour integral 
\be
\label{Leeintnew}
\chi(a_1,\dots a_{K};v) = \oint R_v(\sum_{\ell=1}^{K-1} 
J_\ell )]^{a_{K}} \, \prod_{\ell =1}^{K-1} 
\frac{\de J_\ell }{2\pi\I\, [R_v(J_\ell )]^{a_\ell }} 
\ee
where $R_v(J)$ has been defined in \eqref{edefrvj}.
Changing variables from $J_\ell $ to $R_\ell =R_v(J_\ell )$, we find
\be
\label{Leeintnewer}
\chi(a_1,\dots a_{K};v) = 
\oint \{ R_v[\sum_{\ell =1}^{K-1} R_v^{-1}(R_\ell )]\}^{a_{K}} \, \prod_{\ell =1}^{K-1} 
\frac{\de R_\ell }{2\pi\I\, (1-R_\ell )(1-v R_\ell )R_\ell ^{a_\ell }}\ .
\ee
Thus, the partition function, after carrying out the $R_\ell$ integrals, is found 
\be \label{esschi}
\begin{split}
 \chi(x_1,\dots, x_{K};y) \equiv &
\sum_{a_1,\cdots a_K}
(-y)^{-a_1-\cdots -a_{K-1}+a_K +K-1} \chi(a_1,\dots a_{K};y^2) \,
x_1^{a_1}\dots x_K^{a_K} \non\\
=& \frac{1}{1+x_{K} y R_{y^2}[\sum_{\ell=1}^{K-1} R_{y^2}^{-1}(-x_\ell /y)]}
\prod_{\ell =1}^{K-1} \frac{x_\ell }{(1+x_\ell /y)(1+x_\ell  y)}
\end{split}
\ee
With some work one may express $R_v[\sum_{\ell=1}^{K-1} 
R_v^{-1}(x_\ell )]$ in terms of the $x_\ell $,
\be \label{echeckthis}
R_{y^2}\left[ \sum_{\ell=1}^{K-1} R_{y^2}^{-1}(-x_\ell/y) \right] = 1+ {(y - y^{-1}) \prod_{\ell =1}^{K-1}
(1+x_\ell /y) \over y^{-1} \prod_{\ell =1}^{K-1}
(1+x_\ell   y) - y \prod_{\ell =1}^{K-1}
(1+x_\ell /y)}\, .
\ee
This gives
\be \label{echixx1}
\chi(x_1,\dots, x_{K};y)= \prod_{\ell =1}^{K-1} \frac{x_\ell }{(1+x_\ell /y)(1+x_\ell  y)}\,
{y^{-1} \prod_{\ell=1}^{K-1} (1 + x_\ell y) - y \prod_{\ell=1}^{K-1} (1 + x_\ell/ y)
\over
y^{-1} \prod_{\ell=1}^{K} (1 + x_\ell y) - y \prod_{\ell=1}^{K} (1 + x_\ell /y)
}\, .
\ee
Setting $y=1$, we find
\be \label{echixx2}
\chi(x_1,\dots, x_{K};1)=\frac{1}{1+x_{K}} \prod_{\ell =1}^{K-1} \frac{x_\ell}{(1+x_\ell)^2}\,
+ D(x_1,\dots, x_{K})\, 
\prod_{\ell =1}^{K} \frac{x_\ell}{1+x_\ell}\ ,
\ee
where 
\be
\label{echixx3}
D(x_1,\dots, x_{K}) = 
\left( 1-\sum_{k=1}^K \frac{x_k}{1+x_k}\right)^{-1}\, \prod_{\ell =1}^{K} \frac{1}{1+x_\ell}\, .
\ee
The function \eqref{echixx3} is recognized as the generating function of the number 
$D(a_1,\dots a_K)$ of derangements
 of a set of $\sum_{\ell=1}^K{a_\ell}$ objets of $K$ different types, with $a_k$ objects of  
type $k$ for $k=1\dots K$. This partition function was computed in \cite{EvenGillis} for
arbitrary $K$, and its relevance for the counting of pure Higgs states 
was noted  in \cite{1205.5023} in the case of 3-node quivers.

On the other hand, the Dolbeault
polynomial of the quiver moduli space is given analogously to \eqref{ecycle2bb} by
\be \label{ecycle2bbss}
\begin{split}
\wt Q(\gamma_1+&\cdots +\gamma_K;y,t) =\wt\Omega^{\rm S}_{\rm ref}(\gamma_1+\cdots +\gamma_K;t)+
H(\{\gamma_1,\cdots \gamma_K\}, \{1,\cdots 1\};y)
\\&
+(-1)^{K-1 + \sum_{\ell} a_\ell} (y-y^{-1})^{-K+1} \,
\sum_{\sigma_1=\pm 1, \sigma_2=\pm 1,
\cdots \sigma_K=\pm 1\atop 
\sign\left[{\sum_{\ell=1}^{K}} a_\ell\sigma_\ell\right] =
- \sigma_K}\,  \left(\prod_{\ell=1}^{K-1} \sigma_\ell\right)\, 
y^{\sum_{\ell=1}^K \sigma_\ell a_\ell} \ .
\end{split}
\ee
Since the second term $H$ vanishes as $y\to 0$, we can ignore it
for the purpose of determining the non-positive powers of 
$y$ in $\wt Q(\gamma_1+\cdots +\gamma_K;y,t)$. 
The constraint $\sign\left[{\sum_{\ell=1}^{K}} a_\ell
\sigma_\ell\right] =
- \sigma_K$ then implies that negative powers of $y$ only come from
terms with $\sigma_K=1$. Moreover,  for such terms
we can drop the constraint $\sign\left[{\sum_{\ell=1}^{K}} a_\ell
\sigma_\ell\right] =
- \sigma_K$ since terms which violate this constraint carry positive powers of $y$.
Thus the generating function
for $\wt Q(\gamma_1+\cdots +\gamma_K;y,t)$ can be written as
\be
\label{egenerate}
\begin{split}
\wh Q&(x_1,\cdots x_K;y,t)  :\simeq\wh Q^{\rm S}(x_1,\cdots x_K;t) 
\\
&\quad +  \sum_{\{a_\ell\}\atop a_\ell\ge 0 \, \forall \, \ell} (x_\ell)^{a_\ell} \,
(-1)^{K-1 + \sum_{\ell} a_\ell} (y-y^{-1})^{-K+1} \!\!\!\!\!
\sum_{\sigma_1=\pm 1, \sigma_2=\pm 1,
\cdots \sigma_{K-1}=\pm 1}\,  \left(\prod_{\ell=1}^{K-1} \sigma_\ell\right)\, 
y^{\sum_{\ell=1}^{K-1} \sigma_\ell a_\ell} y^{a_K}
\\
& :\simeq   \wh Q^{\rm S}(x_1,\cdots x_K;t)
+ {1\over 1+x_K y} \prod_{\ell=1}^{K-1} {x_\ell\over (1+x_\ell y) (1+x_\ell/y)}\, ,
\end{split}
\ee
where $:\simeq$ denotes equality up to additive positive powers of $y$,
and $\wh Q^{\rm S}$ is the generating function for $\wt \Omega^{\rm S}_{\rm ref}$. Now
according to \eqref{PoincaHirz} we can 
equate this at $t=1/y$ to  $\chi(x_1,\dots, x_{K};y)$. This gives
\be \label{efinqsvalue}
\wh Q^{\rm S}(x_1,\cdots x_K;1/y) :\simeq 
{1-y^2\over y^{-1} \prod_{\ell=1}^K (1+x_\ell y) - y \prod_{\ell=1}^K (1 + x_\ell/y)}\, 
\prod_{\ell=1}^K {x_\ell\over 1 + x_\ell y}
\ee
After expanding  in powers of the $x_\ell$'s and picking  the coefficient of 
the monomial $\prod_\ell x_\ell^{a_\ell}$. the right hand side of \eqref{efinqsvalue}
gives the negative and zero powers of $y$
in  $\wt\Omega^{\rm S}_{\rm ref}(\gamma_1+\cdots \gamma_K;1/y)$.
The positive powers of $y$ are found using the $y\to 1/y$ symmetry.

In fact, one may compute the complete partition function of $\wt\Omega^{\rm S}_{\rm ref}(\gamma_1+\cdots \gamma_K;1/y)$, including positive powers of $y$, by 
using the prescription \eqref{uint}. One can exchange the sums over $a_\ell$ with the 
integral over 
$u$ as long as $|x_\ell|<< |u|$ $\forall \ell$.
This gives
\ben \label{eqintegral}
\wh Q^{\rm S}(x_1,\cdots x_K;1/y)  &=& \oint \frac{\de u}{2\pi\I} \frac{(1/u-u)}{(1-uy)(1-u/y)}
{1-u^2\over u^{-1} \prod_{\ell=1}^K (1+x_\ell u) - u \prod_{\ell=1}^K (1 + x_\ell/u)}\, 
\non \\ 
&& \quad \quad \times \prod_{\ell=1}^K {x_\ell\over 1 + x_\ell u}\, ,
\een
where it is understood that the $u$ integration contour encloses all poles
which go to zero as $x_\ell\to 0$ but does not enclose any other poles.
The integral can be evaluated as follows:
\begin{enumerate}
\item We first make a change of variables from $u\to 1/u$. This moves the
integration contour so as to enclose the poles at $y$ and $1/y$.
\item We now deform the integration contour back to the original position,
in that process picking up residues at the poles at $u=y$ and $u=1/y$.
\item We then take the average of the original integral \refb{eqintegral}
and the new result.
\end{enumerate}
At the end of the process one arrives at the result
\ben\label{eqfind1}
\wh Q^{\rm S}(x_1,\cdots x_K;1/y)  &=& -{1\over 2}
{y -1/y \over y^{-1} \prod_{\ell=1}^K (1+x_\ell y) - y \prod_{\ell=1}^K (1 + x_\ell/y)}\, 
\non \\
&& \quad \times\left[  y \prod_{\ell=1}^K {x_\ell\over 1 + x_\ell y} + 
y^{-1} \prod_{\ell=1}^K {x_\ell\over 1 + x_\ell y^{-1}} \right]\non\\
&&-{1\over 2} \, \oint \frac{\de u}{2\pi\I} \frac{(u-u^{-1})^2}{(1-uy)(1-u/y)} 
\prod_{\ell=1}^K {x_\ell \over (1+x_\ell u) (1 + x_\ell/u)}\, .
\een
We can now shrink the last contour to $u=0$, picking up the residues at the
poles at $u=-x_k$ in that process. This gives
\ben \label{eqfind2}
\wh Q^{\rm S}(x_1,\cdots x_K;1/y)  &=& -{1\over 2}
{y -1/y \over y^{-1} \prod_{\ell=1}^K (1+x_\ell y) - y \prod_{\ell=1}^K (1 + x_\ell/y)}\, 
\non \\
&& \quad \times\left[  y \prod_{\ell=1}^K {x_\ell\over 1 + x_\ell y} + 
y^{-1} \prod_{\ell=1}^K {x_\ell\over 1 + x_\ell y^{-1}} \right]\non\\
&& + {1\over 2} \sum_{k=1}^K
\frac{1-x_k^2}{(1+x_k/y)(1+yx_k)}
\prod_{\ell=1\dots K \atop \ell\neq k} \frac{x_\ell}{(1 - x_\ell / x_k) (1- x_\ell x_k)} 
\, .\non\\
\een
This agrees with \refb{QS3t} for $K=3$. Like \refb{QS3t}, 
eq.\refb{eqfind2} is also symmetric under the exchange of the
$x_\ell$'s reflecting the fact that the single centered index 
remains invariant under wall crossing. It is worthwhile noting that 
the poles at $y=-x_\ell$ and $y=-1/x_\ell$ precisely cancel between
the two terms in \eqref{eqfind2}. The partition function \eqref{eqfind2} is also
regular at $y=1$, where it reduces to 
\be \label{eqfind3}
\wh Q^{\rm S}(x_1,\cdots x_K;1)  = 
\frac{ \prod_{\ell=1}^K \frac{x_\ell}{(1+x_\ell)^2} }{1-\sum_{k=1}^K \frac{x_k}{1+x_k}}
+ {1\over 2} \sum_{k=1}^K
\frac{1-x_k}{1+x_k}
\prod_{\ell=1\dots K \atop \ell\neq k} \frac{x_\ell}{(1 - x_\ell / x_k) (1- x_\ell x_k)} 
\, .
\ee
Using the same techniques as in \cite{1205.5023}, it is straightforward to extract
the asymptotic growth of the index of pure Higgs states $\Omega^{\rm S}_{\rm ref}(\gamma_1+\dots+\gamma_K)$ as the arrow multiplicities $a_\ell$ are uniformly scaled to infinity. 
The asymptotics is governed by the pole of the partition function at 
\be
\label{eqfind4}
\sum_{k=1}^K \frac{x_k}{1+x_k} = 1\ .
\ee
Setting for simplicity all $a_\ell$ equal to $a$, the solution to  \eqref{eqfind4} is $x_k=1/(K-1)$,
leading to an exponential growth 
\be
\Omega^{\rm S}_{\rm ref}(\gamma_1+\dots+\gamma_K) \stackrel{a\to\infty}{\sim}  
a^{\frac{1-K}{2}} \, (K-1)^{K a}\ .
\ee
Since $a=\langle \gamma_{\ell},\gamma_{\ell+1}\rangle$ scales like the square of the charges,
the exponential growth of $\Omega^{\rm S}_{\rm ref}$ is consistent with the Bekenstein-Hawking
entropy of a macroscopic single-centered black hole.

\section{Quivers with two loops} \label{sclosed}

So far we have considered quivers for which the links
form a single closed loop.
In this section we shall apply the general algorithm of \S\ref{sintro}
to compute the Poincar\'e polynomial of quivers with more than one
oriented loop.

\subsection{Abelian four-node, two-loop quivers} 
\label{s4node}

We consider the class of Abelian quivers with four nodes and two oriented 
loops represented below, 
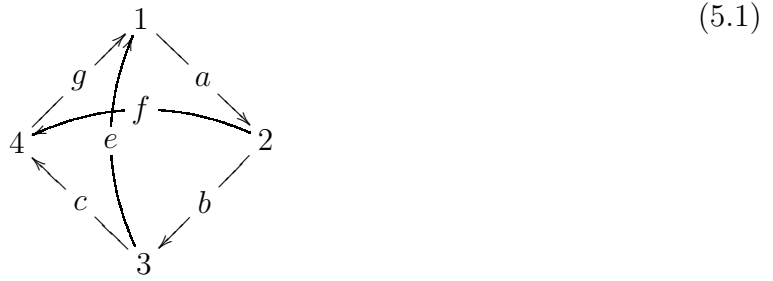
\be
\label{xy4node}
\begin{xy} 0;<1pt,0pt>:<0pt,-1pt>:: 
(47,0) *+{1} ="0",
(94,46) *+{2} ="1",
(48,93) *+{3} ="2",
(0,47) *+{4} ="3",
"0", {\ar|*+{a}"1"},
"2", {\ar@/^1pc/|*+{e}"0"},
"3", {\ar|*+{g}"0"},
"1", {\ar|*+{b}"2"},
"1", {\ar@/_1pc/|*+{f}"3"},
"2", {\ar|*+{c}"3"},
\end{xy}
\ee
We denote by $\gamma_1,\cdots \gamma_4$
the charge vectors carried by the four nodes, and by
$\gamma_{12}=a$, $\gamma_{23}=b$, $\gamma_{34}=c$,
$\gamma_{41}=g$, $\gamma_{31}=e$, $\gamma_{24}=f$
the multiplicities of arrows, which we assume to be 
strictly positive. Using \eqref{essp1}, \eqref{essp2}, \eqref{eone} and that 
$\Omega^{\rm S}_{\rm ref}(\gamma_\ell)=1$ we get
\be
 \label{etwo}
\begin{split}
Q&(\gamma_1+\gamma_2+\gamma_3+\gamma_4;y)= g_{\rm ref} (\gamma_1, \gamma_2, \gamma_3, \gamma_4; y)\\
& +\bigg\{g_{\rm ref}(\gamma_1, \gamma_2+
\gamma_3+\gamma_4; y) 
\left(\Omega^{\rm S}_{\rm ref}(\gamma_2+\gamma_3+ \gamma_4) 
+ H(\{\gamma_2,\gamma_3,\gamma_4\};\{1,1,1\}; y)\right) + {\rm perm}  \bigg\} \\
&+ \Omega^{\rm S}_{\rm ref}(\gamma_1+\gamma_2+\gamma_3+ \gamma_4) 
+ H(\{\gamma_1,\gamma_2,\gamma_3,\gamma_4\};\{1,1,1,1\}; y)\, .
\end{split}
\ee
The coefficients
$H(\{\gamma_i,\gamma_j,\gamma_k\};\{1,1,1\}; y)$'s have been determined 
in \S\ref{s3node}. The coefficient 
$H(\{\gamma_1,\gamma_2,\gamma_3,\gamma_4\};\{1,1,1,1\}; y)$ is
determined by demanding that the right hand side of \eqref{etwo} is a
polynomial in $y$ and that $H$ is invariant under 
$y\to y^{-1}$ and vanishes as $y\to\infty$. Instead of trying to solve the problem for a general set of charges, we
shall illustrate our algorithm for special choices of the $\gamma_{ij}
\equiv \langle \gamma_i, \gamma_j\rangle$ and the $c_i$'s.
We shall consider two examples:

\subsubsection{Example with only 3-center scaling solutions}

We choose multiplicities \footnote{Since scaling the $\gamma_{ij}$'s by an overall constant
$k$ maps a solution to \eqref{epwextrema} to another solution related by simple
rescaling of the $x_i$'s without changing their relative order, the computation
of $g_{\rm ref}$ can be done at one go for a family of quivers labelled
by differerent values of $k$.}
\be
 \label{eg1}
a = 3k, \quad b=4k, \quad c = 7k\ ,\quad g=4k\ , \quad e=5k, \quad f= 4k, \quad 
\ee
where $k$ is an arbitrary positive integer, and choose the FI parameters to be
\be
\label{FI4node}
c_1=2.1\ ,\quad c_2=3\ ,\quad c_3=-1.1\ ,\quad c_4=-4\ .
\ee
 Since the subquivers 134 and 234 have no closed 
loops, the corresponding  $H$ and $\Omega^{\rm S}_{\rm ref}$ must vanish:
\ben \label{ehvanish}
&& H(\{\gamma_1,\gamma_3,\gamma_4\};\{1,1,1\}; y)
= H(\{\gamma_2,\gamma_3,\gamma_4\};\{1,1,1\}; y)=0, \nonumber \\
&&  \Omega^{\rm S}_{\rm ref}(\gamma_1+\gamma_3+\gamma_4)
= \Omega^{\rm S}_{\rm ref}(\gamma_2+\gamma_3+\gamma_4)=0\, .
\een
In contrast, the subquivers 123 and 124 are 3-node quivers with loops of the
type analyzed in \S\ref{sec_3loop}, satisfying the  triangular inequalities \eqref{3trian}.
We can therefore borrow the result from \eqref{eks4},
\ben \label{esubq}
H(\{\gamma_1,\gamma_2,\gamma_3\};\{1,1,1\}; y)
&=& -2 (y-y^{-1})^{-2}, \nonumber \\
H(\{\gamma_1,\gamma_2,\gamma_4\};\{1,1,1\}; y)
&=&  \begin{cases}
(y+y^{-1})\, (y-y^{-1})^{-2} \, \hbox{for $k$ odd} \cr -2 
\, (y-y^{-1})^{-2} \, \hbox{for $k$ even}
\end{cases}\, .
\een
The two-center Coulomb indices 
$g_{\rm ref}(\gamma_4, \gamma_1+\gamma_2+\gamma_3;y)$ and 
$g_{\rm ref}(\gamma_3, \gamma_1+\gamma_2+\gamma_4;y)$
can be computed from \eqref{e2node} using $\{\hat c_i\}=
\{c_4,c_1+c_2+c_3\}=\{-4, 4\}$ and  $\{c_3,c_1+c_2+c_4\}
=\{-1.1, 1.1\}$, respectively, 
\be
g_{\rm ref}(\gamma_4, \gamma_1+\gamma_2+\gamma_3; y)
=  (-1)^{k+1}\, {y^{7k} - y^{-7k}\over y - y^{-1}}\, ,\qquad
g_{\rm ref}(\gamma_3, \gamma_1+\gamma_2+\gamma_4; y)=0\ .
\ee
Finally, an  explicit analysis of the solutions to \eqref{epwextrema} gives
\be \label{eGsol}
g_{\rm ref}(\gamma_1, \gamma_2,\gamma_3,\gamma_4; y)
= (-1)^{k+1}\, (y-y^{-1})^{-3} (y^{9k} - y^{-9k} + y^{5k} - y^{-5k})\, ,
\ee
with the contribution to $g_{\rm ref}(\gamma_1, \gamma_2,\gamma_3,\gamma_4;y)$ arising from the orderings
\be \label{earr1}
\{1,2,3,4;+\}, \quad \{4,1,2,3;-\}
\ee
and their reverse (the last entries in \eqref{earr1} give the associated signs 
$s(p)$). Substituting these into \eqref{etwo} we get
\be
 \label{etwoexplicit}
 \begin{split}
 Q(\gamma_1+\gamma_2&+\gamma_3+\gamma_4;y)\\
=  &(-1)^{k+1} \{(y-y^{-1})^{-3} (y^{9k} - y^{-9k} + y^{5k} - y^{-5k})
- 2 (y-y^{-1})^{-3}  (y^{7k} - y^{-7k})  \\
& + (y-y^{-1})^{-1} (y^{7k} - y^{-7k}) \Omega^{\rm S}_{\rm ref}
(\gamma_1+\gamma_2+\gamma_3) \}  \\ 
& + H(\{\gamma_1,\gamma_2,\gamma_3,\gamma_4\};\{1,1,1,1\}; y)
+ \Omega^{\rm S}_{\rm ref}(\gamma_1+\gamma_2+\gamma_3+\gamma_4)\, .
\end{split}
\ee
Requiring this to be a polynomial in $y$, $y^{-1}$, and $H$ to be even
under $y\to y^{-1}$ and vanish as $y\to\infty$, we get
\be \label{ehvalue}
H(\{\gamma_1,\gamma_2,\gamma_3,\gamma_4\};\{1,1,1,1\}; y)
= 0 \, .
\ee
This gives
\be
\label{egfin} 
\begin{split}
Q(\gamma_1+\gamma_2+\gamma_3+\gamma_4;y)=&\Omega^{\rm S}_{\rm ref}(\gamma_1+\gamma_2+\gamma_3+\gamma_4) \\
&+  (-1)^{k+1}\, 
(y^{-7k+1}+ y^{-7k+3} + \cdots + y^{7k-1}) \\
& \times  \Big\{
\Omega^{\rm S}_{\rm ref}
(\gamma_1+\gamma_2+\gamma_3)  
+ (y^{-k+1}+y^{-k+3}+\cdots
+ y^{k-1})^2 \Big\} \, .
\end{split}
\ee
The coefficient $\Omega^{\rm S}_{\rm ref}(\gamma_1+\gamma_2+\gamma_3)$ can be
determined from the generating function of pure Higgs states given in \eqref{QS3t} for
$(a,b,c)=(3k,4k,5k)$ and $t=1$.

The vanishing of $H(\{\gamma_1,\gamma_2,\gamma_3,\gamma_4\};\{1,1,1,1\}; y)$ 
indicates that in this case there are no 4-center scaling solutions. This can be verified 
by noting that there exist no choice of  $\vec r_1,\cdots \vec r_4$ for the four
centers such that the total angular momentum
\be \label{etotang}
\vec J = {1\over 2} \sum_{i<j} \alpha_{ij} \, {\vec r_{ij}\over |\vec r_{ij}|} \, ,
\ee
vanishes. As a consequence 
$\Omega^{\rm S}_{\rm ref}(\gamma_1+\gamma_2+\gamma_3+\gamma_4)$
also vanishes. In \S\ref{4node2} we consider another example where
there is a genuine 4-center scaling solution.

Let us now compute 
$Q(\cM;y)$ by a direct analysis of the cohomology of the quiver moduli  
space $\cM$.   
For the multiplicities \eqref{eg1} and FI parameters \eqref{FI4node}
the D-term equations take the form:
\ben \label{edt1}
\phi_{12,\alpha}^* \phi_{12,\alpha} - \phi_{31,\alpha}^* \phi_{31,\alpha}
- \phi_{41,\alpha}^* \phi_{41,\alpha} &=& 2.1 \, , \nonumber \\
  -\phi_{12,\alpha}^* \phi_{12,\alpha} + \phi_{23,\alpha}^* \phi_{23,\alpha}
+ \phi_{24,\alpha}^* \phi_{24,\alpha} &=& 3\, , \nonumber \\
 - \phi_{23,\alpha}^* \phi_{23,\alpha} 
+ \phi_{31,\alpha}^* \phi_{31,\alpha}
+ \phi_{34,\alpha}^* \phi_{34,\alpha} &=& -1.1\, , \nonumber \\
  \phi_{41,\alpha}^* \phi_{41,\alpha} - \phi_{24,\alpha}^* \phi_{24,\alpha}
- \phi_{34,\alpha}^* \phi_{34,\alpha} &=& -4\, .
\een
Note that the last equation follows from the first three. In the absence of superpotential,
the variables $\{\phi^{\rm NC}\}=\{\phi_{31,\alpha},\phi_{41,\alpha}\}$ may become arbitrarily
large, but for fixed values of those the remaining variables 
$\{\phi^{\rm C}\}=\{\phi_{12,\alpha},\phi_{23,\alpha},\phi_{24,\alpha},\phi_{34,\alpha}\}$
lie in a compact domain. 

Due to the existence of the closed loops
123, 124 and 1234, the generic superpotential
takes the form
\be \label{est1}
W= C^{(1)}_{\alpha\beta\gamma} \phi_{12,\alpha} \phi_{23,\beta}
\phi_{31,\gamma} 
+C^{(2)}_{\alpha\beta\gamma} \phi_{12,\alpha} \phi_{24,\beta}
\phi_{41,\gamma} + C^{(3)}_{\alpha\beta\gamma\delta} 
\phi_{12,\alpha} \phi_{23,\beta}
\phi_{34,\gamma} \phi_{41,\delta}\, ,
\ee
where $C^{(i)}$'s are arbitrary constants. A family of solutions to the F-term
and D-term equations
can be found by setting:
\ben \label{eso1}
&& \phi_{41,\alpha}=\phi_{31,\alpha} =0\, , \nonumber \\
&& \phi_{12,\alpha}^* \phi_{12,\alpha} =2.1\ ,\quad
\phi_{23,\alpha}^* \phi_{23,\alpha} = 1.1 + \phi_{34,\alpha}^* \phi_{34,\alpha}\ ,\quad
\phi_{34,\alpha}^* \phi_{34,\alpha} +  \phi_{24,\alpha}^* \phi_{24,\alpha} 
= 4,
\nonumber \\ &&
C^{(1)}_{\alpha\beta\gamma} \phi_{12,\alpha} \phi_{23,\beta}=0, 
\quad 
C^{(2)}_{\alpha\beta\delta} \phi_{12,\alpha} \phi_{24,\beta}
+ C^{(3)}_{\alpha\beta\gamma\delta} \,
\phi_{12,\alpha} \phi_{23,\beta}
\phi_{34,\gamma}=0\, .
\een
Since the Poincar\'e polynomial remains unchanged under a
deformation of the superpotential as long as the moduli space
does not become singular or non-compact, we can simplify the
problem by choosing the superpotential appropriately. Let us set
the coefficients $C^{(3)}_{\alpha\beta\gamma\delta}$ to zero.
In that case the last set of equations, $\gamma_{41}=4k$ in number,
can be solved by setting the $\gamma_{24}=4k$ 
components $\phi_{24,\alpha}$
to zero. The equations now simplify to
\ben \label{eso2}
&& \phi_{41,\alpha}=\phi_{31,\alpha} =\phi_{24,\alpha}=0\, , \nonumber \\
&& \phi_{12,\alpha}^* \phi_{12,\alpha} =2.1, \quad
 \phi_{23,\alpha}^* \phi_{23,\alpha} = 5.1, \quad
\quad \phi_{34,\alpha}^* \phi_{34,\alpha} = 4,
\nonumber \\ &&
C^{(1)}_{\alpha\beta\gamma} \phi_{12,\alpha} \phi_{23,\beta}=0
\, .
\een
After quotienting by the complexified gauge group $(\IC^\times)^4$, 
the moduli space of classical solutions factorizes into a product 
of $\CP^{c-1}$, parametrized by the variables $\phi_{34,\alpha}$,
and of a complete intersection of $e$ degree (1,1) hypersurfaces in 
$\CP^{a-1}\times \CP^{b-1}$, 
parametrized by $\phi_{12,\alpha}$ and $\phi_{23,\alpha}$ 
The cohomology of the complete intersection can be analysed
using the Lefschetz hyperplane theorem as explained in \S\ref{sec_abcohom},
or simply borrowed from our previous analysis of 3-node quivers in \eqref{efin3node} with
$(\gamma_{12},\gamma_{23},\gamma_{31})=(a,b,e)$. Since the Poincar\'e polynomial
is multiplicative, we arrive at 
\be \label{egfina}
\begin{split}
Q(\cM;y)=&
(-1)^{k+1}\, (y^{-7k+1}+ y^{-7k+3} + \cdots + y^{7k-1}) \\
& \times \left\{
\Omega^{\rm S}
(\gamma_1+\gamma_2+\gamma_3) + (y^{-k+1}+y^{-k+3}+\cdots
+ y^{k-1})^2 \right\} 
\end{split}
\ee
The value of $\Omega^{\rm S}_{\rm ref}(\gamma_1+\gamma_2+\gamma_3)$ can be
determined from the generating function of pure Higgs states in \eqref{QS3t}. 
Eq.\eqref{egfina} is in perfect agreement with \eqref{egfin}
with $\Omega^{\rm S}_{\rm ref}(\gamma_1+\gamma_2+\gamma_3+\gamma_4)=0$.

Finally, let us compare the cohomology of the vacuum moduli space $\cM_0$ in absence of 
superpotential with  the results of the HN recursion method explained in \S\ref{srepresent}.
As noted below \eqref{edt1},  the variables $\phi^{\rm NC}_i=\phi_{31,\alpha}, \phi_{41,\alpha}$ can vary freely over  $\IC^e\times \IC^g$, while, for fixed values of those, 
the remaining variables parametrize the compact manifold $\CP^{a-1}\times \CP^{b-1}\times \CP^{c+f-1}$. The topology of  $\cM_0$ is therefore  $\CP^{a-1}\times \CP^{b-1}\times \CP^{c+f-1} \times {\mathbb C}^e \times {\mathbb C}^g$, with 
\be \label{eq4node}
Q(\cM_0;y)
= (-1)^{a+b+c+g+e+f+1} y^{-g-e} (y-y^{-1})^{-3} (y^a - y^{-a})
(y^b - y^{-b}) (y^{c+f} - y^{-c-f})\, .
\ee 
On the other hand, using the fact that the charge vectors are 
ordered according to
\ben
&& \gamma_4 < \gamma_3+\gamma_4 < 
\gamma_3 < \gamma_1+\gamma_3+\gamma_4 <
\gamma_1+\gamma_4 < \gamma_2+\gamma_3+\gamma_4 < \gamma_2+\gamma_4 
<\gamma_1+\gamma_2+\gamma_3+\gamma_4  \nonumber \\ &&
< \gamma_1+\gamma_2+\gamma_4 < \gamma_1+\gamma_3 <\gamma_2+\gamma_3 <
\gamma_1+\gamma_2+\gamma_3 
< \gamma_1  < \gamma_1+\gamma_2 < \gamma_2\ ,
\een
where we have used the shorthand notation $\alpha<\beta$ to denote $\mu(\alpha)<\mu(\beta)$, 
the HN recursion method yields %
\begin{eqnarray}
&&\CI(\gamma_1+\gamma_2;w)=\CI(\gamma_1+\gamma_3;w)=
\CI(\gamma_1+\gamma_4;w)=0,\non \\
&& \CI(\gamma_2+\gamma_3;w)=\frac{w^b-w^{-b}}{(w-w^{-1})^2},\qquad
\CI(\gamma_2+\gamma_4,w)=\frac{w^f-w^{-f}}{(w-w^{-1})^2}\ ,\\
&& \CI(\gamma_3+\gamma_4;w)=\frac{w^c-w^{-c}}{(w-w^{-1})^2},\qquad  \CI(\gamma_1+\gamma_3+\gamma_4;w)=0,\non\\
&& \CI(\gamma_1+\gamma_2+\gamma_3;w)=w^e
(w-w^{-1})^{-3}(w^b-w^{-b})(w^a-w^{-a}),\non\\
&& \CI(\gamma_1+\gamma_2+\gamma_4;w)=w^g
(w-w^{-1})^{-3}(w^f-w^{-f})(w^a-w^{-a}),\non\\
&& \CI(\gamma_2+\gamma_3+\gamma_4;w)=(w-w^{-1})^{-3}(w^{c+f}-w^{-c-f})(w^b-w^{-b})\ .
\non
\end{eqnarray}
Using these results and
eq.\refb{eq:CIrecurs}, we have, for the total charge vector $\gamma_1+\gamma_2+\gamma_3+\gamma_4$,
\ben
\CI(\gamma;w) &=&
 h(\gamma;w)
-\cF(\gamma_2, \gamma_1, \gamma_3+\gamma_4;w)
-\cF(\gamma_1,\gamma_2+\gamma_3+\gamma_4;w) \non\\ &&
-\cF(\gamma_1,\gamma_2+\gamma_3,\gamma_4;w)
-\cF(\gamma_1+\gamma_2+\gamma_3,\gamma_4;w)
-\cF(\gamma_2, \gamma_1, \gamma_3, \gamma_4;w) \non\\
&& - \cF(\gamma_1, \gamma_2+ \gamma_4, \gamma_3;w) 
- \cF(\gamma_1+\gamma_2+ \gamma_4, \gamma_3;w)\, . \non\\
&=& w^{e+g}\, (w-w^{-1})^{-4}\, (w^a-w^{-a})\, (w^b-w^{-b})\, (w^{c+f}-w^{-c-f})\ .
\een
Using this and
\eqref{eq:stackinv}, we precisely reproduce 
\eqref{eq4node}. This bolsters
our hypothesis that the HN method is applicable to quivers with loops, as long as
the superpotential vanishes.

\subsubsection{Example with  4-center scaling solutions  \label{4node2}}
We now consider a 4-node quiver with the same topology \eqref{xy4node} but with
multiplicities
\be
\label{eg1nn}
a = 15k, \quad b=20k, \quad c = 35k\ ,\quad g=10k\ , \quad e=5k, \quad f= 2k, \quad 
\ee
where $k$ is a positive integer, and with FI parameters
\be
c_1=2\ ,\quad c_2=3\ ,\quad c_3=-6\ ,\quad c_4=1\ .
\ee
In this case only the subquiver 123 and 124 have closed loops,  but the subquiver 
124 does not satisfy the triangle inequalities \eqref{3trian}. Hence we expect  $H$ and 
$\Omega^{\rm S}$
to vanish for the subquivers 124, 234 and 134:
\ben \label{ehvanishnn}
&& H(\{\gamma_1,\gamma_3,\gamma_4\};\{1,1,1\}; y)
= H(\{\gamma_2,\gamma_3,\gamma_4\};\{1,1,1\}; y)=
H(\{\gamma_1,\gamma_2,\gamma_4\};\{1,1,1\}; y) = 0, \nonumber \\
&&  \Omega^{\rm S}_{\rm ref}(\gamma_1+\gamma_3+\gamma_4)
= \Omega^{\rm S}_{\rm ref}(\gamma_2+\gamma_3+\gamma_4)=
 \Omega^{\rm S}_{\rm ref}(\gamma_1+\gamma_2+\gamma_4)= 0\, .
\een
The analog of \eqref{esubq} now has the form:
\be \label{eh3n}
H(\{\gamma_1,\gamma_2,\gamma_3\};\{1,1,1\}; y)
= -2 (y-y^{-1})^{-2}\, .
\ee
Finally an explicit analysis of \eqref{epwextrema} give
\ben \label{eGsolnew}
&& g_{\rm ref}(\gamma_4, \gamma_1+\gamma_2+\gamma_3; y)=0\, ,
\nonumber \\
&& g_{\rm ref}(\gamma_1, \gamma_2,\gamma_3,\gamma_4;y)
=(-1)^{k+1}\, (y-y^{-1})^{-3} (y^{3k} - y^{-3k})\, ,
\een
with the contribution to $g_{\rm ref}(\gamma_1, \gamma_2,\gamma_3,\gamma_4;y)$ arising from the
arrangements
\be \label{earr2}
\{4,1,2,3;+\},
\ee
and its reverse. 
Substituting these into \eqref{etwo} we get
\be \label{etwoexplicitnew}
\begin{split}
Q(\gamma_1+\gamma_2+\gamma_3+\gamma_4;y)
=  & (-1)^{k+1} \, (y-y^{-1})^{-3} (y^{3k} - y^{-3k})
+ H(\{\gamma_1,\gamma_2,\gamma_3,\gamma_4\};\{1,1,1,1\}; y)
\\ &
+ \Omega^{\rm S}_{\rm ref}(\gamma_1+\gamma_2+\gamma_3+\gamma_4)\, .
\end{split}
\ee
The unique choice of 
$H(\{\gamma_1,\gamma_2,\gamma_3,\gamma_4\};\{1,1,1,1\}; y)$
consistent with the requirements is
\be \label{ehnow}
 H(\{\gamma_1,\gamma_2,\gamma_3,\gamma_4\};\{1,1,1,1\}; y)
 = \begin{cases}  {3\over 2} \,k\,  (y - y^{-1})^{-2} (y+y^{-1})\, \, \hbox{for $k$ even} \cr
 - 3\,k\,  (y - y^{-1})^{-2}\, \, \hbox{for $k$ odd}
 \end{cases}\, .
 \ee
The fact that $H(\{\gamma_1,\gamma_2,\gamma_3,\gamma_4\};\{1,1,1,1\}; y)$ does not
vanish is consistent with the existence of scaling solutions where all four centers come together
(i.e. the existence of four vectors $\vec r_1, \vec r_2, \vec r_3, \vec r_4$ such that the
angular momentum \eqref{etotang} vanishes).
 Substituting this into \eqref{etwoexplicitnew} we get
\be
\label{etwofinG}
\begin{split}
Q(\gamma_1&+\gamma_2+\gamma_3+\gamma_4;y) \\
=& \begin{cases} 
\Omega^{\rm S}_{\rm ref}(\gamma_1+\gamma_2+\gamma_3+\gamma_4) - 
\left(y-y^{-1}\right)^{-3} \, \left\{y^{3k}-y^{-3k} - {3\over 2} k (y^2 - y^{-2})\right\} \, 
\hbox{for $k$ even}\cr
\Omega^{\rm S}_{\rm ref}(\gamma_1+\gamma_2+\gamma_3+\gamma_4)+\, 
\left(y-y^{-1}\right)^{-3} \, \left\{y^{3k}-y^{-3k} - 3k (y - y^{-1})\right\}\, \hbox{for $k$ odd}
\end{cases}
\end{split}
\ee

Let us now compute 
$Q(\cM;y)$ by a direct analysis of the cohomology of the quiver moduli space $\cM$. 
In this case the D-term equations take the form:
\ben \label{edt1e2}
 \phi_{12,\alpha}^* \phi_{12,\alpha} - \phi_{31,\alpha}^* \phi_{31,\alpha}
- \phi_{41,\alpha}^* \phi_{41,\alpha} &=& 2\, , \nonumber \\
  -\phi_{12,\alpha}^* \phi_{12,\alpha} + \phi_{23,\alpha}^* \phi_{23,\alpha}
+ \phi_{24,\alpha}^* \phi_{24,\alpha} &=& 3\, , \nonumber \\
 - \phi_{23,\alpha}^* \phi_{23,\alpha} 
+ \phi_{31,\alpha}^* \phi_{31,\alpha}
+ \phi_{34,\alpha}^* \phi_{34,\alpha} &=& -6\, , \nonumber \\
  \phi_{41,\alpha}^* \phi_{41,\alpha} - \phi_{24,\alpha}^* \phi_{24,\alpha}
- \phi_{34,\alpha}^* \phi_{34,\alpha} &=& 1\, .
\een
In the absence of a superpotential, the variables $\{\phi^{\rm NC}\}=\{\phi_{31,\alpha},\phi_{34,\alpha}, \phi_{24,\alpha}\}$ may become arbitrarily
large, but for fixed values of those the remaining variables 
$\{\phi^{\rm C}\}=\{\phi_{12,\alpha},\phi_{23,\alpha},\phi_{41,\alpha}\}$
lie in a compact domain. In general however,  the superpotential
is given by \eqref{est1}, 
where $C^{(i)}$'s are arbitrary constants. A family of solutions to the F-term
and D-term equations
can be found by setting
\ben \label{eso1e2}
&& \phi_{34,\alpha}=\phi_{31,\alpha} =\phi_{24,\alpha}=0\, , \nonumber \\
&& \phi_{12,\alpha}^* \phi_{12,\alpha} =3, \quad \phi_{23,\alpha}^* \phi_{23,\alpha} 
 = 6,
\quad \phi_{41,\alpha}^* \phi_{41,\alpha} 
= 1,
\nonumber \\ &&
C^{(1)}_{\alpha\beta\gamma} \phi_{12,\alpha} \phi_{23,\beta}=0, 
\quad 
C^{(2)}_{\alpha\beta\gamma} \phi_{12,\alpha} \phi_{41,\gamma} =0, \quad
C^{(3)}_{\alpha\beta\gamma\delta} \,
\phi_{12,\alpha} \phi_{23,\beta}
\phi_{41,\delta}=0\, .
\een
Now the variables $\phi_{12,\alpha}$, $\phi_{23,\alpha}$  and
$ \phi_{41,\alpha}$ subject to the constraints given in the second
line of \eqref{eso1e2}, and the identification under the complexified gauge transformations
$(\IC^\times)^4$,
describe the product manifold $\cME=\CP^{a-1}\times \CP^{b-1}\times
\CP^{g-1}$. The constraints in the last line of
\eqref{eso1e2} describe a codimension $e+f+c$ submanifold $\cM$ inside
$\cME$. Thus the quiver moduli space $\cM$ has dimension $d=a+b+g-e-f-c=3k-3$. 
By repeated use of Lefschetz hyperplane theorem one can argue that
the Betti numbers $b_p(\cM)$ for $p<d$ coincide with that of the ambient space $\cME$. Thus the negative powers of $y$ in the 
Laurent polynomial  of $\cM$ are given by 
\be \label{eneg1kk}
\begin{split}
Q(\cM;y) \simeq & \ 
y^{-3k+3} (1+y^2 +\cdots y^{30k-2}) (1+y^2 +\cdots y^{40k-2}) (1+y^2 +\cdots y^{20k-2}) \\
\simeq & \ y^{-3k+3} (1-y^2)^{-3}\, ,
\end{split}
\ee
where $\simeq$ denotes equality of terms involving negative powers of $y$.
This is in perfect agreement with \eqref{etwofinG}. The $y\to y^{-1}$ symmetry ensures that
the positive powers of $y$ in  $Q(\cM;y)$  also 
agree with that given in \eqref{etwofinG}. To determine the constant term
$\Omega^{\rm S}_{\rm ref}(\gamma_1+\gamma_2+\gamma_3+\gamma_4)$ in
\eqref{etwofinG}, it suffices to compute the Euler number of the 
complete intersection manifold described by eq.\eqref{eso1e2}. Using  the method of \S\ref{sec_abcohom}, we find 
\be
\Omega^{\rm S}_{\rm ref}(\gamma_1+\gamma_2+\gamma_3+\gamma_4) = (-1)^{k+1} \chi(\cM) 
- \begin{cases} 
\frac{k}{8}(4-9k^2) \  \hbox{for $k$ even}\cr
\frac{k}{8}{(9k^2-1)} \  \hbox{for $k$ odd}
\end{cases}
\ee
where
\be
\begin{split}
\chi(\cM) = &\oint \prod_{i=1}^3 \frac{\de J_i}{2\pi\I}\left(\frac{1+J_1}{J_1}\right)^{15k}
\left(\frac{1+J_2}{J_2}\right)^{20k}
\left(\frac{1+J_2}{J_3}\right)^{10k} \\
&\times \left(\frac{J_1+J_2}{1+J_1+J_2}\right)^{5k}
\left(\frac{J_1+J_3}{1+J_1+J_3}\right)^{2k}
\left(\frac{J_1+J_2+J_3}{1+J_1+J_2+J_3}\right)^{35k}\ .
\end{split}
\ee
The contour integral can be easily evaluated for any $k$ using the method 
of \S\ref{sec_abcohom}.

\subsection{A 5-node quiver with nested scaling configurations}
\label{snested}

We now consider the 5-node, 2-loop Abelian quiver depicted below,
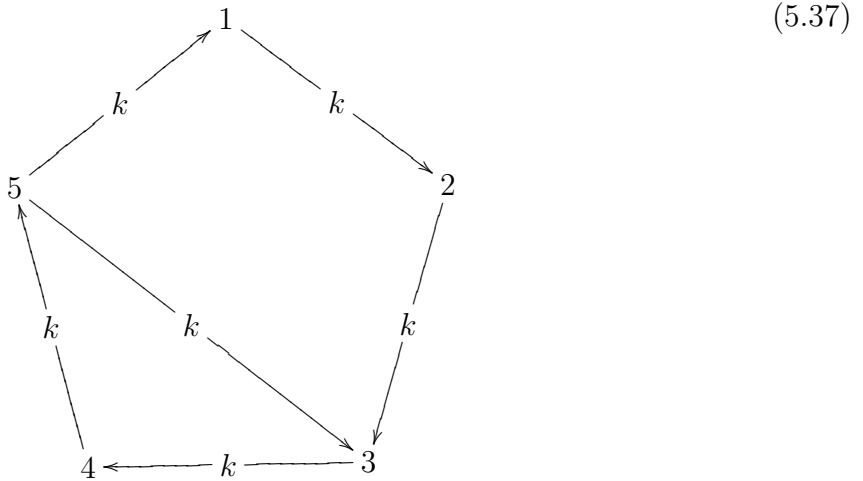
\be
\begin{xy} 0;<1pt,0pt>:<0pt,-1pt>:: 
(80,0) *+{1} ="0",
(164,63) *+{2} ="1",
(134,168) *+{3} ="2",
(28,170) *+{4} ="3",
(0,64) *+{5} ="4",
"0", {\ar|*+{k}"1"},
"4", {\ar|*+{k}"0"},
"1", {\ar|*+{k}"2"},
"2", {\ar|*+{k}"3"},
"4", {\ar|*+{k}"2"},
"3", {\ar|*+{k}"4"},
\end{xy}
\ee
We choose the FI parameters to be
\be \label{enest1}
\quad c_1 = c_2 = c_3=c_4=1, \quad c_5=-4\, .
\ee
In this case the nodes $345$ form a subquiver satisfying triangle
inequality and hence $\Omega^{\rm S}_{\rm ref}(\gamma_3+\gamma_4+
\gamma_5)$ and $H(\{\gamma_3,\gamma_4,\gamma_5\}, \{1,1,1\}; y)$
are non vanishing. Thus we have
\be
 \label{enest2}
 \begin{split}
Q(\gamma_1+&\cdots +\gamma_5;y) = \Omega^{\rm S}_{\rm ref}(\gamma_1+\cdots +\gamma_5) \\
& + g_{\rm ref}(\gamma_1, \gamma_2, \gamma_3+\gamma_4+\gamma_5;y)
\, \left[ \Omega^{\rm S}_{\rm ref}(\gamma_3+\gamma_4+
\gamma_5) + H(\{\gamma_3,\gamma_4,\gamma_5\}, \{1,1,1\}; y)\right]
\\
& +g_{\rm ref}(\gamma_1,\cdots \gamma_5;y) +
H(\{\gamma_1,\cdots \gamma_5\}, \{1,\cdots 1\};y)
\\
&+ H(\{\gamma_1, \gamma_2, \gamma_3+\gamma_4+\gamma_5\},
\{1,1,1\}; y) \Omega^{\rm S}_{\rm ref}(\gamma_3+\gamma_4+\gamma_5)
\, . 
\end{split}
\ee
The coefficients
$g_{\rm ref}(\gamma_1, \gamma_2, \gamma_3+\gamma_4+\gamma_5;y)$,
$H\left(\{\gamma_3,\gamma_4,\gamma_5\}, \{1,1,1\}; y\right)$  and
$H(\{\gamma_1, \gamma_2, \gamma_3+\gamma_4+\gamma_5\},
\{1,1,1\}; y)$ 
can be
read off from the results of \S\ref{s3node}. We have
\ben \label{enest3}
&& 
g_{\rm ref}(\gamma_1, \gamma_2, \gamma_3+\gamma_4+\gamma_5;y)
= (-1)^k \left(y - y^{-1}\right)^{-2}\, (y^k + y^{-k})\, , \nonumber \\
&& H\left(\{\gamma_3,\gamma_4,\gamma_5\}, \{1,1,1\}; y)\right)
= \begin{cases} - 2   \,  (y-y^{-1})^{-2}\, \, \hbox{for $k$ even}\cr
(y + y^{-1}) \, (y-y^{-1})^{-2} \, \, \hbox{for $k$ odd}
\end{cases} \nonumber \\ &&
H(\{\gamma_1, \gamma_2, \gamma_3+\gamma_4+\gamma_5\},
\{1,1,1\}; y) 
=  \begin{cases} - 2   \,  (y-y^{-1})^{-2}\, \, \hbox{for $k$ even}\cr
(y + y^{-1}) \, (y-y^{-1})^{-2} \, \, \hbox{for $k$ odd}
\end{cases}\, . \nonumber \\
\een
Finally the contribution to $g_{\rm ref}(\gamma_1,\cdots \gamma_5;y)$ turns out
to arise from the following arrangement of the nodes
\be \label{enest4}
\{1,2,3,4,5;+\}, \quad \{1,2,5,4,3;+\}, 
\ee
and their reverse. This gives
\be \label{enest5}
g_{\rm ref}(\gamma_1,\cdots \gamma_5;y)= (y - y^{-1})^{-4}\, 
\left( y^{2k} + y^{-2k} +2 \right)
\ee
First consider the case where $k$ is even. In this case \eqref{enest2}-\eqref{enest5}
gives
\be \label{enest6}
\begin{split}
Q(\gamma_1+\cdots +\gamma_5;y)
= & \Omega^{\rm S}_{\rm ref}(\gamma_1+\gamma_2+\gamma_3+\gamma_4+\gamma_5)+ H(\{\gamma_1,\cdots \gamma_5\}, 
\{1,\cdots 1\};y) \\
&+ (y - y^{-1})^{-2} \, (y^{k/2} - y^{-k/2})^2 \, 
\Omega^{\rm S}_{\rm ref}(\gamma_3+\gamma_4+\gamma_5)  \\
& + \, (y - y^{-1})^{-4}\, \left( y^{2k} - 2 y^k  +2 - 2 y^{-k} 
+ y^{-2k} \right)\, .
\end{split}
\ee
{}From this we get
\be \label{enest7}
H(\{\gamma_1,\cdots \gamma_5\}, 
\{1,\cdots 1\};y) = -{k^2\over 2} \, (y - y^{-1})^{-2}\, ,
\ee
and hence
\be
\label{enest8}
\begin{split}
Q(\gamma_1+\cdots +\gamma_5;y) 
=& \Omega^{\rm S}_{\rm ref}(\gamma_1+\gamma_2+
\gamma_3+\gamma_4+\gamma_5) \\
&+ (y - y^{-1})^{-2} \, (y^{k/2} - y^{-k/2})^2 \, 
\Omega^{\rm S}_{\rm ref}(\gamma_3+\gamma_4+\gamma_5)  + 
\\ &
+ \, (y - y^{-1})^{-4}\, \big( y^{2k} - 2 y^k  +2 - 2 y^{-k} 
+ y^{-2k} - {k^2 \over 2} (y - y^{-1})^2\big)
 \, .
\end{split}
\ee
Next we consider the case where $k$ is odd. In this case 
\eqref{enest2}-\eqref{enest5}
gives
\be
\begin{split}
 \label{enest9}
Q(\gamma_1+\cdots +\gamma_5;y) 
=&\Omega^{\rm S}_{\rm ref}(\gamma_1+\gamma_2+\gamma_3+\gamma_4+
\gamma_5)+ H(\{\gamma_1,\cdots \gamma_5\}, 
\{1,\cdots 1\};y)
\\ &
 -(y - y^{-1})^{-2} \, (y^{k } + y^{-k } -y - y^{-1}) \, 
\Omega^{\rm S}_{\rm ref}(\gamma_3+\gamma_4+
\gamma_5) 
\\& 
+ \, (y - y^{-1})^{-4}\, \left( y^{2k}  +2  
+ y^{-2k} - (y + y^{-1})(y^k + y^{-k})\right)
 \, .
\end{split}
\ee
{}From this we get
\be \label{enest10}
H(\{\gamma_1,\cdots \gamma_5\}, 
\{1,\cdots 1\};y) = -{k^2-1\over 2} \, (y - y^{-1})^{-2}\, ,
\ee
and hence
\be \label{enest11}
\begin{split}
Q&(\gamma_1+\cdots +\gamma_5;y)
=   \Omega^{\rm S}_{\rm ref}(\gamma_1+\gamma_2+
\gamma_3+\gamma_4+
\gamma_5)\\
 &-(y - y^{-1})^{-2} \, (y^{k } + y^{-k } -y - y^{-1}) \, 
\Omega^{\rm S}_{\rm ref}(\gamma_3+\gamma_4+
\gamma_5)  
\\&
+ (y - y^{-1})^{-4}\,\big( y^{2k}  +2  
+ y^{-2k} - (y + y^{-1})(y^k + y^{-k}) - {k^2-1 \over 2} (y - y^{-1})^2\big)
 \, .
\end{split}
\ee

Let us now compare these predictions with the result of direct computation
of the cohomology of the quiver moduli space.
The D-term equations \eqref{emodi1} now take the form:
\ben \label{edt1e3}
 \phi_{12,\alpha}^* \phi_{12,\alpha} - \phi_{51,\alpha}^* \phi_{51,\alpha}
& =& 1\, , \nonumber \\
  -\phi_{12,\alpha}^* \phi_{12,\alpha} + \phi_{23,\alpha}^* \phi_{23,\alpha}
&=& 1\, , \nonumber \\
 - \phi_{23,\alpha}^* \phi_{23,\alpha} 
+ \phi_{34,\alpha}^* \phi_{34,\alpha}
- \phi_{53,\alpha}^* \phi_{53,\alpha} &=& 1\, , \nonumber \\
  \phi_{45,\alpha}^* \phi_{45,\alpha} - \phi_{34,\alpha}^* \phi_{34,\alpha}
 &=& 1\, , \nonumber \\
  - \phi_{45,\alpha}^* \phi_{45,\alpha} 
+ \phi_{53,\alpha}^* \phi_{53,\alpha}
+ \phi_{51,\alpha}^* \phi_{51,\alpha} &=& -4\, ,\, .
\een
In the absence of a superpotential,  the variables $\phi_i^{\rm NC} = \{ \phi_{51,\alpha},\phi_{53,\alpha} \}$ may vary freely but for a fixed value of these variables, the
remaining variables live in a compact space. 
Due to the existence of the oriented closed loops
12345 and 345, the generic superpotential
takes the form
\be \label{est1e3}
W= C^{(1)}_{\alpha\beta\gamma} \phi_{34,\alpha} \phi_{45,\beta}
\phi_{53,\gamma} 
+C^{(2)}_{\alpha\beta\gamma\delta\sigma} \phi_{12,\alpha} \phi_{23,\beta}
\phi_{34,\gamma}  \phi_{45,\delta}
\phi_{51,\sigma} \, ,
\ee
where $C^{(i)}$'s are arbitrary constants. A family of solutions to the F-term
and D-term conditions can be found by setting:
\ben \label{eso1e3}
&& \phi_{51,\alpha}=\phi_{53,\alpha} =0\, , \nonumber \\
&& \phi_{12,\alpha}^* \phi_{12,\alpha} =1, \quad \phi_{23,\alpha}^* \phi_{23,\alpha} 
 = 2,
\quad \phi_{34,\alpha}^* \phi_{34,\alpha} 
= 3, \quad \quad \phi_{45,\alpha}^* \phi_{45,\alpha} 
= 4\, ,
\nonumber \\ &&
C^{(1)}_{\alpha\beta\gamma} \phi_{34,\alpha} \phi_{45,\beta}=0, 
\quad 
C^{(2)}_{\alpha\beta\gamma\delta\sigma} \phi_{12,\alpha} \phi_{23,\beta}
 \phi_{34,\gamma} \phi_{45,\delta} =0
\, .
\een
The variables $\phi_{12,\alpha}, \phi_{23,\alpha}, \phi_{34,\alpha}$ and
$\phi_{45,\alpha}$ satisfying the constraints in the second line describe
a product manifold $\CP^{k-1}\times\CP^{k-1}\times\CP^{k-1}
\times\CP^{k-1}$. The first equation in the third line
describes a codimension $k$ manifold embedded in the product
of the last two $\CP^{k-1}$ factors. 
Let us denote the resulting $k-2$ dimensional manifold by
$\bar\MM$. The cohomology of $\bar\MM$ is
in fact identical to that associated with a three node quiver carrying charges
$\gamma_3$, $\gamma_4$ and $\gamma_5$ and is given by \eqref{efin3node} with
$a=b=c=k$. Thus 
\be \label{eqmy}
Q({\bar\MM};y) = \begin{cases}
\Omega^{\rm S}_{\rm ref}(\gamma_3+\gamma_4+\gamma_5) 
+ (y - y^{-1})^{-2} (y^k + y^{-k} -2) \quad \hbox{for $k$ even} \cr
\Omega^{\rm S}_{\rm ref}(\gamma_3+\gamma_4+\gamma_5) 
- (y - y^{-1})^{-2} (y^k + y^{-k} -y - y^{-1}) \quad \hbox{for $k$ odd}
\end{cases}\, .
\ee
The last equation in the third line of \eqref{eso1e3} now describes a
codimension $k$ subspace embedded in 
$\CP^{k-1}\times\CP^{k-1}\times\bar\MM$.
The resulting manifold $\MM$ 
has complex dimension $d=2(k-1)+(k-2)-k=2k-4$,
and by the Lefschetz hyperplane theorem its Betti numbers 
$b_p$ are given by those of 
$\CP^{k-1}\times\CP^{k-1}\times\bar\MM$ for $p<d$.
This in turn means that the negative powers of $Q(\MM;y)$ are
given by 
\be \label{enegfin}
Q(\MM;y)\simeq (-y)^{-2k+4} (1 + y^2 + y^4+\cdots + y^{2k-2})^2 \, (-y)^{k-2} \, Q({\bar\MM};y)\, .
\ee
Using \eqref{eqmy}, and throwing away terms involving non-negative powers of $y$, 
we find
\be
 \label{enegfin2}
\begin{split}
Q(\MM;y)\simeq& (-y)^{-k+2} (1-y^2)^{-2} \Omega^{\rm S}_{\rm ref}(\gamma_3+\gamma_4+\gamma_5) 
\\
&+\begin{cases}
 y^{-k+2} (1-y^2)^{-2} (y - y^{-1})^{-2} (y^k + y^{-k} -2) \ \quad  \hbox{for $k$ even} \\
 y^{-k+2} (1-y^2)^{-2} (y - y^{-1})^{-2} (y^k + y^{-k} -y - y^{-1}) \ \quad\hbox{for $k$ odd}\ .
 \end{cases}
 \end{split}
 \ee
 It is easy to see that the negative powers of $y$ in this expression match 
 those in \eqref{enest8},
\eqref{enest11}. By $y\to 1/y$ symmetry the positive powers of $y$ in the 
polynomial $Q(\MM;y)$  also match.
The constant term is determined by the Euler number of $\MM$, which can be computed
using the method of \S\ref{sec_abcohom},
\be
\label{enegfin3}
\chi(\cM) = \oint
\prod_{i=1}^4 \frac{\de J_i}{2\pi\I}\,
\left( \frac{(1+J_1)(1+J_2)(1+J_3)(1+J_4) (J_3+J_4) (J_1+J_2+J_3+J_4)}
               {J_1 J_2 J_3 J_4 (1+J_3+J_4) (1+J_1+J_2+J_3+J_4)} \right)^k\ .
\ee
Equating \eqref{enegfin3} and  \eqref{enest11} at $y=1$ allows to determine
the pure Higgs state degeneracy $\Omega^{\rm S}_{\rm ref}(\gamma_1+\cdots+
\gamma_5)$. 

\section{Non-Abelian quivers} \label{snonabelian}

So far we have only considered quivers for which each node carries
a $U(1)$ factor. We shall now analyze some examples of non-Abelian
quivers.

\subsection{Rank (1,1,2) \label{sec112}}
We consider again the 3-node quiver \eqref{xy3node},  but now
allow for a $U(2)$ gauge group at node 3, keeping 
$U(1)$ gauge groups at node 1 and 2. We assume that the multiplicities
$a,b,c$ are positive integers satisfying 
\be \label{erange}
 a<2c, \quad b< c\, , \quad k\equiv a+2b-2c>0\ .
\ee
We choose the FI terms such that 
\be 
\label{ena1}
\quad c_1>0, \quad c_1+c_2>0, \quad c_2<0, \quad c_3\to 0^-.
\ee
As mentioned in \S\ref{sintro}, when dealing with non-Abelian quivers it is important
not to enforce the  $y$ independence of
$\Omega^{\rm S}_{\rm ref}$ until we determine the relevant $H$'s.
Using the fact that the only combination of charges for which the
scaling solutions  exist are
$\gamma_1+\gamma_2+\gamma_3$ and $\gamma_1+\gamma_2
+2\gamma_3$, we can express 
Eqs.\eqref{essp1}, \eqref{essp2} in the form:
\be\label{ena2}
\begin{split}
Q&(\gamma_1+\gamma_2+2\gamma_3;y)  
= \frac12 g_{\rm ref}(\gamma_1,\gamma_2,\gamma_3, \gamma_3;
y)  \, \Omega^{\rm S}_{\rm ref}(\gamma_1;y)
\, \Omega^{\rm S}_{\rm ref}(\gamma_2;y) \, \Omega^{\rm S}_{\rm ref}(\gamma_3;y)^2
\\
&+ \frac12 {y - y^{-1}\over y^2 - y^{-2}} \, g_{\rm ref}(\gamma_1,\gamma_2,2\gamma_3;y)\, 
\Omega^{\rm S}_{\rm ref}(\gamma_1;y)\, 
\Omega^{\rm S}_{\rm ref}(\gamma_2;y) \, \Omega^{\rm S}_{\rm ref}(\gamma_3;y^2)
\\ 
&+  g_{\rm ref}(\gamma_1+\gamma_2+\gamma_3, \gamma_3;y) \, \Omega^{\rm S}_{\rm ref}(\gamma_3;y) \\
&\quad \times
\left[ \Omega^{\rm S}_{\rm ref}(\gamma_1+\gamma_2+\gamma_3;y)
+ H(\{\gamma_1,\gamma_2,\gamma_3\}, \{1,1,1\};y)\,
\Omega^{\rm S}_{\rm ref}(\gamma_1;y)\, 
\Omega^{\rm S}_{\rm ref}(\gamma_2;y) \, 
\Omega^{\rm S}_{\rm ref}(\gamma_3;y)
\right]
\\
& + \Omega^{\rm S}_{\rm ref}(\gamma_1+\gamma_2+2\gamma_3;y)
+ H( \{\gamma_1,\gamma_2,\gamma_3,\gamma_3\},\{1,1,1,1\};y)\,
\Omega^{\rm S}_{\rm ref}(\gamma_1;y)\,
\Omega^{\rm S}_{\rm ref}(\gamma_2;y) \,
\Omega^{\rm S}_{\rm ref}(\gamma_3;y)^2
 \\
 &+ H( \{\gamma_1,\gamma_2,\gamma_3\},\{1,1,2\};y)
 \Omega^{\rm S}_{\rm ref}(\gamma_1;y)\,
\Omega^{\rm S}_{\rm ref}(\gamma_2;y) \,
\Omega^{\rm S}_{\rm ref}(\gamma_3;y^2)
\end{split}
\ee
Using \eqref{erange}, \eqref{e2node} and the result of \S\ref{sec_3loop} we get
\be
\label{ena3}
g_{\rm ref}(\gamma_1+\gamma_2+\gamma_3, \gamma_3;y)
=0\, , \qquad
  g_{\rm ref}(\gamma_1,\gamma_2,2\gamma_3;y)
= (-1)^k {y^k + y^{-k}\over (y - y^{-1})^2} \, .
\ee
Finally to find $g_{\rm ref}(\gamma_1,\gamma_2,\gamma_3, \gamma_3;
y)$ we label the coordinates of the charges 
$\gamma_1,\gamma_2,\gamma_3, \gamma_3$ by $x_1,x_2,x_3,x_4$.
\eqref{epwextrema} now gives 
\be \label{echararrange}
{a\over |x_{12}|} - {c\over |x_{13}|} - {c\over |x_{14}|} = c_1, 
\quad {c\over |x_{13}|} - {b\over |x_{23}|} = c_3, \quad
{c\over |x_{14}|} - {b\over |x_{24}|} = c_3\, .
\ee
Note that the last two equations, regarded as equations for $x_3$ and
$x_4$ respectively, are identical equations and hence we can try to solve
them simultaneously for fixed $x_1,x_2$. Using
translation invariance and reversal symmetry of the $x$ axis
 we can take $x_1=0$,  $x_2>0$.
In the $c_3\to 0$ limit the last two equations in \eqref{echararrange} give
\be \label{elast1}
b |x_a| = c |x_a-x_2| \quad \hbox{for $a=3,4$}\, .
\ee
Since $b<c$, this equation has two possible solutions -- one 
solution $x_{\rm m}$ in the range
$x_1< x_a<x_2$, and another $x_{\rm r}$ in the range $x_a>x_2$,
\be\label{elast2}
0=x_1 < x_{\rm m}\equiv \tfrac{c}{b+c}x_2 < x_2 < x_{\rm r}\equiv \tfrac{c}{c-b}x_2 \ .
\ee
Consider now the solution where $n_A$ of the $\gamma_3$'s 
sit at $x_{\rm m}$ and $n_B$ of the  $\gamma_3$ sit at $x_{\rm r}$.
Here $n_A, n_B=0,1,2$ subject to the restriction
$n_A+n_B=2$. Substituting the corresponding values of $x_i$ into the first
equation in
\eqref{echararrange} we get
\be \label{elast4}
c_1 x_2 = a - b(n_A-n_B) - c(n_A+n_B) \, .
\ee
Since $c_1>0$ and $x_2>0$, the right hand side of \eqref{elast4} must be
positive. The condition $a<2c$ in \eqref{erange} now shows that
this is possible only for the choice $n_A=0, n_B=2$. 
Furthermore one finds that for this case $s(p)=1$ \cite{1103.1887}.
Thus the contribution
to $g_{\rm ref}(\gamma_1,\gamma_2,\gamma_3, \gamma_3;
y)$
comes  from the permutation $\{1,2,3,4;+\}$ and its
reverse.\footnote{For this solution the locations $x_3$ and $x_4$ coincide and hence
the same solution also appears in the permutation $\{1,2,4,3\}$. But following our
prescription we count the solution only once.}
This gives
\be \label{ena4}
 g_{\rm ref}(\gamma_1,\gamma_2,\gamma_3, \gamma_3;
y) = (-1)^{k+1}\, (y - y^{-1})^{-3} \, (y^k - y^{-k})\, ,
\ee
Eq.\eqref{ena2} now gives
\be
 \label{ena5}
\begin{split}
Q&(\gamma_1+\gamma_2+2\gamma_3;y) = \Omega^{\rm S}_{\rm ref}(\gamma_1+\gamma_2+2\gamma_3;y)\\
&
+ \left\{H( \{\gamma_1,\gamma_2,\gamma_3\},\{1,1,2\};y)
+{(-1)^k \over 2(y - y^{-1})^{2}}{y^k + y^{-k}\over y + y^{-1}}\right\}
\Omega^{\rm S}_{\rm ref}(\gamma_1;y)\,
\Omega^{\rm S}_{\rm ref}(\gamma_2;y) \,
\Omega^{\rm S}_{\rm ref}(\gamma_3;y^2)
\\
&
+ \left\{H( \{\gamma_1,\gamma_2,\gamma_3,\gamma_3\}\{1,1,1,1\};y)
+ {1\over 2}\,  (-1)^{k+1}\, (y - y^{-1})^{-3} (y^k - y^{-k})\right\} \\
&
\qquad \qquad
 \times 
\Omega^{\rm S}_{\rm ref}(\gamma_1;y)\,
\Omega^{\rm S}_{\rm ref}(\gamma_2;y) \,
\Omega^{\rm S}_{\rm ref}(\gamma_3;y)^2
\, .
\end{split}
\ee
Requiring that the coefficients of 
$\Omega^{\rm S}_{\rm ref}(\gamma_1;y)
\Omega^{\rm S}_{\rm ref}(\gamma_2;y) \Omega^{\rm S}_{\rm ref}(\gamma_3;y^2)$
and
$\Omega^{\rm S}_{\rm ref}(\gamma_1;y)
\Omega^{\rm S}_{\rm ref}(\gamma_2;y)$ $ \Omega^{\rm S}_{\rm ref}(\gamma_3;y)^2$
be  polynomials in $y$, $y^{-1}$ we get
\be
\label{ena5.5}
\begin{split}
H&( \{\gamma_1,\gamma_2,\gamma_3,\gamma_3\}\{1,1,1,1\};y)
= \begin{cases} {1\over 4}\, \, k\, (y - y^{-1})^{-2} (y+y^{-1})\, \quad
\hbox{for $k$ even}\cr
-{1\over 2}\, \, k\, (y - y^{-1})^{-2} \, \quad \hbox{for $k$ odd}
\end{cases}\, 
\\
H&( \{\gamma_1,\gamma_2,\gamma_3\},\{1,1,2\};y) \\
& = \begin{cases}{1\over 4} (y - y^{-1})^{-2} (y+y^{-1})^{-1}
\left\{-(y+y^{-1})^2 + (-1)^{k/2} (y-y^{-1})^2 \right\}\, \, \quad \hbox{for $k$ even}\cr
{1\over 2}  \, (y-y^{-1})^{-2} \, \, \quad \hbox{for $k$ odd}\end{cases} 
\end{split}
\ee
Once the $H$'s have been determined we can drop the $y$ dependence
of $\Omega^{\rm S}_{\rm ref}(\gamma_1+\gamma_2+2\gamma_3)$ 
and set $\Omega^{\rm S}_{\rm ref}(\gamma_\ell;y)=1$.
This gives
\be \label{ena6}
\begin{split}
Q&(\gamma_1+\gamma_2+2\gamma_3;y)\\
=& \Omega^{\rm S}_{\rm ref}(\gamma_1+\gamma_2+2\gamma_3)
+ (y - y^{-1})^{-3}\, (y+y^{-1})^{-1}\, \,
\bigg\{ y^{-k+1} -  y^{k-1} \\
& \qquad \qquad  + {1\over 4} (k-1) (y + y^{-1})^2 (y-y^{-1})
+{1\over 4} (-1)^{k/2} (y - y^{-1})^3 \bigg\}
\quad \hbox{for $k$ even}
\\
=&\Omega^{\rm S}_{\rm ref}(\gamma_1+\gamma_2+2\gamma_3)
+ (y - y^{-1})^{-3}\, (y+y^{-1})^{-1}\, \bigg\{ y^{k-1} - y^{-k+1}
\\
& \qquad \qquad - {1\over 2}(k-1) (y^2 - y^{-2})\bigg\}  
\quad \hbox{for $k$ odd}
\end{split}
\ee
We note that both for $k$ even and odd the negative powers of $y$ in this
expression are given by
\be \label{ena6A}
Q(\gamma_1+\gamma_2+2\gamma_3;y) \simeq 
(-1)^{k+1} \, y^{-k+5} (1-y^2)^{-3} (1+y^2)^{-1}\, .
\ee

Let us now compare this prediction with an explicit computation of 
the cohomology of the Higgs branch.
Since the node $3$ carries an $U(2)$ gauge group, 
the fields $\phi_{23,\alpha}$ and $\phi_{31,\alpha}$ carry an extra
$U(2)$ index which we shall label by $s$.\footnote{Even though
we use the same symbol $s$ it should be understood that for $\phi_{23}$ it labels
the anti-fundamental representation 
of $SU(2)$ while for $\phi_{31}$ it labels the fundamental representation of
$SU(2)$.} The D-term equations for the $U(1)$ factors take the form
\ben \label{edt1e4}
\phi_{12,\alpha}^* \phi_{12,\alpha} - \phi_{31,\alpha,s}^* \phi_{31,\alpha,s}
 &=& c_1\, , \nonumber \\
  -\phi_{12,\alpha}^* \phi_{12,\alpha} + \phi_{23,\alpha,s}^* \phi_{23,\alpha,s}
&=& c_2\, , \nonumber \\
 - \phi_{23,\alpha,s}^* \phi_{23,\alpha,s} 
+ \phi_{31,\alpha,s}^* \phi_{31,\alpha,s}
&=&2c_3\ ,
\een
while the D-term equations for the $SU(2)$ gauge group further require
\be
\label{Dsu2}
  \phi_{23,\alpha,s}^* T^a_{ss'} \phi_{23,\alpha,s'}
-  \phi_{31,\alpha,s}^* T^a_{ss'} \phi_{31,\alpha,s'} 
= 0\ ,
\ee
where $T^a$ for $1=1,2,3$ are the Lie algebra generators  
(Pauli matrices in this case). The superpotential is given by
\be \label{esope4}
W= C_{\alpha\beta\gamma} \phi_{12,\alpha} \phi_{23,\beta,s}
\phi_{31,\gamma,s}\, ,
\ee
where $C_{\alpha\beta\gamma}$ are constants. 
If we ignore the last set of equations \eqref{Dsu2} then solutions
to \eqref{edt1e4} can be found by choosing:
\be \label{esole4}
\phi_{31,\alpha,s}=0, \quad \phi_{12,\alpha}^* \phi_{12,\alpha}=c_1>0, \quad
\phi_{23,\alpha.s}^* \phi_{23,\alpha,s}
= c_1+c_2>0, \quad C_{\alpha\beta\gamma} \phi_{12,\alpha} \phi_{23,\beta,s} =0\, .
\ee
This describes the complete intersection of $2c$ hypersurfaces  of degree (1,1) inside
$\CP^{a-1}\times\CP^{2b-1}$, generating a manifold of complex dimension 
$a+2b-2c-2=k-2$. At generic points on this space, the $SU(2)$ gauge symmetry
is completely broken. The space of solutions to the $SU(2)$ D-term equations
\eqref{Dsu2} modulo the action of the compact gauge group is isomorphic to the 
quotient of the semi-stable locus by the complexified gauge group $SL(2,\IC)$,
and is a complex manifold $\MM$ of dimension $k-5$. This agrees with the 
maximal negative power of $y$ in \eqref{ena6A}.
Our goal is to compute the cohomology
of this manifold $\MM$ and compare it with \eqref{ena6A}.

For this purpose, we shall first consider the cohomology of the vacuum moduli
space $\cM_0$ in the absence of superpotential, i.e. the space of solutions to the
$U(1)$ and $SU(2)$ D-term constraints \eqref{edt1e4} and \eqref{Dsu2} modulo
the gauge group $U(1)\times $U(1)$\times U(2)$.  To compute the cohomology
of $\cM_0$, we shall use 
the HN recursion method described in 
\S\ref{srepresent}.\footnote{Alternatively, one can use Maxwell-Boltzmann
statistics to compute $Q_0(\gamma; y)$ 
from a set of Abelian
quivers \cite{1011.1258}.}
Under the same assumptions as in \eqref{erange}, \eqref{ena1}, we find
that the slopes are ordered according to 
\be
\gamma_2 < \gamma_2+\gamma_3 < \gamma_2 + 2\gamma_3 < \gamma_3 <
\gamma_1+\gamma_2+2\gamma_3 < \gamma_1+\gamma_2+\gamma_3 < \gamma_1+\gamma_2 <
\gamma_1+2\gamma_3 < \gamma_1+\gamma_3 < \gamma_1\ .
\ee
Using \eqref{eq:totalset}, \eqref{eq:CIrecurs} we arrive at
\ben \label{eivalues}
&& \CI (\gamma_1+\gamma_2;w) = {(w^a - w^{-a}) (w - w^{-1})^{-2}}\, , \quad 
\CI(\gamma_2+\gamma_3;w)=0, \quad \CI(\gamma_1+\gamma_3;w)=0, \non\\
&& \CI(2\gamma_3;w)= {w^{-1} (w - w^{-1})^{-1} (w^2- w^{-2})^{-1}}\, ,\quad
\CI(\gamma_1+2\gamma_3;w)=0, \quad \CI(\gamma_2+2\gamma_3;w)=0, \non\\
&& \CI (\gamma_1+\gamma_2+\gamma_3;w)= w^{a+b+c} (w - w^{-1})^{-3}
(w^a - w^{-a}) (w^b - w^{-b})
\een
hence, for the total charge vector $\gamma=\gamma_1+\gamma_2+2\gamma_3$,
\be
\begin{split}
 \CI (\gamma;w)= & 
h(\gamma;w) - \cF(\gamma_1+\gamma_2,2\gamma_3;w)
-\cF(\gamma_1+\gamma_2+\gamma_3,\gamma_3;w)
-\cF(\gamma_1,2\gamma_3,\gamma_2;w) \\
=& w^{2c} \, (w-1/w)^{-4} (w+1/w)^{-1}\, (w^a-w^{-a})(w^b-w^{-b}) (w^{b-1}-w^{1-b})
\end{split} 
\ee
and therefore 
\be \label{eqvalue2}
\begin{split} 
Q_0(\gamma;y) =   (-1)^{a+1}\, 
y^{5-a-2b-2c} (1-y^2)^{-3} (1+y^2)^{-1} (1- y^{2a}) (1 - y^{2b}) (1-y^{2b-2}) \ .
\end{split} 
\ee
This is recognized as the Laurent polynomial of $\MM_0=\IP^{a-1}\times G(2,b)\times \IC^{2c}$, where $G(k,n)$ is the Grassmaniann of $k$ planes inside $\IC^n$, a compact $k(n-k)$-dimensional variety with Laurent polynomial given by the $q$-deformed binomial coefficient
\be \label{egiv1}
Q( G(k,n); y) =  \frac{(-y)^{-k(n-k)} [n,y]!}{[k,y]!\, [n-k,y]!}\ .
\ee
Here $[N,y]!=[1,y][2,y]\dots [N,y]$ is the $q$-deformed factorial, with $[N,y]=(1-y^{2N})/(1-y^2)$.
The three factors in $\MM_0$ correspond to the $a$ chiral fields $\phi_{12,\alpha}$ modulo $\IC^\times$, the $2b$ chiral fields $\phi_{23,\beta,s}$ modulo $GL(2)$, and the $2c$ chiral fields 
$\phi_{31,\gamma,s}$, respectively (after fixing the $GL(2)$ symmetry using the 
$\phi_{23,\beta,s}$'s).
The effect of the F-term constraints is to remove the $\IC^{2c}$ factor 
by setting the fields $\phi_{31,\gamma,s}$ to zero
and impose $2c$ equations $C_{\alpha\beta\gamma} \phi_{12,\alpha} \phi_{23,\beta,s} =0$ on the gauge-invariant 
degrees of freedom of $\phi_{23,\beta,s}$ described by $G(2,b)$.
Assuming that these $2c$ equations form a complete intersection,  the Laurent polynomial 
of  the quiver moduli space $\cM$ is therefore, up to constants and positive powers of $y$,
\be \label{eqynon}
Q(\cM;y) \simeq (-1)^{k+1} y^{-k+5} (1-y^2)^{-3} (1+y^2)^{-1} (1 - y^{2a}) (1 - y^{2b}) (1 - y^{2b-2}
)\, .
\ee
Now it follows from the inequalities \eqref{erange} that we can drop the
$y^{2a}$, $y^{2b}$ and $y^{2b-2}$
terms from \eqref{eqynon} without affecting the negative powers of $y$.
This gives
\be \label{eqynonmod}
Q(\cM;y) \simeq (-1)^{k+1} y^{-k+5} (1-y^2)^{-3} (1+y^2)^{-1} \, ,
\ee
 in perfect agreement with \eqref{ena6A}.

Before leaving this example we should draw the readers' attention
to a subtle point. We could solve
the $U(1)$ D-term constraints \eqref{edt1e4} as well as the F-constraint 
coming from the superpotential \eqref{esope4} by choosing:
\be \label{esole4alt}
\phi_{31,\alpha,s}=0, \ \  \phi_{12,\alpha}^* \phi_{12,\alpha}=c_1,\ \  
 \phi_{23,\alpha, s} = f_\alpha u_s, \ \ u_s^* u_s=1,\ \ 
f_\alpha^* f_\alpha 
= c_1+c_2, \ \  C_{\alpha\beta\gamma} \phi_{12,\alpha} f_\beta =0\, .
\ee
This gives a codimension $c$ hypersurface in 
$\CP^{a-1}\times\CP^{b-1}\times\CP^1$
spanned by $\phi_{12,\alpha}$, $f_\alpha$ and $u_s$ respectively and
has dimension $(a+b-c-1)$. This seems to be larger that the dimension 
$(a+2b-2c-2)$
of the manifold we found earlier, since we have $b<c$.
However, these solutions do not satisfy the $SU(2)$ D-term constraint
\eqref{Dsu2}, as they would require $u_s^* T^a_{ss'} u_{s'}=0$. 
This will give $u_s=0$ and hence is
inconsistent with the normalization of $u_s$ given in \eqref{esole4alt}.
Thus, the set of solutions \eqref{esole4alt} does not belong to the semi-stable
locus.

\subsection{Rank $(1,1,N)$} 

We now generalize the previous example to allow for a $U(N)$
gauge group at the third node, keeping 
$U(1)$ gauge groups at the first two nodes. We choose the FI parameters 
as in \eqref{ena1} 
and assume, for reasons to become apparent shortly, that the 
arrow multiplicities satisfy 
\be \label{easy4}
(c-b)N < a < (c-b)N+ 2b\ ,\quad b<c.
\ee
In this case it is easy to see that
\ben \label{easy1}
&& g_{\rm ref} (\gamma_1+ \gamma_2+ k_0\gamma_3, k_1\gamma_3, k_2 \gamma_3, \cdots;y ) = 0\ ,
\nonumber \\
&& \Omega^{\rm S}_{\rm ref} (\gamma_1+p\gamma_3)=0, \quad \Omega^{\rm S}_{\rm ref}
(\gamma_2+q\gamma_3)=0 \, ,
\een
for positive integers $p,q,k_1, k_2, \cdots$ and non-negative integer $k_0$. 
Recalling that $\Omega^{\rm S}_{\rm ref}(\gamma_\ell;y)=1$ for $\ell=1,2,3$, we find that
\eqref{essp1} takes the form
\be
\label{easy2}
\begin{split}
Q&(\gamma_1+\gamma_2+N\gamma_3;y) = 
\Omega^{\rm S}_{\rm ref}(\gamma_1+\gamma_2+N\gamma_3;y)
\\ &
+ \sum_{s_1, s_2, \cdots\atop \sum m s_m = N} \left(\prod_m {1\over s_m!} \right)
g_{\rm ref} (\gamma_1, \gamma_2, \gamma_3, \cdots \gamma_3, 2\gamma_3, \cdots 2\gamma_3,
\cdots; y) \prod_{m=1}^\infty \left( {1\over m}{y - y^{-1}\over y^m - y^{-m}}\right)^{s_m}
\\
& + \sum_{k_1, k_2, \cdots \atop \sum_i k_i = N} H(\{\gamma_1, \gamma_2, \gamma_3, \gamma_3,
\cdots 
\gamma_3\}, \{1, 1, k_1, k_2, \cdots\}; y) \, .
\end{split}
\ee
In the second line $s_1$ represents the number of $\gamma_3$'s, $s_2$ represents 
the number of $2\gamma_3$'s etc. in the argument of $g_{\rm ref}$. Now this form
is not suitable for determining the individual $H$'s since we have set the
$\Omega^{\rm S}_{\rm ref}(\gamma_3;y)=1$ from the beginning. These will be needed
for analyzing bigger systems which include the current quiver as a subsystem. 
However for the purpose of finding $Q(\gamma_1+\gamma_2+N\gamma_3;y)$ itself, 
we can proceed as follows.
Since by construction the $H$'s vanish as $y\to \infty$ and $y\to 0$, we see that they
do not contribute negative powers of $y$ or constant term
in a Laurent series expansion of
\eqref{easy2} around $y=0$. Thus we have
\be \label{easy3}
\begin{split}
Q&(\gamma_1+\gamma_2+N\gamma_3;y) \simeq\\
&
\sum_{s_1, s_2, \cdots\atop \sum m s_m = N} \left(\prod_m {1\over s_m!} \right)
g_{\rm ref} (\gamma_1, \gamma_2, \gamma_3, \cdots \gamma_3, 2\gamma_3, \cdots 2\gamma_3,
\cdots; y) \prod_{m=1}^\infty \left( {1\over m}{y - y^{-1}\over y^m - y^{-m}}\right)^{s_m}
\, , 
\end{split}
\ee
where $\simeq$ denotes equality of negative powers of $y$. The positive powers of $y$
in the Laurent polynomial $Q(\gamma_1+\gamma_2+N\gamma_3;y)$ 
are then found using the $y\to 1/y$ symmetry, and
the constant term is given by $\Omega^{\rm S}_{\rm ref}(\gamma_1+\gamma_2+N\gamma_3)$.

To proceed further we need to compute the Coulomb index $g_{\rm ref} (\gamma_1, \gamma_2, 
\gamma_3, \cdots, 2\gamma_3,
\cdots; y)$. For this we can proceed as in \eqref{echararrange}. Since the centers with
charge $k_i\gamma_3$ do not interact among themselves, they must sit at one of the two
possible locations  $x_{\rm m}$ and $x_{\rm r}$ in \eqref{elast2} . If we assume that the 
inequalities \eqref{easy4} are satisfied,
then the analog of \eqref{elast4} shows that all the centers carrying charge
proportional to $\gamma_3$ must in fact sit   at
$x_{\rm r}$, \i.e.\ the centers are arranged in the order 
$\{\gamma_1, \gamma_2, k_1\gamma_3,
k_2\gamma_3,
\cdots \}$ with all the $k_i\gamma_3$'s coincident,
and its reverse. This gives
\be \label{easy5}
\begin{split}
g_{\rm ref} (\gamma_1, \gamma_2, \gamma_3, & \cdots \gamma_3, 2\gamma_3, \cdots 2\gamma_3,
\cdots; y) \\
&= (-1)^{k+ \sum_m s_m -1} (y-y^{-1})^{-\sum_m s_m-1} 
\left(y^k - (-1)^{\sum_m s_m} y^{-k}\right) \, ,
\end{split}
\ee
where as before $s_m$ denotes the number of $m\gamma_3$'s in the argument
of $g_{\rm ref}$, and
\be \label{edefkagain}
k \equiv a + N(b-c) \, .
\ee
We now note that in a series expansion of \refb{easy3} around $y=0$, the
contribution from the first term inside the parantheses in \refb{easy5},
$y^k$, produces only positive powers of $y$. Thus we can drop this term
for the purpose of computing the negative powers of $y$ in \refb{easy3}.
This gives
\be \label{eqp1}
Q(\gamma_1+\gamma_2+N\gamma_3;y) \simeq (-1)^{k+1} y^{-k+1}
(1-y^2)^{-1} Q'(N;y)\, ,
\ee
where
\be \label{eqp2}
Q'(N;y) = \sum_{s_1, s_2, \cdots\atop \sum_m m s_m = N}
\prod_{m=1}^\infty \left\{{1\over s_m!} \left({1\over m} {1\over y^m - y^{-m}}\right)^{s_m}
\right\}\, .
\ee
Introducing the generating 
function\footnote{Note that since the expression for $g_{\rm ref}$ used in
\eqref{easy5} is valid only when \eqref{easy4} holds, we can use this generating
function to compute $Q(\gamma_1+\gamma_2+N\gamma_3;y)$ only
in this range.}
\be \label{easy6}
F(z;y) \equiv \sum_{N=0}^\infty z^N Q'(N;y)\, ,
\ee
and using \refb{eqp2}, we find that $F(z;y)$ is given by the $q$-deformed 
Pochhammer symbol,
\ben \label{easy7}
F(z;y) &=& \exp\left[ \sum_{m=1}^\infty z^m {1\over m} {1\over y^m - y^{-m}}\right]=
 \exp\left[ - \sum_{m=1}^\infty {1\over m}\,  z^m y^m \sum_{n=0}^\infty y^{2mn}\right]
\non\\
&=& \exp\left[\sum_{n=0}^\infty \ln (1 - z y^{2n+1})\right] = \prod_{n=1}^\infty (1 - z y^{2n+1}) 
\, .
\een
To find $Q'(N;y)$
we need to extract the coefficient of $z^N$ in \eqref{easy7}.
The coefficient of the $z^N$ term in the Pochhammer symbol is
given by $(-1)^N y^{N^2} / \{(1-y^2) (1-y^4) \cdots (1-y^{2N})\}$. 
Using \refb{eqp1} we now get
\be \label{easy8}
Q(\gamma_1+\gamma_2+N\gamma_3;y)
\simeq (-1)^{k+N+1} y^{-k+N^2+1} (1-y^2)^{-2} \prod_{n=2}^N (1-y^{2n})^{-1}
\, .
\ee

The analysis of the Higgs branch  proceeds as in \S\ref{sec112}. We arrive at the same 
set of equations 
\eqref{edt1e4}-\eqref{esole4} with the only difference that the index $s$ carried by $\phi_{31}$
labels the fundamental representation of $U(N)$ and the index $s$ carried by
$\phi_{23}$ runs over the anti-fundamental representation of $U(N)$.
The analog of \eqref{esole4} now gives a complete intersection 
of codimension $Nc$ in $\CP^{a-1}\times \CP^{Nb-1}$, of dimension $a+Nb-Nc-2=k-2$. The 
$SU(N)$ D-term constraints reduces this to $k-N^2-1$, 
in agreement with the maximum negative power of $y$ in \eqref{easy8}. 
As explained below \S\ref{eqvalue2}, $\cM$ can be alternatively described as a complete
intersection of codimension $Nc$ in $\CP^{a-1} \times G(N,b)$,
which predicts, up to constant and positive powers of $y$,
\be
Q(\gamma_1+\gamma_2+N\gamma_3;y)
\simeq  (-y)^{Nc}\, \frac{(-1)^{a-1}(y^a-y^{-a})}{y-y^{-1}}\, 
\frac{(-y)^{-N(b-N))} [b,y]!}{[N,y]!\, [b-N,y]!}\ .
\ee
This is indeed in perfect agreement with the Coulomb branch result \eqref{easy8}.

\subsection{Rank (1,2,2)}

In this final example, we consider again the same 3-node quiver \eqref{xy3node} but with a $U(1)$ factor at node
1 and  $U(2)$ factors at nodes 2 and 3. For definiteness we take
\be \label{ena1AA}
\gamma_{12}=a=3k, \quad \gamma_{23}=b=2k, \quad \gamma_{31}=c=3k,
\quad c_1= 2, \quad c_2=-2.9, \quad c_3 = 1.9\, ,
\ee
where $k$ is a positive integer. Besides \eqref{eone}, in this case we also
have the relations
\ben \label{eex2a}
&&
\Omega^{\rm S}_{\rm ref}(\gamma_1+2\gamma_2) = 
\Omega^{\rm S}_{\rm ref}(\gamma_1+2\gamma_3)
= 0\, ,  \nonumber \\
&&\Omega_{\rm scaling}(\gamma_1+2\gamma_2) = 
\Omega_{\rm scaling}(\gamma_1+2\gamma_3)
= 0\, , 
\een
since the corresponding subquivers do not have closed oriented loops.
This leads to
\be \label{eex2b}
\begin{split}
Q&(\gamma_1+2\gamma_2+2\gamma_3;y) 
= \Omega^{\rm S}_{\rm ref}(\gamma_1+2\gamma_2+2\gamma_3;y)
+ \Omega_{\rm scaling}(\gamma_1+2\gamma_2+2\gamma_3;y) \\
&
 + g_{\rm ref}(\gamma_1+\gamma_2+\gamma_3, \gamma_2, \gamma_3;y)
\left[ \Omega^{\rm S}_{\rm ref}(\gamma_1+\gamma_2+\gamma_3;y)
+ \Omega_{\rm scaling}(\gamma_1+\gamma_2+\gamma_3;y)
\right]\\
&
\qquad \qquad \qquad \qquad \times \Omega^{\rm S}_{\rm ref}(\gamma_2;y)
\, \Omega^{\rm S}_{\rm ref}(\gamma_3;y) \\
&
+ g_{\rm ref}(\gamma_1+2\gamma_2+\gamma_3, \gamma_3;y)
\left[ \Omega^{\rm S}_{\rm ref}(\gamma_1+2\gamma_2+\gamma_3;y)
+ \Omega_{\rm scaling}(\gamma_1+2\gamma_2+\gamma_3;y)
\right]\, \Omega^{\rm S}_{\rm ref}(\gamma_3;y) \\
&
 + g_{\rm ref}(\gamma_1+\gamma_2+2\gamma_3, \gamma_2;y)
\left[ \Omega^{\rm S}_{\rm ref}(\gamma_1+\gamma_2+2\gamma_3;y)
+ \Omega_{\rm scaling}(\gamma_1+\gamma_2+2\gamma_3;y)
\right] \Omega^{\rm S}_{\rm ref}(\gamma_2;y)\\
&
 + \frac{1}{4(y+1/y)}   \, 
 g_{\rm ref}(\gamma_1,\gamma_2, \gamma_2, 2\gamma_3;y)\, 
\Omega^{\rm S}_{\rm ref}(\gamma_1;y) \, 
\Omega^{\rm S}_{\rm ref}(\gamma_2;y)^2\,
\Omega^{\rm S}_{\rm ref}(\gamma_3;y^2)
\\
&
 +  \frac{1}{4(y+1/y)}  g_{\rm ref}(\gamma_1,2\gamma_2, \gamma_3, \gamma_3;y)\, \Omega^{\rm S}_{\rm ref}(\gamma_1;y) \,
 \Omega^{\rm S}_{\rm ref}(\gamma_2;y^2)\,
\Omega^{\rm S}_{\rm ref}(\gamma_3;y)^2\\
& 
+ \frac{1}{4(y+1/y)^2}  \, 
g_{\rm ref}(\gamma_1,2\gamma_2, 2\gamma_3;y)\,
\Omega^{\rm S}_{\rm ref}(\gamma_1;y)\, 
\Omega^{\rm S}_{\rm ref}(\gamma_2;y^2) \,
\Omega^{\rm S}_{\rm ref}(\gamma_3;y^2) \\
&
 + {1\over 4} g_{\rm ref}(\gamma_1,\gamma_2,\gamma_2,\gamma_3, \gamma_3;y)\,
 \Omega^{\rm S}_{\rm ref}(\gamma_1;y) \,
 \Omega^{\rm S}_{\rm ref}(\gamma_2;y)^2\,
 \Omega^{\rm S}_{\rm ref}(\gamma_3;y)^2\, .
\end{split}
\ee
In this case from \eqref{e2node} we have
\be \label{extrab}
g_{\rm ref}(\gamma_1+2\gamma_2+\gamma_3, \gamma_3;y) =0, 
\qquad g_{\rm ref}(\gamma_1+\gamma_2+2\gamma_3, \gamma_2;y) =0\, .
\ee
Using \eqref{extrab}, \eqref{ess2} and that 
$\Omega^{\rm S}_{\rm ref}(2\gamma_2;y)=
\Omega^{\rm S}_{\rm ref}(2\gamma_3;y)=0$, we can reduce \eqref{eex2b} to
\be
\begin{split}
\label{eex2c}
Q&(\gamma_1+2\gamma_2+2\gamma_3;y)=
\Omega^{\rm S}_{\rm ref}(\gamma_1+2\gamma_2+2\gamma_3;y)\\ 
&
+H(\{\gamma_1, \gamma_2, \gamma_2, \gamma_3, \gamma_3\},\{1,1,1,1,1\}; y) 
\Omega^{\rm S}_{\rm ref}(\gamma_1;y)\, 
\Omega^{\rm S}_{\rm ref}(\gamma_2;y)^2\,
 \Omega^{\rm S}_{\rm ref}(\gamma_3;y)^2 \\ 
 &
+ H(\{\gamma_1, \gamma_2, \gamma_3, \gamma_3\},\{1,2,1,1\}; y) \,
\Omega^{\rm S}_{\rm ref}(\gamma_1;y) \,
\Omega^{\rm S}_{\rm ref}(\gamma_2;y^2)\,
 \Omega^{\rm S}_{\rm ref}(\gamma_3;y)^2\\ 
 &
 + H(\{\gamma_1, \gamma_2, \gamma_2, \gamma_3\},\{1,1,1,2\}; y) 
 \Omega^{\rm S}_{\rm ref}(\gamma_1;y)\, 
\Omega^{\rm S}_{\rm ref}(\gamma_2;y)^2\,
 \Omega^{\rm S}_{\rm ref}(\gamma_3;y^2)\, \\ 
 &
 + H(\{\gamma_1, \gamma_2, \gamma_3\},\{1,2,2\}; y)\,
 \Omega^{\rm S}_{\rm ref}(\gamma_1;y)\, 
\Omega^{\rm S}_{\rm ref}(\gamma_2;y^2)\,
 \Omega^{\rm S}_{\rm ref}(\gamma_3;y^2)\\
&
 + H(\{\gamma_1+\gamma_2+\gamma_3, \gamma_2, \gamma_3\},\{1,1,1\}; y) 
 \Omega^{\rm S}_{\rm ref}(\gamma_1+\gamma_2+\gamma_3; y)\, 
\Omega^{\rm S}_{\rm ref}(\gamma_2;y)\,
 \Omega^{\rm S}_{\rm ref} (\gamma_3;y)\\
& 
+ g_{\rm ref}(\gamma_1+\gamma_2+\gamma_3, \gamma_2, \gamma_3;y)\,
\Omega^{\rm S}_{\rm ref}(\gamma_1+\gamma_2+\gamma_3;y)\,
\Omega^{\rm S}_{\rm ref}(\gamma_2;y)\,
\Omega^{\rm S}_{\rm ref}(\gamma_3;y)\, \\ 
&
 + g_{\rm ref}(\gamma_1+\gamma_2+\gamma_3, \gamma_2, \gamma_3;y)\,
H(\{\gamma_1, \gamma_2, \gamma_3\},\{1,1,1\}; y)\, 
\Omega^{\rm S}_{\rm ref}(\gamma_1;y)\,
\Omega^{\rm S}_{\rm ref}(\gamma_2;y)^2\,
 \Omega^{\rm S}_{\rm ref}(\gamma_3;y)^2
  \\
& + \frac{1}{4(y+1/y)}  \, 
g_{\rm ref}(\gamma_1,\gamma_2, \gamma_2, 2\gamma_3;y)\,
\Omega^{\rm S}_{\rm ref}(\gamma_1;y) \,
\Omega^{\rm S}_{\rm ref}(\gamma_2;y)^2\,
\Omega^{\rm S}_{\rm ref}(\gamma_3;y^2)\\
&
 + \frac{1}{4(y+1/y)}  \, 
 g_{\rm ref}(\gamma_1,2\gamma_2, \gamma_3, \gamma_3;y)\,
\Omega^{\rm S}_{\rm ref}(\gamma_1;y) \,
\Omega^{\rm S}_{\rm ref}(\gamma_2;y^2)\,
\Omega^{\rm S}_{\rm ref}(\gamma_3;y)^2
\\
& +  \frac{1}{4(y+1/y)^2}  \, 
g_{\rm ref}(\gamma_1,2\gamma_2, 2\gamma_3;y)\, 
\Omega^{\rm S}_{\rm ref}(\gamma_1;y)\,
 \Omega^{\rm S}_{\rm ref}(\gamma_2;y^2)\,
  \Omega^{\rm S}_{\rm ref}(\gamma_3;y^2) \\
  & 
+ {1\over 4} g_{\rm ref}(\gamma_1,\gamma_2,\gamma_2,\gamma_3, \gamma_3;y)\,
\Omega^{\rm S}_{\rm ref}(\gamma_1;y)\, \Omega^{\rm S}_{\rm ref}(\gamma_2;y)^2\,
 \Omega^{\rm S}_{\rm ref}(\gamma_3;y)^2\, .
\end{split}
\ee
The coefficients
$g_{\rm ref}(\gamma_1+\gamma_2+\gamma_3, \gamma_2, \gamma_3;y)$,
$H(\{\gamma_1+\gamma_2+\gamma_3, \gamma_2, \gamma_3\},
\{1,1,1\}; y)$,  $H(\{\gamma_1, \gamma_2, \gamma_3\},$ 
$\{1,1,1\}; y)$ and 
$g_{\rm ref}(\gamma_1,2\gamma_2, 2\gamma_3;y)$
can all be evaluated from the results of \S\ref{s3node}
using the assignments $(a,b,c)=(k,k,2k)$ for the first two cases,
$(3k,3k,2k)$ in the third case and
$(6k,6k,8k)$ in the last case (note that some permutations of the nodes
are necessary in order to satisfy  \eqref{eeight}).
Thus from
\eqref{eks2}, \eqref{eks4} and the analysis of \S\ref{s3.1cou} we have
\ben \label{eghnon}
&& g_{\rm ref}(\gamma_1+\gamma_2+\gamma_3, \gamma_2, \gamma_3;y)
= 0\, , \nonumber \\
&& H(\{\gamma_1+\gamma_2+\gamma_3, \gamma_2, \gamma_3\},
\{1,1,1\}; y)
= 0\, , \nonumber \\
&& H(\{\gamma_1, \gamma_2, \gamma_3\},
\{1,1,1\}; y) = -2 (y-y^{-1})^{-2}\, , \nonumber \\
&& g_{\rm ref}(\gamma_1,2\gamma_2, 2\gamma_3;y) = (y-y^{-1})^{-2}\, 
\left(y^{4k} + y^{-4k}\right)\, .
\een
Note that for the first two cases we cannot directly apply
\refb{eks2}, \refb{eks4} since we have
$2k=k+k$ and the triangle inequality is saturated.
Instead we use
the analysis given at the end of \S\ref{s3.1cou} which leads to
the vanishing of $g_{\rm ref}$ and hence also $H$.\footnote{Alternatively
we could deform $\gamma_{12}$, $\gamma_{23}$ and $\gamma_{31}$ slightly
away from those given in \refb{ena1AA} -- {\it e.g.} by adding small
even integers to them for large $k$ -- so that  for the triple
$(\gamma_1+\gamma_2+\gamma_3, \gamma_2, \gamma_3)$ strict  triangle
inequality holds, and take the limit back to the
original values of $\gamma_{12}$, $\gamma_{23}$ and $\gamma_{31}$
at the end of the calculation. In that case we could use \refb{eks2}, \refb{eks4} 
for the triple $(\gamma_1+\gamma_2+\gamma_3, \gamma_2, \gamma_3)$.
The final result is unaffected by this.}
Finally a direct computation gives
\ben \label{egrest}
&& g_{\rm ref}(\gamma_1,\gamma_2, \gamma_2, 2\gamma_3;y) = - (y - y^{-1})^{-3}
(y^{4k} - y^{-4k}) \, ,
\nonumber \\
&& g_{\rm ref}(\gamma_1, 2\gamma_2, \gamma_3, \gamma_3;y) = 
-(y - y^{-1})^{-3} (y^{4k} - y^{-4k})\, , \nonumber \\
&& g_{\rm ref}(\gamma_1,\gamma_2, \gamma_2, \gamma_3, \gamma_3;y) = 
(y - y^{-1})^{-4} (y^{4k} + y^{-4k})\, .
\een
with the contributions coming from the permutations
$\{4,1,2,3;+\}$ and its reverse for the first term, 
$\{2,1,3,4;-\}$ and its reverse for the second term and
$\{2,3,1,4,5;+\}$ and its reverse for the last term.

Requiring that the coefficients of 
$\Omega^{\rm S}_{\rm ref}(\gamma_1;y) 
\Omega^{\rm S}_{\rm ref}(\gamma_2;y)^2
 \Omega^{\rm S}_{\rm ref}(\gamma_3;y)^2$, 
$\Omega^{\rm S}_{\rm ref}(\gamma_1;y) 
\Omega^{\rm S}_{\rm ref}(\gamma_2;y^2)$
 $\Omega^{\rm S}_{\rm ref}(\gamma_3;y)^2$,
 $\Omega^{\rm S}_{\rm ref}(\gamma_1;y) 
\Omega^{\rm S}_{\rm ref}(\gamma_2;y)^2
 \Omega^{\rm S}_{\rm ref}(\gamma_3;y^2)$ and
 $\Omega^{\rm S}_{\rm ref}(\gamma_1;y) 
\Omega^{\rm S}_{\rm ref}(\gamma_2;y^2)
 \Omega^{\rm S}_{\rm ref}(\gamma_3;y^2)$ are Laurent polynomials
 in $y$, we get
 \ben \label{ehvalueAA}
&& H(\{\gamma_1, \gamma_2, \gamma_2, \gamma_3, \gamma_3\},
\{1,1,1,1,1\}; y)=  (y - y^{-1})^{-4} \left\{-k^2 (y - y^{-1})^2 - {1\over 2}  \right\}
\nonumber \\ &&
H(\{\gamma_1, \gamma_2, \gamma_3, \gamma_3\},
\{1,2,1,1\}; y) =  {k\over 2} \, (y - y^{-1})^{-2}\, ,
\nonumber \\ &&
H(\{\gamma_1, \gamma_2, \gamma_2, \gamma_3\},
\{1,1,1,2\}; y) =  {k\over 2} \, (y - y^{-1})^{-2}\, ,
\nonumber \\ &&
H(\{\gamma_1, \gamma_2,  \gamma_3\},
\{1,2,2\}; y) = -{1\over 2} \, (y+y^{-1})^{-2} (y - y^{-1})^{-2}\, .
\een
Once  the $H$'s  have been determined we can set $\Omega(\gamma_\ell;y)=1$, and
get, from \eqref{eex2c}, 
\ben \label{eex2h}
&& Q(\gamma_1+2\gamma_2+2\gamma_3;y)
= \Omega^{\rm S}_{\rm ref}(\gamma_1+2\gamma_2+2\gamma_3) +
 (y - y^{-1})^{-4} (y + y^{-1})^{-2} \nonumber \\ && \qquad \qquad \times
\left[-k^2 y^4-{k^2}{y^{-4}}+2 k^2-2
   k+y^{2-4 k}+y^{4 k-2}+k y^4+{k}{y^{-4}}-y^2-{y^{-2}}
\right]\, . \nonumber \\
\een
It is easy to verify that the term inside the square bracket has 
$(y-y^{-1})^4 (y+y^{-1})^2$ as a factor and hence \eqref{eex2h}
describes a Laurent polynomial in $y$. The negative powers of
$y$ in this expansion are given by 
\be \label{enegA}
Q(\gamma_1+2\gamma_2+2\gamma_3;y) \simeq 
y^{-4k+8} (1-y^2)^{-4} (1+y^2)^{-2}\, .
\ee 

Let us now compare this result with an explicit computation of the cohomology
of the Higgs branch.
In this case the nodes 2 and $3$ carry  $U(2)$ gauge groups. As a result
the fields $\phi_{12,\alpha}$ and $\phi_{31,\alpha}$ carry an extra
$U(2)$ index each, 
and $\phi_{23,\alpha}$
carries an extra pair of $U(2)$ indices. The $U(1)$ 
D-term equations take the form:
\ben \label{edt1e4B}
 \phi_{12,\alpha,s}^* \phi_{12,\alpha,s} - \phi_{31,\alpha,s'}^* \phi_{31,\alpha,s'}
 &=& 2\, , \nonumber \\
  -\phi_{12,\alpha,s}^* \phi_{12,\alpha,s} + \phi_{23,\alpha,s,t}^* 
\phi_{23,\alpha,s,t}
&=& -5.8\, , \nonumber \\
 - \phi_{23,\alpha,s,t}^* \phi_{23,\alpha,s,t} 
+ \phi_{31,\alpha,t}^* \phi_{31,\alpha,t}
&=&3.8
\een
while the $SU(2)\times SU(2)$ D-term equations  require
\ben 
\label{Dsu2su2}
  \phi_{23,\alpha,s,t}^* T^a_{ss'} \phi_{23,\alpha,s',t}
-  \phi_{12,\alpha,s}^*  T^a_{ss'} \phi_{12,\alpha,s'} &=& 0\,
\nonumber \\
 \phi_{23,\alpha,s,t}^* T^a_{tt'} \phi_{23,\alpha,s,t'}
-  \phi_{31,\alpha,t}^* T^a_{tt'} \phi_{31,\alpha,t'} &=& 0\, .
\een
The superpotential is given by
\be \label{esope4B}
W= C_{\alpha\beta\gamma} \phi_{12,\alpha,s} \phi_{23,\beta,s,t}
\phi_{31,\gamma,t}\, .
\ee
If we ignore \eqref{Dsu2su2}  then a solution to \eqref{edt1e4B} and
\eqref{esope4B} can be found by choosing:
\be \label{esole4B}
\phi_{23,\alpha,s,t}=0, \quad \phi_{12,\alpha,s}^* \phi_{12,\alpha,s}=6, \quad
\phi_{31,\alpha,t}^* \phi_{31,\alpha,t}
= 4, \quad C_{\alpha\beta\gamma} \phi_{12,\alpha,s} \phi_{31,\gamma,t} =0\, .
\ee
This describes a complete intersection of $8k$ hypersurfaces of 
degree $(1,1)$ in
$\CP^{6k-1}\times\CP^{6k-1}$, hence is a 
manifold of complex dimension $4k-2$. The $SU(2)\times SU(2)$
D-term constraints \eqref{Dsu2su2} together with the identification under
gauge transformations lead to a manifold $\cM$ of
complex dimension $4k-8$. 
This agrees with the maximal negative power of $y$ in
\eqref{enegA}.
Our goal is to compute the cohomology
of this manifold and compare it with \eqref{enegA}.

For this purpose, we shall first consider the cohomology of the vacuum moduli
space $\cM_0$ in the absence of superpotential, i.e. the space of solutions to the
 D-term constraints \eqref{edt1e4B} and \eqref{Dsu2su2} modulo
the gauge group $U(1)\times U(2)\times U(2)$.  To compute the cohomology
of $\cM_0$, we shall use 
the HN recursion method described in \S\ref{srepresent}.
Under the same assumptions \eqref{ena1AA}, we find that the  slopes are ordered
according to 
\ben
&&\gamma_2<2\gamma_2+\gamma_3 < 
\gamma_1 + 2\gamma_2 
<\gamma_2+\gamma_3 < 
\gamma_1+2\gamma_2+\gamma_3 < \gamma_1+\gamma_2<
\gamma_1+2\gamma_2+2\gamma_3<\gamma_2+2\gamma_3 \non\\
&& 
<\gamma_1+\gamma_2+\gamma_3 < \gamma_1+\gamma_2+2\gamma_3<
\gamma_3 < \gamma_1+2\gamma_3 <
\gamma_1+\gamma_3<\gamma_1\ .
\een
Thus we find
\ben
&& \CI(\gamma_1+\gamma_3;w) = \CI(\gamma_2+\gamma_3;w)
=\CI(\gamma_1+2\gamma_3;w)=\CI(\gamma_2+2\gamma_3;w)= 0\, ,\non\\
&&
\CI(2\gamma_2+\gamma_3;w)=\CI(2\gamma_2+2\gamma_3;w)=0 \, ,\non\\
&& \CI(\gamma_1+\gamma_2;w) = (w^a - w^{-a}) / (w - w^{-1})^2, \non\\
&& \CI(\gamma_1+2\gamma_2;w) = (w^a - w^{-a}) (w^{a-1} - w^{1-a}) (w - w^{-1})^{-3}
(w + w^{-1})^{-1}\, ,\non \\
&& \CI(\gamma_1+\gamma_2+\gamma_3;w) = w^b \, (w^a - w^{-a}) (w^c - w^{-c}) (w - w^{-1})^{-3}
\, ,\non \\ 
&& \CI(\gamma_1+2\gamma_2+\gamma_3;w) =w^{2b} (w^a - w^{-a}) (w^{a-1} - w^{1-a})
(w^c - w^{-c}) (w - w^{-1})^{-4}(w + w^{-1})^{-1}\, ,\non \\
&&  \CI(\gamma_1+\gamma_2+2\gamma_3;w) = w^{2b} (w^a - w^{-a})
(w^c - w^{-c}) (w^{c-1} - w^{1-c})(w - w^{-1})^{-4}(w + w^{-1})^{-1} \, ,\non \\
\een
and, for the total charge vector $\gamma=\gamma_1+2\gamma_2+2\gamma_3$,
\ben
\CI(\gamma;w)&=&h(\gamma;w)
-\cF(\gamma_1,2\gamma_3,2\gamma_2;w) 
- \cF(\gamma_3,\gamma_1+2\gamma_2+\gamma_3;w)\non\\
&& -\cF(\gamma_3,\gamma_1+\gamma_2+\gamma_3,\gamma_2;w)
-\cF(2\gamma_3,\gamma_1+2\gamma_2;w) \non\\
&&-\cF(2\gamma_3,\gamma_1+\gamma_2,\gamma_2;w)
-\cF(\gamma_1+\gamma_2+2\gamma_3,\gamma_2;w)
\een
finally arriving at 
\be \label{elast}
Q_0(\gamma;y)
= y^{-4b-2a-2c+8} (1 - y^{2a})(1 - y^{2a-2})
(1-y^{2c}) (1-y^{2c-2})(1-y^2)^{-4}(1+y^2)^{-2}\, .
\ee
It follows that the Betti numbers of $\cM_0$ are given by
\be \label{egiv1A}
\sum_p b_p (-y)^p = (1-y^2)^{-4} (1+y^2)^{-2} (1 - y^{2a})(1 - y^{2a-2}
) (1 - y^{2c}) (1 - y^{2c-2}
)\, ,
\ee
where
\be \label{eabcdef}
a=\gamma_{12} = 3k, \quad b = \gamma_{23}=2k, \quad c=\gamma_{31}=3k\, .
\ee
This is recognized as the Laurent polynomial of $\cM_0=G(2,a)\times G(2,c)\times \IC^{4b}$,
where each factor is parametrized by $\phi_{12,\alpha,s}, \phi_{31,\gamma,s},\phi_{23,\beta,s,t}$,
respectively, giving further evidence that the HN recursion method works for quivers with loops
but zero superpotential. The effect of the F-term constraints is to remove the factor $\IC^{4b}$
and impose $4b$ constraints $C_{\alpha\beta\gamma} \phi_{12,\alpha,s} \phi_{31,\gamma,t} =0$
in $G(2,a)\times G(2,c)$. Assuming that these constraints are in complete intersection, we find
\ben \label{eqynonA}
Q(\cM;y) &\simeq&  y^{-4k+8} (1-y^2)^{-4} (1+y^2)^{-2} (1 - y^{6k}) (1 - y^{6k-2})
(1 - y^{6k}) (1 - y^{6k-2})\non\\
&\simeq& y^{-4k+8} (1-y^2)^{-4} (1+y^2)^{-2}\, ,
\een
 in perfect agreement with \eqref{enegA}.

\acknowledgments

We are grateful to M. Berkooz, A. Collinucci and S. El-Showk for valuable discussions. 
J.M. thanks the Amsterdam Summer Workshop on String 
Theory for its hospitality during part of this work. 
B.P. and A.S. 
wish to express their gratitude to the ICTS in Bangalore for generous hospitality 
during part of this work. The work of A.S.  was
supported in part by the project 11-R\&D-HRI-5.02-0304
and the J. C. Bose fellowship of 
the Department of Science and Technology, India.

\end{document}